\documentclass[a4paper,onecolumn,showpacs,notitlepage,floatfix,nofootinbib,superscriptaddress,prd]{revtex4-2}

\pdfoutput=1

\usepackage[utf8]{inputenc}

\usepackage{xcolor}

\usepackage{graphicx}
\usepackage{dcolumn}
\usepackage{bm}
\usepackage{mathtools}
\usepackage{amsfonts}
\usepackage{graphicx, caption, subcaption}
\usepackage{ulem}

\newcommand{\nn}{\nonumber}
\newcommand{\nl}{\nonumber \\}

\newcommand{\be}{\begin{eqnarray}}
\newcommand{\ee}{\end{eqnarray}}

\newcommand{\logt}{\log_{(2)}}
\newcommand{\ppd}{\partial}

\newcommand{\lsim}{\raisebox{-0.13cm}{~\shortstack{$<$ \\[-0.07cm]
      $\sim$}}~}

\begin{document}

\author{Syo Kamata}
\email{skamata11phys@gmail.com}
\affiliation{National Centre for Nuclear Research, 02-093 Warsaw, Poland}
\author{Jakub Jankowski}
\email{Jakub.Jankowski@fuw.edu.pl}
\affiliation{Institute of Theoretical Physics, University of Wroc{\l}aw, 50-204 Wroc{\l}aw, Poland}
\affiliation{Polish-Japanese Academy of Information Technology,02-008 Warsaw Poland}
\author{Mauricio Martinez}
\email{mmarti11@ncsu.edu}
\affiliation{Department of Physics, North Carolina State University, Raleigh, NC 27695, USA}

\title{Novel features of attractors and transseries in non-conformal Bjorken flows}
\begin{abstract} 
In this work we investigate the impact of conformal symmetry breaking on hydrodynamization of a far-from-equilibrium fluid. We find a new kind of transseries solutions for the non-conformal hydrodynamic equations of a longitudinal boost invariant expanding plasma. The new transseries solutions unveil a rich physical structure which arises due to the interplay of different physical scales. In the perfect fluid case the non-conformal speed of sound
slows down the cooling of the temperature due to the emergence of logarithmic corrections that depends on the mass of the particle. These terms propagate into the perturbative and non-perturbative sectors of the transseries once viscous corrections are included. The logarithmic mass contributions increase the asymptotic value of the Knudsen number while decreasing the damping rate of the transient non-hydrodynamic modes and thus, yielding to an extremely slow hydrodynamization process where flow lines merge to their forward attractor at extremely late times. The early time free streaming expansion is modified and receives \-lo\-ga\-rithmic mass corrections induced by the shear-bulk couplings. The global flow structure and numerical analyses carried out in our work demonstrate the existence of the early and late-time attractors for the shear viscous tensor and bulk viscous pressure.
\end{abstract}

\maketitle

\tableofcontents

\section{Introduction} 

The success of hydrodynamical models in high energy nuclear collisions has attracted the interest of the theoretical physics community to understand the foundations of relativistic fluid dynamics at extreme conditions. One of the most relevant advances towards determining the validity of hydrodynamics in far-from-equilibrium regimes was made by of Heller and Spalinski~\cite{Heller:2015dha}. In their seminal work the authors investigated the convergence properties of the hydrodynamic gradient expansion for a conformal fluid which undergoes longitudinal Bjorken boost invariant expansion~\cite{Bjorken:1982qr}. Within their approach, the authors showed that this perturbative expansion has zero radius of convergence and thus, it opened to the possibility of a new systematic resummation scheme. The authors also provided strong evidence of the existence of an attractor since most of the initial numerical data merged rapidly into a unique single universal line. The late-time behavior of this universal line is described by a few terms of the perturbative gradient expansion. Surprisingly, it was noticed that the Navier Stokes regime is reached out rapidly when the pressure anisotropies are extremely large.
All together, these results opened the possibility to understand the emergent fluid behavior from an initially prepared non-equilibrium state. This non-linear relaxation process is usually known as hydrodynamization. Moreover, these results strongly support the idea of a new theory for far-from-equilibrium fluids whose physical properties are still unknown~\cite{Romatschke:2017vte}. Subsequent works have verified and extended these results to more generic conformal setups in different coupling regimes (see recent reviews in this subject~\cite{Florkowski:2017olj,Soloviev:2021lhs,Romatschke:2017ejr,Berges:2020fwq} and references therein). 

The resummation scheme of the divergences introduced in Ref.~\cite{Heller:2015dha} is based on the mathematical theory of resurgence and transseries~\cite{costin2008asymptotics,Marino:2015yie,edgar2010transseries,Dorigoni:2014hea,delabaere2016divergent,sauzin2014introduction,Aniceto:2018bis}. Transseries are defined as a sort of generalizations of asymptotic perturbative expansions  where functions with stronger/weaker convergence, such as exponential~\cite{costin1998} or logarithms, are introduced together with the usual power law perturbative expansions of the coupling constant~\cite{Schiappa:2013opa,Garoufalidis:2010ya,Aniceto:2011nu,Aniceto:2018bis}. These mathematical objects have been not only extremely useful to give very accurate numerical approximations to exact solutions of ordinary and partial differential equations (ODEs and PDEs respectively) but these also tells us about the physics not encoded in the usual perturbative expansion. For instance, the decay rate at which the non-hydro modes die off is associated with  non-perturbative contributions around the hydrodynamic gradient expansion. When these non-perturbative contributions are resummed completely to all orders, non-equilibrium constitutive relations emerge in far-from-equilibrium regimes such that the transport coefficients get effectively renormalized~\cite{Behtash:2018moe,Behtash:2019txb,Behtash:2020vqk,Blaizot:2020gql,Blaizot:2021cdv}. In these situations the fluid features a neat transient non-newtonian behavior while hydrodynamizing~\cite{Behtash:2018moe,Behtash:2019txb,Behtash:2020vqk}. These type of non-perturbative effects become important when going backward in the flow time to intermediate regimes where the gradient expansion fails and thus, these terms are regarded as needed. Furthermore, in the context of resurgence it is also known that perturbative and non-perturbative terms which belong to a particular class of transseries are related to each other through a relationship called a resurgent relation~\cite{costin2008asymptotics,Marino:2015yie,edgar2010transseries,Dorigoni:2014hea,delabaere2016divergent,sauzin2014introduction,Aniceto:2018bis}. In this sense, transseries gives us the relevant physical information which goes beyond the perturbative asymptotic series from the perturbative data itself.

On the other hand, it has been noticed that the transseries solutions are intrinsically connected with the existence of an attractor. In the field of dynamical systems the concept of an attractor is very well understood~\cite{kloeden2011nonautonomous,caraballo2016applied}. Within this framework the attractor is an invariant subspace of the phase space of dynamical variables, e.g. temperature and dissipative fluid variables, to which neighboring flows attract monotonically and sharply~\cite{Behtash:2017wqg,Behtash:2018moe,Behtash:2019txb,Behtash:2019qtk,Du:2022bel,Heller:2020anv,Heller:2020anv}. This view of the attractor does not depend on the particular model or coupling regime since it has solid mathematical and physical foundations~\cite{kloeden2011nonautonomous,caraballo2016applied}. For instance, for a Bjorken expanding fluid the resurgence analysis~\cite{Heller:2015dha,Aniceto:2018uik,Spalinski:2018mqg,
Spalinski:2017mel,Heller:2016rtz,Aniceto:2015mto,Basar:2015ava,
Casalderrey-Solana:2017zyh,Romatschke:2017vte,Blaizot:2020gql,Blaizot:2021cdv} showed that the set of initial conditions located in the basin of attraction of the attractor drop off as the flow time $w=\tau T$ (being $T$ the temperature) increases, and the dynamical variables acquire interesting universal properties which characterize the attractor. The flow time variable $w$ is related with the inverse Knudsen number $Kn^{-1}$ in this case. The pullback attraction starts at very small values of the flow time $w$ (UV) where the longitudinal pressure almost vanishes. In this region of the phase space, the rapid expansion of the fluid dominates over the collisions of the particles so most of the non-hydro modes are present and the plasma expands ala free streaming. The transseries solutions in this regime are power law series of the inverse Knudsen number with a finite radius of convergence~\cite{Behtash:2019txb,Kurkela:2019set,Blaizot:2019scw,Behtash:2020vqk,Blaizot:2020gql,Blaizot:2021cdv}. The radius of convergence of this particular UV transseries grows linearly with the ratio of the shear viscosity over entropy density~\cite{Behtash:2020vqk}. Now, although the attractor is up to some extent a global object in the phase space, it can also be partially understood from the transseries structure in the late flow time regime (IR)~\footnote{Through this paper we shall call to the early flow time and late flow time  asymptotic limits as the UV and IR kinematic regimes respectively.}. As a matter of fact, it is suggested that the IR sharp convergent property for neighboring flows arises from the rapid decay of the non-hydrodynamic modes in the transseries solutions~\cite{Aniceto:2018uik,Spalinski:2018mqg,
Spalinski:2017mel,Buchel:2016cbj,Heller:2016rtz,Aniceto:2015mto,Basar:2015ava,
Casalderrey-Solana:2017zyh,Romatschke:2017vte,Blaizot:2020gql,Blaizot:2021cdv,Du:2021fok}.

All these sharp convergent properties of the non-equilibrium attractor are valid only for kinetic conformal setups where the inverse Knudsen number $w$ is chosen as the flow time. The resurgent analysis in terms of $w$ as the flow time helps to reduce the number of variables of the dynamical system~\cite{Aniceto:2018uik,Spalinski:2018mqg,
Spalinski:2017mel,Heller:2016rtz,Basar:2015ava,
Casalderrey-Solana:2017zyh,Blaizot:2020gql,Blaizot:2021cdv}. From the physics point of view, this helps to understand the universal properties that characterize the system. However, one might lose sight of some other properties of the dynamical system like its stability. The reason comes from the fact that the boost invariant time $\tau$ and the inverse Knudsen number $w$ are not globally one-to-one to each other.
In this sense, the flow structure in the phase space of dynamical variables as well as the convergent properties of its solutions depend on the choice of the flow time~\cite{Behtash:2019qtk,Behtash:2019txb,Behtash:2020vqk}. 

Most of the lessons learned recently on the convergence of the hydrodynamic gradient series in rapidly expanding fluids in weakly and strongly coupled models assume that either the particles are massless or the system being invariant under the conformal group respectively~\cite{Florkowski:2017olj,Soloviev:2021lhs,Romatschke:2017ejr,Berges:2020fwq}. In those scenarios the only dynamical variables are the energy density (or temperature) and the shear viscous tensor. The breaking of conformal symmetry plays a relevant role in the hydrodynamical modeling of heavy ion collisions (see discussion in Ref.~\cite{Monnai:2021kgu} and references therein). Motivated by this, we develop a novel theoretical approach to analyze different asymptotic regimes in non-conformal fluids. To the best of our knowledge these mathematical tools are discussed for first time in the literature and generalize the well known transasymptotic techniques introduced in previous works~\cite{Basar:2015ava,Behtash:2018moe,Behtash:2019qtk,Behtash:2019txb,Behtash:2020vqk,costin1998,costin2008asymptotics}. 

In this work we study the second order Chapman-Enskog non-conformal fluid dynamics~\cite{2014ARM,2014SWRM,2014SSC}. Our approach does not focus necessarily on the possible unphysical features of this particular truncation scheme~\cite{Jaiswal:2021uvv}). Rather, we present an unified systematic understanding of these non-linear differential equations (ODEs) from the perspective of transasymptotics and dynamical systems. In our case, the conformal symmetry is explicitly breaking by considering that the plasma constituents have a constant mass $m$ which consequently leads to a complex equation of state. More remarkable,  a new dynamical variable emerges when the conformal symmetry is broken: the viscous bulk pressure. The presence of the bulk pressure increases the dimensionality of the phase space so the structure of the dynamical system, its stability properties and its physical description are expected to dramatically change. Furthermore, when comparing with the massless case, the conformal symmetry breaking induces highly non-trivial changes in the flow structure in the UV and IR regimes. This fact strongly suggests that the transseries structure must suffer dramatic changes. In other words, one can expect that transseries structure is determined by symmetry through structure of differential equations. For instance, in our case the presence of massive particles breaks explicitly the conformal symmetry and thus, physical properties of the non-hydrodynamic modes change. The mathematical and physical analysis carried out in this work is to the best of our knowledge entirely new and thus, is an important step since it can also lead to new milestones when analyzing more generic setups due to symmetry breaking.

This article is organized as follows: In Sect.~\ref{sec:ResSumm} we present the summary of the main new findings of this work.  In Sec.\ref{sec:setup}, we describe the theoretical setup considered in this paper. The discussion of the flow structure of the phase space of dynamical system is presented in Sec.\ref{sec:find_ISOF}. This section serves prepare the reader for the transseries analysis in the UV and IR which is discussed thoroughly in Sec.\ref{sec:formal_trans}. In Sec.\ref{sec:attractor}, we investigate the meaning of the attracting behavior of the attractor when the flow time of the dynamical system is the proper time. For completeness, we present a good discussion on the difference of the massive and the massless Bjorken flows from the viewpoint of the global flow structure and the formal transseries. Sect.~\ref{sec:summary} is devoted to the conclusions and outlook. Technical derivation of the transseries construction among other details are summarized in appendices.

\section{Summary of main results}
\label{sec:ResSumm}
Many of the results described in this article are technical so we summarize in this section the main concepts, approximations and results derived in our work in order to provide a compact overview. We derive a new kind of transseries solutions to non-conformal fluid equations~\cite{2014ARM,2014SWRM,2014SSC} by means of superasymptotics and hyperasymptotic techniques~\cite{boyd:1999,costin2008asymptotics}. The fluid under consideration undergoes Bjorken expansion and the explicit breaking of the conformal symmetry is introduced by assuming that the particle constituents have a physical constant mass $m$. This model is reviewed in Sect.~\ref{sec:setup}. The new solutions to the non-conformal fluid equations  are written as multiparameter transseries. The derivation of the new UV and IR transseries solutions have a physical origin which simply reflects the effects related with the mass effects on the corresponding thermodynamic properties encoded in the speed of sound and effective temperature. In this sense, the new transseries solutions differ from the ones studied in conformal Bjorken fluids~\cite{Aniceto:2018uik,Spalinski:2018mqg,
Spalinski:2017mel,Buchel:2016cbj,Heller:2016rtz,Aniceto:2015mto,Basar:2015ava,
Casalderrey-Solana:2017zyh,Romatschke:2017vte,Blaizot:2020gql,Blaizot:2021cdv}.

We determine the global flow structure of the fluid dynamical model in Sect.~\ref{sec:find_ISOF}. In hydrodynamic models it has been common to assume regular initial conditions at some early time $\tau_0$. Nonetheless, when taking $\tau_0=0$ such a generic regular configuration turns out to be divergent for the Bjorken flow. While the temperature always diverges at $\tau=0$, we show that there are very special, fine tuned configurations where $\pi(0)$ and $\Pi(0)$ are regular, as can be seen from the non-linear equations of the dynamical system, Eqs.~\eqref{eq:fluideqs}. 
We find five such points, that we name {\it convergent points}~\footnote{The dynamical system of fluid dynamical equations, Eqs.~\eqref{eq:fluideqs}, is non-autonomous~\cite{kloeden2011nonautonomous,caraballo2016applied,Behtash:2019qtk,Behtash:2018moe,Behtash:2019txb,Behtash:2020vqk}. For this type of dynamical systems the notion of fixed points either in the UV or IR emerge as limits of the solutions for the ODEs~\cite{caraballo2016applied}. In order to avoid confusion we call these limits of the flow solutions as \textit{convergent points}.}, three of which are present in the conformal case, and two appear purely due to non zero mass (cf. Eq.~ (\ref{eq:UV_cp}))~\footnote{For the two additional points, the temperature divergence must be fine tuned as in Eq.~(\ref{eq:T_asym_UV}).}.
A small perturbation around each convergent points can be either attracted or repelled for each basis, a behaviour which is determined by the eigenvalues of the corresponding Jacobian matrix, implying a global behaviour of our dynamical system in question. 
Beyond that, detailed structure of solutions around the regular points determines the corresponding transseries solutions and the way in which initial conditions are propagated forward in the time. 
As a consequence possible connections between convergent points at $\tau=0$ and $\tau=\infty$ determines {\it invariant subspace of flows} in the system.
The invariant subspace of flows control the motion of flows in the entire flow time, like critical points in autonomous systems such as renormalization group flows.
Therefore, the existences of convergent points and invariant subspace of flows are essential not only to investigate the global flow structure of a dynamical system but also for application of transseries analysis to investigate physics in the UV and IR regime~\cite{Janig_2019}. A schematic illustration which summarizes the global flow structure of the massive and massless fluids undergoing Bjorken flow is shown in Fig.~\ref{fig:flow_fig}.

Building upon these results, we determine the invariant subspace of flows when the dynamics is described in terms of two flow time different variables, {\it i.e.}, the proper time $\tau$ and the inverse Knudsen number $w=\tau/\tau_R$. 
In this paper, for convenience, we call an invariant subspace of flows `attractor solution' if it enforces neighboring flows around associated UV and IR convergent points to attract to itself on the viscosities-time plane without the temperature axis.
In both cases of $\tau$ and $w$, we show that the attractor solution corresponds to the flow line connecting the UV convergent point $\mathfrak{R}_1$ and the unique IR convergent point $\mathfrak{A}$ (see Sect.~\ref{sec:find_ISOF}). 
The later convergent point corresponds to the thermodynamical equilibrium state of the system. The former convergent point corresponds to the case when the longitudinal pressures gets its minimum value which it is frequently called the free streaming.  
Similar to the massless Bjorken expanding case, the mechanism that dominates the early time behavior is the expansion rate which drives the system towards the attractor. For example, in the free streaming case with no interactions, different kinetic models indicate that the energy density behaves like $\varepsilon\sim 1/\tau$~\cite{Giacalone:2019ldn}, and such behaviour is modified by conformally invariant interactions giving rise to a general power law $\varepsilon\sim1/\tau^\beta$ \cite{Jankowski:2020itt}. However, in the non-conformal case considered here there are two different regions with different exponents that represent the early-time behavior of shear and bulk viscous degrees of freedom present in the theory whose specific values are summarized in Tab.~\ref{table:para_UV}. Additionally, initial conditions fall into two classes: for low temperatures compared to the mass scale $m$, the memory of initial conditions is preserved for longer times than the set of initial values when the temperature is larger than the mass $m$. Finally, the language of dynamical systems allows to {\it precisely} define the notion of attractor, which can be extended at more complicated systems admitting such a description.


\begin{figure}[tb]
\begin{center}
\includegraphics[width=0.6\textwidth]{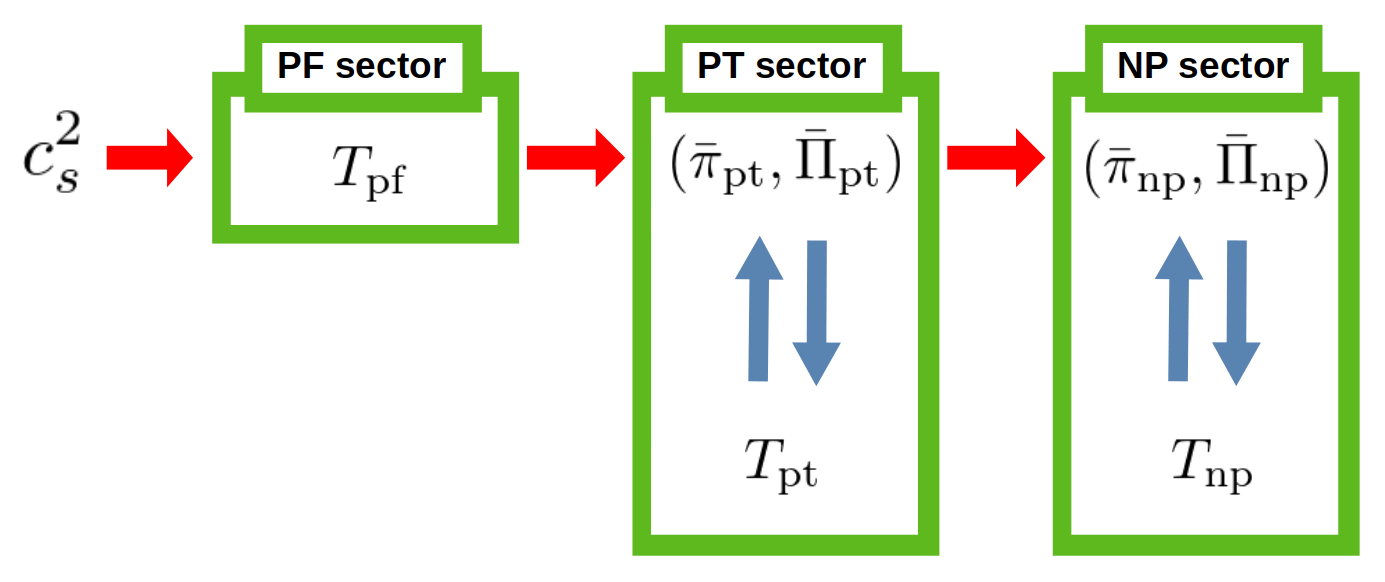}
\caption{
Schematic figure illustrating the determination of each physical sectors of considered model.
The speed of sound $c_s^2$ determines the PF sector, $T_{\rm pf}$.
At the same time, $T_{\rm pf}$ gives the leading order of $(\bar{\pi}_{\rm pt},\bar{\Pi}_{\rm pt})$, and they determine the leading order of $T_{\rm pt}$ which gives the sub-leading order of $(\bar{\pi}_{\rm pt},\bar{\Pi}_{\rm pt})$.
By repeating of this procedure, the PT sectors are completely determined.
The NP sector is also derived in the similar way to the case of the PT sector after obtaining the PT sector.
}
\label{fig:det_sectors}
\end{center}
\end{figure}

New solutions to the perfect fluid behavior for a non-trivial equation of state are found and discussed in Sect.~\ref{subsect:PF}.
The asymptotic behavior of perfect fluid (PF) is determined entirely by the speed of sound $c_s^2(m/T)$, which now exhibits a non trivial temperature dependence (cf. Fig.~\ref{fig:cs2_PFT}). As a result, the breaking of the conformal symmetry changes the simple power law $T(\tau)\sim \tau^{-c_s^2}$ to a non trivial, convergent series in terms of $\log(m\tau)$ and $\log_{(2)}(m\tau) := \log \log(m\tau)$ terms. The proof of the convergence property of this transseries solution to the perfect fluid case includes a generalization of the Borel transform and it is discussed in Sec. \ref{subsect:PF}.  The leading order contribution to perfect fluid behaviour is $T_{\rm pf}\sim 1/\log(m\tau)$~\footnote{Through the paper we shall be discussing the physics of different sectors of the transseries. When referring to each of these sectors we shall denote with a subscript. For instance, the perfect fluid, perturbative and non-perturbative sectors of the transseries solutions for the temperature are denoted as $T_{\rm pf}$, $T_{\rm pt}$ and $T_{\rm np}$. Clearly, the bulk pressure and shear viscous tensor do not have perfect fluid sector. However, these fluid variables are influenced by the perfect fluid sector of the temperature due to their coupling at the level of the ODEs, see Eqs.~\eqref{eq:fluideqs}.}. As a result, the system reaches its thermal equilibrium state at extremely long time scales. This is due to the mass gap present in the system and suppression of all possible excitations with energy lower than the mass $m$. 

\begin{figure}[h]
\begin{center}
\includegraphics[width=0.60\textwidth]{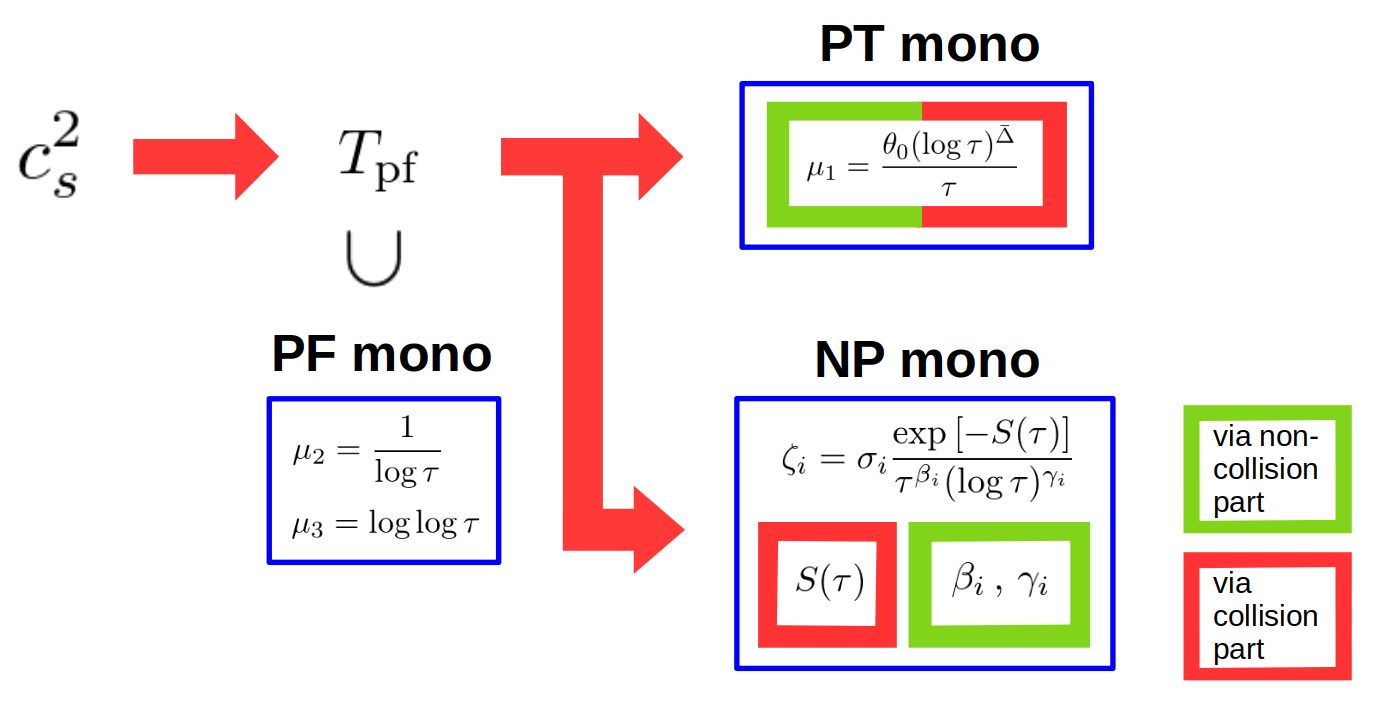}
\caption{
The schematic figure of correlations and relations among transmonomials in the IR transseries, {\it i.e.}, at late times.
A more detailed explanation is presented in Sec.~\ref{sec:attractor}.
}
\label{fig:corr_mono}
\end{center}
\end{figure}

The non-trivial effects induced by the mass into the equation of state in the perfect fluid case affect the perturbative sector (PT) and transient non-perturbative (NP) sectors of the transseries solutions for the temperature, bulk pressure and shear viscous component, ($T_{\rm pt},\pi_{\rm pt},\Pi_{\rm pt}$) and ($T_{\rm np},\pi_{\rm np},\Pi_{\rm np}$) respectively. In Fig.~\ref{fig:det_sectors} we outline the relations among different sectors of the transseries solutions for the fluid variables $T$, $\Pi$ and $\pi$. The notion of perturbative or non-perturbative is determined by the corresponding dependence on the parameter $\theta_0$ controlling the strength of the relaxation time, {\it i.e.}, $\tau_R\sim\theta_0/T^{\bar{\Delta}}$ (cf. Eq.~(\ref{eq:tauR})), which is polynomial for the former and exponential for the later sectors, see Fig.~\ref{fig:corr_mono} where the relations among transmonomials of different sectors are schematically described. It should be clear that $\theta_0$ does not play any fundamental role in the perfect fluid transseries as expected. The non-perturbative terms, apart from exponentially damped terms, are modified with logarithmic corrections, $\log(m\tau)$ and $\log_{(2)}(m\tau) := \log \log(m\tau)$ respectively, in a similar manner as it is seen perfect fluid. However, these modifications have relevant physical consequences in the decay of the non-hydrodynamic modes towards the attractor since the logarithmic corrections enhance the Knudsen number asymptotically while slowing down the decay of the non-hydrodynamic modes. As a result, the non-conformal fluid hydrodynamizes at long time scales which is in contrast with the conformal Bjorken cases. Similarly to the massless case, at late times the information about initial conditions is encoded in the transseries parameters. In the non-conformal fluid model studied here the information about the initial temperature stays present even at late times as a temperature integration constant, due to the existence of a preferred energy scale in the system. The dependence on the initial UV data at late times is an indication that the IR physics of the transport coefficients depends on the deformation history of the fluid and thus, its rheology~\cite{Behtash:2018moe,Behtash:2019txb,Behtash:2020vqk}. It is also important to mention that the limits of late time and mass going to zero {\it do not} commute. 


In a conformally invariant model, the bulk viscosity vanishes exactly, and one finds that the transseries solutions are given in terms of power law of the inverse Knudsen number at early times~\cite{Kurkela:2019set,Behtash:2020vqk}. In the non-conformal case the UV transseries is also written as a power law series but it receives non-trivial contributions from the shear bulk coupling which depends explicitly on the logarithm of the mass $\log(m\tau)$ and $\log \log(m\tau)$. As a result, the UV transseries gets strongly modified by the breaking of conformality due to the mass of the particle. 

Knowing the transseries structure of the solutions as well as global structure we proceed with a discussion of an attractor concept. To that end we define measure on a reduced phase space $(\bar{\pi},\bar{\Pi},\tau)$ that allows us to quantify the distance between two flow lines. Using the knowledge of the global structure we define the attractor as an invariant subspace of flows connecting proper UV convergent point with an equilibrium IR convergent point.
We numerically check the pullback and forward attracting properties identifying two attraction mechanisms known from the conformal case \cite{Blaizot:2017ucy}, which get non-trivially modified by the presence of the mass effects. 
The expansion dominated phase exhibits two sub-phases with different power laws, while the transient mode decay dominated phase gets altered by a plateau phase for some initial conditions. The details of this picture depend on the choice of time variable between the proper time $\tau$ and the inverse Knudsen number $w$.
Eventually we present a detailed comparison between the non-conformal model studied in the present paper with known results for conformal hydrodynamical models.

\section{Non-conformal hydrodynamics of a boost invariant massive plasma} \label{sec:setup}

In this section, we briefly review the second order non-conformal fluid model studied in this work. The technical details of this non-conformal fluid model are fully described in Refs.~\cite{2014ARM,2014SWRM,2014SSC}. In our analysis  we employ the Milne coordinates, $(\tau,x,y,\zeta) $, with a metric $g_{\mu \nu }(x) = {\rm diag}(1,-1,-1,-\tau^2)$, where $\tau = \sqrt{t^2-z^2}$ is the proper time and $\zeta= {\rm arctanh} (z/t)$ is space-time rapidity. We consider a longitudinal boost invariant expanding plasma whose particle constituents have  a constant mass $m$. In the Landau frame the energy momentum tensor of the fluid reads as
\begin{equation}
\label{eq:EMtensor}
    T^{\mu\nu}= \text{diag.}\left(\varepsilon(\tau),P_T(\tau),P_T(\tau),P_L(\tau)\right)\,,
\end{equation}
where $P_T$ and $P_L$ denote the transverse and longitudinal pressures respectively while $\epsilon$ is the energy density. $P_T$ and $P_L$ are defined in terms of the thermodynamic pressure $P$, the bulk pressure $\Pi$ and the independent shear viscous component $\pi$ as follows:
\begin{equation}
    P_T = P+\Pi +\frac{\pi}{2}\,,\qquad P_L = P+\Pi-\pi\,.
\end{equation}
The evolution equations for the energy density $\epsilon$, the bulk pressure $\Pi$ and shear viscous independent component $\pi$ are~\cite{2014ARM,2014SWRM,2014SSC}

\begin{subequations}
\label{eq:fluideqs}
\begin{align}
& \frac{d \varepsilon}{d \tau} = - \frac{1}{\tau} \left( \varepsilon + P - \pi  + \Pi \right), \label{eq:depsdt} \\
& \frac{d \pi}{d\tau} = - \frac{\pi}{\tau_{R}} - \frac{1}{\tau} \left[ \left( \delta_{\pi \pi} + \frac{\tau_{\pi \pi}}{3} \right) \pi - \frac{2 \lambda_{\pi \Pi}}{3} \Pi -\frac{4 \beta_{\pi}}{3} \right], \label{eq:dpidtau0} \\
& \frac{d \Pi}{d \tau} = - \frac{\Pi}{\tau_{R}} - \frac{1}{\tau} \left( \delta_{\Pi \Pi} \Pi - \lambda_{\Pi \pi} \pi + \beta_{\Pi}\right). \label{eq:dPIdtau0}
\end{align}
\end{subequations}
Previous evolution equations are obtained through the second order Chapman-Enskog method of relativistic kinetic theory~\cite{DeGroot:1980dk} and within the relaxation time approximation of the collisional kernel of the Boltzmann equation~\cite{2014ARM,2014SWRM,2014SSC}. 

In Eqs.~\eqref{eq:fluideqs} the thermodynamic transport coefficients s depend solely on $z=m/T = \beta m$. These are given by \cite{2014ARM}
\begin{equation} 
\label{eq:transportcoeff}
\begin{split}
&\delta_{\pi \pi}(z) = \frac{5}{3} + \frac{7}{3}\frac{\alpha_3(z)}{\alpha_5(z)} \qquad
\tau_{\pi \pi}(z) = 2 +  4 \frac{\alpha_3(z)}{\alpha_5(z)} ,\qquad
\lambda_{\pi \Pi}(z) = - \frac{2}{3} \chi(z), \\
&\beta_{\pi}(z) = m^4 z \alpha_5(z), \qquad
\lambda_{\Pi \pi}(z) = \frac{2}{3} + \frac{7}{3} \frac{\alpha_3(z)}{\alpha_{5}(z)}  - c_s^2(z),\qquad
\delta_{\Pi \Pi}(z) = - \frac{5}{9} \chi(z) - c_s^2(z),\\
&\beta_{\Pi}(z) = \frac{5}{3} m^4 z \alpha_{5}(z) - \frac{m^4 \alpha_1(z)}{z^2} c_s^2(\beta), \qquad
\chi(z) = \frac{m^4 z}{\beta_\Pi(z)} \left[ \left(1-3 c_s^2(z) \right)\left( \alpha_5(z) -  \frac{\alpha_1(z)}{z^3} \right)-\alpha_4(z) \right].
\end{split}
\end{equation}
 In the previous expressions the functions $\alpha_{n}(z)$ are expressed in terms of the modified Bessel functions $K_n(x)$ and the Bickley-Naylor functions ${\rm Ki}_n(x)$ (see appendix \ref{sec:tpcoeff_alpha}). For completeness, we quote below the functions $\alpha_{n}(z)$ entering in Eqs.~\eqref{eq:transportcoeff}
\begin{equation}
\label{eq:alphafunc}
\begin{split}
 &\alpha_1(z) = \frac{1}{2 \pi^2}  \left[ 4 {\rm K}_2(z) + z {\rm K}_1(z) \right],\qquad \alpha_2(z) = \frac{1}{2 \pi^2} \left[ (12+z^2){\rm K}_2(z)+ 3z {\rm K}_1(z) \right],\\
 &\alpha_3(z) = -\frac{1}{3360 \pi^2} \left[ {\rm K}_5(z) - 11 {\rm K}_3(z) + 58 {\rm K}_1(z) + 16 {\rm Ki}_3(z) - 64 {\rm Ki}_1(z) \right], \\
 &\alpha_4(z) = -\frac{1}{30 \pi^2} \left[ {\rm K}_3(z) -4 {\rm K}_1(z)  + {\rm Ki}_3(z) + 2 {\rm Ki}_1(z)  \right], \\
 &\alpha_5(z)= \frac{1}{480 \pi^2} \left[ {\rm K}_5(z) -7 {\rm K}_3(z) + 22 {\rm K}_1(z) - 16 {\rm Ki}_1(z) \right], \\
 &\alpha_6(z) = \frac{1}{2 \pi^2} \left[  (48 + 5 z^2) {\rm K}_2(z) + z (12 + z^2) {\rm K}_1(z) \right].
 \end{split}
\end{equation}
In Eq.~\eqref{eq:depsdt} the energy density and the thermodynamic pressure are expressed in terms of the modified Bessel function $K_n(x)$ as
\be
&& \varepsilon(z) = \frac{m^4}{2 \pi^2 z^2} \left[ 3 {\rm K}_2(z) + z {\rm K}_1(z) \right], \qquad P(z) = \frac{m^4}{2 \pi^2 z^2} {\rm K}_2 (z),  \label{eq:def_epsP} 
\ee
These physical quantities are related to each other via the equation of state which determines the speed of sound $c_s^2$, i.e.,
 \be
c_s^2(z) &=& \left. \frac{dP(\varepsilon)}{d \varepsilon} \right|_{\varepsilon=\varepsilon(z)}
= \frac{\alpha_1(z)}{\alpha_2(z)}, \label{eq:cs2}
\ee
Furthermore, in Eqs.~\eqref{eq:fluideqs} we parametrize the relaxation time $\tau_R$ as follows:
\be
\tau_R(T)= \frac{\theta_0}{T^{\bar{\Delta}}}~,
\label{eq:tauR}
\ee
where $T$ is the temperature of the system, $\theta_0>0$ and $0 \le \bar{\Delta} \le 1$ respectively. When $\bar{\Delta}=0$ the relaxation time is constant. Notice that the traceless condition, $T^{\mu}_{\ \mu}=0$, is satisfied if both the mass of the particle as well as the bulk pressure vanish exactly at all times, i.e., $z(\tau) = 0$ and $\Pi(\tau) = 0$. 

The combination of Eqs.(\ref{eq:depsdt}) and (\ref{eq:cs2}) yields to the following ODE for the temperature
\be
 \frac{d T}{d \tau} &=& -\frac{T}{\tau} \left[ \bar{\Pi} - \bar{\pi}  + c_s^2(z) \right] =: F_{T}(T,\bar{\pi},\bar{\Pi},\tau), \label{eq:dTdtau_fm}
\ee
where we introduce the effective inverse Reynolds numbers for the shear component $\bar{\pi}$ and bulk pressure $\bar{\Pi}$ which are defined as respectively~\footnote{Our definitions for the inverse Reynolds numbers of the dissipative variables do not agree with the ones presented in recent publications~\cite{Chattopadhyay:2021ive,Jaiswal:2021uvv,Chen:2021wwh}. 
In these works, the authors considered the inverse Reynolds numbers of the shear viscous tensor and bulk pressure defined as $Re^{-1}_\pi = \pi/(\varepsilon + P)$ and $Re^{-1}_\Pi =\Pi/(\varepsilon + P)$  respectively. These variables are related with ours by multiplying these by the speed of sound~\eqref{eq:cs2}, i.e.,  as $\bar{\pi} = c_s^2 Re^{-1}_\pi$ and $\bar{\Pi}=c_s^2 Re^{-1}_\Pi$. Our definitions turn out to be more convenient when performing the transasymptotic analysis discussed in Sect.~\ref{sec:formal_trans}.}
\begin{equation}
    \label{eq:effRey}
    \bar{\pi}=z^2 \frac{\pi} {m^4\alpha_2(z)}\,,\qquad
    \bar{\Pi}=z^2 \frac{\Pi}{m^4 \alpha_2(z)}\,.
\end{equation}
Instead of studying Eqs.(\ref{eq:dpidtau0}) and (\ref{eq:dPIdtau0}), we analyze the set of ODEs for $\bar{\pi}$ and $\bar{\Pi}$, i.e. 
\begin{subequations}
\label{eq:odediss}
\begin{align}
\frac{d \bar{\pi}}{d \tau} 
 &=   - \frac{T^{\bar{\Delta}}}{\theta_0} \bar{\pi}  - \frac{1}{\tau} \left[ C_\pi(z)
   \bar{\pi} -  \frac{2 \lambda_{\pi \Pi}(z)}{3}  \bar{\Pi} -  \frac{4 \bar{\beta}_{\pi}(z)}{3}  - D (z) \left( \bar{\Pi} - \bar{\pi} \right) \bar{\pi}   \right] =: F_{\bar{\pi}}(T,\bar{\pi},\bar{\Pi},\tau),  \label{eq:dpidtau_fm} \\
 \frac{d \bar{\Pi}}{d \tau} &=  - \frac{T^{\bar{\Delta}}}{\theta_0}  \bar{\Pi} - \frac{1}{\tau} \left[ C_{\Pi}(z)  \bar{\Pi} -  \lambda_{\Pi \pi}(z)  \bar{\pi} +  \bar{\beta}_{\Pi}(z)  - D (z) \left( \bar{\Pi} - \bar{\pi} \right) \bar{\Pi} \right]=: F_{\bar{\Pi}}(T,\bar{\pi},\bar{\Pi},\tau), \label{eq:dPIdtau_fm} 
 \end{align}
\end{subequations}
where 
\begin{equation}
\begin{split}
&\bar{\beta}_{\pi,\Pi}:=z^2 \beta_{\pi,\Pi}/(m^4 \alpha_2(z))\,,\qquad
D(z) = \frac{\alpha_6(z)}{\alpha_2(z)}\,, \\
& 
C_\pi(z) = \delta_{\pi \pi}(z) + \frac{\tau_{\pi \pi}(z)}{3} - D (z) c_s^2(z)\,,\qquad
C_\Pi(z) =  \delta_{\Pi \Pi}(z) - D (z)  c_s^2(z), 
\end{split}
\end{equation}
Furthermore, the transverse pressure and the longitudinal pressure can be also expressed in terms of $\bar{\pi}$ and $\bar{\Pi})$ as follows:
\be
&& P_T= P+\left(\bar{\Pi}+\frac{\bar{\pi}}{2}\right)(\varepsilon+P)/c_s^2~, \qquad P_L = P+\left(\bar{\Pi}-\bar{\pi}\right)(\varepsilon+P)/c_s^2~ . 
\label{eq:PTPL2}
\ee
Our fundamental object of study will be the solutions of the macroscopic quantities  $(T,\bar{\pi},\bar{\Pi})$ and any other observable related with these physical variables. 

Let us first discuss the kinematic region where the particle mass becomes relevant. Firstly, different phenomenological models of the QGP assume that during its early stages the temperature of the system is extremely high relative to the light quark masses~\footnote{It is important to mention here that these assumptions have been recently questioned when considering different quark flavors with physical masses, see Refs.~\cite{Martinez:2018tuf,Martinez:2018ygo,Martinez:2019jbu,Carzon:2019qja}.}. Then, $z = m/T \ll 1$. In this case, the speed of sound~\eqref{eq:cs2} and the transport coefficients~\eqref{eq:transportcoeff} become constants (see App.~\ref{sec:tpcoeff_alpha}), so the mass effect is irrelevant.
Now, when the system hydrodynamizes at $\tau/\tau_R \rightarrow +\infty$ the condition under which the mass effect can be considered negligible can be roughly estimated as
\be
\frac{T}{m} \ \gg \  \frac{\tau}{\tau_R(T)} \ \gg \ 1 \qquad \Rightarrow \qquad m \tau \ \ll \  \theta_0 T^{1-\bar{\Delta}} \ \ll \  T \tau,
\ee
where we used explicitly Eq.~\eqref{eq:tauR}. Thus, the mass of the particle constituents plays a role in the dynamics of the plasma in any kinematic regime different than these two cases previously described.

Now in the small mass limit $z 
\rightarrow0$, the ODEs~\eqref{eq:dTdtau_fm} and~\eqref{eq:odediss} read as 
\begin{subequations}
\label{eq:massless_limit}
\begin{align} 
& \frac{d T}{d \tau} = - \frac{T}{\tau} \left[ \bar{\Pi} - \bar{\pi} + \frac{1}{3} \right], \label{eq:dTdtau_ml} \\
& \frac{d \bar{\pi}}{d \tau} 
 =   - \frac{T^{\bar{\Delta}}}{\theta_0} \bar{\pi}  - \frac{1}{\tau} \left[ \frac{10}{21}
   \bar{\pi} -  \frac{4}{5}  \bar{\Pi} -  \frac{4}{45}  - 4 \left( \bar{\Pi} - \bar{\pi} \right) \bar{\pi}   \right],  \label{eq:dpidtau_ml} \\
& \frac{d \bar{\Pi}}{d \tau} =  - \frac{T^{\bar{\Delta}}}{\theta_0}  \bar{\Pi} + \frac{1}{\tau} \left[ \frac{2}{3}  \bar{\Pi} + 4 \left( \bar{\Pi} - \bar{\pi} \right) \bar{\Pi} \right]. \label{eq:dPIdtau_ml} 
\end{align}
\end{subequations}
In the RHS of Eqs.~\eqref{eq:dpidtau_ml} and~\eqref{eq:dPIdtau_ml}, the terms are proportional to either the relaxation time scale scale~\eqref{eq:tauR} or the thermodynamic transport coefficients~\eqref{eq:transportcoeff} in the limit considered here. We shall call the former terms collisional and the later ones as non-collisional. Let us introduce the following scale transformation 
\begin{equation}
\label{eq:scaling}
    (T,\bar{\pi},\bar{\Pi},\tau) \mapsto (\lambda^{-\Delta_T} T,\lambda^{-\Delta_{\bar{\pi}}} \bar{\pi} ,\lambda^{-\Delta_{\bar{\Pi}}} \bar{\Pi},\lambda \tau)\,,\qquad\text{with}\,\,
    (\Delta_T,\Delta_{\bar{\pi}},\Delta_{\bar{\Pi}})=(1/\bar{\Delta},0,0)\,,
\end{equation}
with a constant $\lambda>0$~\footnote{
When $\bar{\Delta}=0$ and $z=0$, $\Delta_T$ appears to become singular.
In that case, however, this singularity just means that $\Delta_T$ is ambiguous and is in fact arbitrary.
Therefore, $\Delta_{\bar{\pi},\bar{\Pi}}=0$, and this symmetry still exists for $\bar{\Delta}=0$. 
}. When the equations of motion satisfy the scaling~\eqref{eq:scaling}, it implies that the temperature decouples independently from Eqs.~\eqref{eq:odediss}. Therefore, the physics of any observable $\cal{O}$ with scaling dimension $\Delta_{\cal O}=0$ can be described by a scale invariant variable, such as $w=\tau/\tau_R$, without being affected explicitly by the temperature.
However, when considering the kinematic region where the particle mass is relevant, the temperature does not decouple from Eqs.~\eqref{eq:odediss} because the scaling~\eqref{eq:scaling} breaks down explicitly. It is important to emphasize that this scale symmetry can be defined independently from the traceless condition of the energy momentum tensor as long as $z = 0$ even when $\bar{\Pi} \ne 0$ (at the level of the ODE). In order to avoid confusion with the traceless condition, we shall call in this work the scaling~\eqref{eq:scaling} as \textit{D-scale symmetry} or simply \textit{D-symmetry}. In the non-conformal model studied here (Eqs.~\eqref{eq:dTdtau_fm} and~\eqref{eq:odediss}), the D-symmetry is a weaker condition compared with the traceless condition because it is manifested only in the small mass limit (or equivalently the high temperature limit) while the traceless condition requires $\bar{\Pi} = 0$ at all times together with the D-symmetry \footnote{The D-symmetry is analogous to the dilation symmetry of the scale invariant theory. In the high temperature limit, it arises when having a non-vanishing value for $\bar{\Pi}$.}. The D-symmetry becomes relevant when analyzing the particle mass effects as we shall see in the next sections.

In the rest of this work we analyze systematically  Eqs.(\ref{eq:dTdtau_fm})-(\ref{eq:dPIdtau_fm}). In order to simplify notation and without losing  generality, we set hereafter the particles of the plasma to have a unit mass, i.e. $m=1$, which implies that the transport coefficients and the speed of sound become solely functions of the inverse temperature $\beta$.
Moreover, we shall consider flows solutions where $(\bar{\pi}(\tau),\bar{\Pi}(\tau))$ are finite for any $\tau \geq 0$.

\section{Global flow structure} \label{sec:find_ISOF}
We begin our considerations by presenting the description of the global flow structure of the non-conformal hydrodynamical equations. The relevant physical significance of global flow structure is the identification of independent physical components of the system at early and late times. This is achieved by determining the {\it invariant subspaces of flows} (ISOF), which is a decomposition of the flow space of dynamical variables into physically distinct, coupled components characterized by specific properties of relevant observables. In turn, this decomposition determines possible transseries solutions, one for each specific configuration of the dynamical variables in the phase space at $\tau=0$ or $\tau=\infty$ - the so called {\it convergent point} (CP)~\footnote{It is often common to call fixed points to the limits of the solutions of the ODEs in the phase space in the IR and UV regimes. 
However, in the case of nonautonomous dynamical systems this choice of words is somehow misleading because these emergent limits are not 'fixed' in the time evolution~\cite{caraballo2016applied}. 
In order to avoid confusion we decide to call those limits as convergent points.}. Therefore, it is needed to characterize first the global flow structure of the dynamical system before carrying any transseries analysis. The later is tightly restricted by the former. Furthermore, any physical property of the theory derived from the corresponding transseries can be inferred in a more straightforward manner. 

In order to determine the global flow structure of the non-conformal Bjorken fluid we shall follow these steps:
(1) find all CPs at the UV and IR regimes, (2) perform detailed stability analysis around each CP, 
and (3) find flow lines connecting them.   We shall also make use of well known mathematical tools in the theory
of non-autonomous dynamical systems (see Appendix A for short review of the ideas worked out in this section). Here we just outline the most important
concepts and highlight their utility. 

As it was mentioned above convergent points are very specific configurations of
the dynamical variables in the UV and IR limits. More precisely we introduce those as specific limits in the phase space of variables as follows:

\begin{subequations}
\begin{align}
     \mbox{UV CP} &: \quad \mathfrak{R} := \lim_{\tau \rightarrow 0_+} (T(\tau),\bar{\pi}(\tau),\bar{\Pi}(\tau)), \\
     \mbox{IR CP} &: \quad \mathfrak{A} := \lim_{\tau \rightarrow +\infty} (T(\tau),\bar{\pi}(\tau),\bar{\Pi}(\tau)). 
\end{align}
\end{subequations}
The set of initial conditions of the flows which merge asymptotically to the same IR CP determine the basin of attraction of the forward attractor. However, flows carry information about the UV region where these emerge so one can establish differences among these~\footnote{Mathematically this means that each subset in the basin of attraction has an associated homotopy class.}.
For instance, in conformal kinetic systems~\cite{Kurkela:2019set,Behtash:2017wqg,Behtash:2019qtk,Behtash:2019txb,Behtash:2020vqk} the energy density decays faster (slower) near the UV CP where the transverse (longitudinal) pressure reaches its minimum value. At the same time, fluctuations at early times might grow or decay rapidly depending on how close these are near a given UV CP.
In the non-conformal case studied here, the situation is more cumbersome since we shall find multiple UV CPs as we show below. Therefore, in order to establish a difference between the elements of the basin of attraction, it is needed to introduce the concept of an ISOF. ISOF is a subspace of the phase space with the property that all flows which fall into it stay there eternally. We will focus on a special case when flow lines connect UV and IR CPs {\it i.e.},
\begin{equation}
\mbox{ISOF} : \quad {\cal S}({\mathfrak R},{\mathfrak A}) := \{ \mbox{all flows converging to ${\mathfrak R}$ and ${\mathfrak A}$ }\}.
\end{equation}
Flows in a manifold $\mathcal{M}$ connecting the UV and IR convergent points determine the decomposition of the basin of attraction into invariant subspaces of flows. Then observables in each ISOF have their own unique asymptotic expansions around the UV/IR respectively. More importantly, the dimensionality of ISOF's gives the global information of the dynamical system such as the {\it attractor} and its corresponding stability analysis. The later is of vital importance for the convergence of numerical solutions as it is very well known in more complex cases~\cite{thompson2002nonlinear}.

\begin{figure}[tb]
\begin{center}
\includegraphics[width=.42\textwidth]{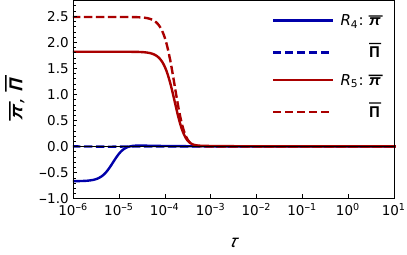} \ \ \
\includegraphics[width=.42\textwidth]{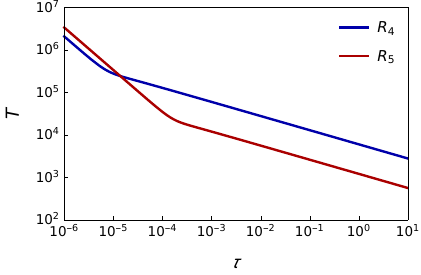}
\caption{
The flow element of ${\cal S}({\frak R}_4, {\frak A})$ (dashed) and ${\cal S}({\frak R}_5, {\frak A})$ (solid).
The parameters are taken as $\theta_0=1$ and $\bar{\Delta}=1$.
}
\label{fig:R45}
\end{center}
\end{figure}

When seeking for the global flow structure of the non-conformal fluid, especially the location of CPs in phase space, the asymptotic behavior of the temperature needs to be taken care of, which generally changes the dominance between the collision part and the non-collision part of the ODEs~\eqref{eq:dTdtau_fm} and~\eqref{eq:odediss} in the UV/IR limit.
For the massive Bjorken flow the asymptotic behaviours of the temperature gives us five UV CPs and one IR CP which read as~\footnote{The UV CPs found in the effective second order Chapman-Enskog expansion, Eq.~\eqref{eq:CPs} do not necessarily coincide with the ones obtained from the underlying microscopic Boltzmann equation. As a matter of fact, the stability analysis of different kinetic theory models in terms of moments~\cite{Behtash:2019txb,Behtash:2019qtk,Behtash:2020vqk,Blaizot:2020gql,Blaizot:2017lht,Blaizot:2019scw,Jaiswal:2022udf} has shown that for a given truncation scheme the UV phase space structure changes despite having a better control over the dynamics at early times by adding further dissipative corrections. In this sense, the UV CPs associated to the second order CE expansion, Eq.~\eqref{eq:CPs} do not necessarily encode the phase structure inherent to the exact Boltzmann equation. It is nonetheless important to analyze the emergence of these UV CPs in order to have a better systematic understanding of the limitations of optimized hydrodynamic schemes.}
\begin{subequations}
\label{eq:CPs}
\begin{align}
\mbox{\underline{UV CPs}} & \nl
 {\frak R}_1 &: \ (T_{\rm UV},\bar{\pi}_{\rm UV}, \bar{\Pi}_{\rm UV}) =
\left(+\infty, - \frac{7}{54}, -\frac{8}{27} \right), \nl 
{\frak R}_2 &: \ (T_{\rm UV},\bar{\pi}_{\rm UV}, \bar{\Pi}_{\rm UV}) = \left(+\infty, - \frac{25 - 3 \sqrt{505}}{420}, 0 \right), \nl
{\frak R}_3 &: \ (T_{\rm UV},\bar{\pi}_{\rm UV}, \bar{\Pi}_{\rm UV}) = \left(+\infty, - \frac{25 + 3 \sqrt{505}}{420}, 0 \right), \label{eq:UV_cp} \\
 {\frak R}_4 &: \ (T_{\rm UV},\bar{\pi}_{\rm UV}, \bar{\Pi}_{\rm UV}) = \left(+\infty, \frac{1}{3} - \frac{1}{\bar{\Delta}},0 \right), \nl
 {\frak R}_5 &: \ (T_{\rm UV},\bar{\pi}_{\rm UV}, \bar{\Pi}_{\rm UV}) =
 \left(+\infty, -\frac{14}{27} + \frac{7}{3 \bar{\Delta}}, - \frac{23}{27} + \frac{10}{3\bar{\Delta}}  \right), \nl \nl
 \mbox{\underline{IR CP}} & \nl
 {\frak A} &: \ (T_{\rm IR},\bar{\pi}_{\rm IR}, \bar{\Pi}_{\rm IR}) =  \left(0, 0, 0 \right),
\end{align}
\end{subequations}
where we recognize ${\frak R}_2$ as the so-called free-streaming limit since the longitudinal pressure is minimized, and ${\frak A}$ corresponds to the local equilibrium. It is important to emphasize that the location of the UV CPs of $\bar{\pi}$ and $\bar{\Pi}$ do not depend on the mass of the particle. As we will see in Sec.\ref{sec:formal_trans}, the temperature $T \sim (\log \tau)^{-1}$ in the IR regime so it decays very slowly near the equilibrium point ${\frak A}$ and converges to zero asymptotically. The UV CPs, ${\frak R}_{1,\cdots,5}$, fall into two classes depending on the dominance of (non-)collisional components of the ODEs in the the UV limit. When the non-collisional part overcomes the collisional one one finds that ${\frak R}_{1,2,3}$ emerge as UV limits of the dynamical variables.
On the other hand, ${\frak R}_{4,5}$ appear as UV limits under the condition that the collision part and the non-collision part are balanced with each other (see Appendix~\ref{sec:trans_early} for details).
The former class indeed gives ISOFs reaching ${\frak A}$ in the IR limit and it is discussed in Ref.\cite{Jaiswal:2021uvv}. The later class can be found by fine-tuning the initial conditions of the temperature such that the UV asymptotic behavior is~\footnote{
Around the UV CPs, $\frak{R}_{4,5}$, the collisional components of the ODEs contribute non-trivially. Given the scaling of the temperature with the proper time around $\frak{R}_{4,5}$~\eqref{eq:T_asym_UV}, i.e. $T(\tau)\sim\tau^{1/\bar{\Delta}}$, as well as the relation between the temperature and the relaxation time scale~\eqref{eq:tauR}, i.e. $\tau_R\sim T(\tau)^{\bar{\Delta}}$, then one gets that in this region of the phase space $\tau_R$ grows linearly with the proper time $\tau$.
}
\be
T \sim 
\begin{dcases}
    \left[\frac{6 \left(70 -55 \bar{\Delta} + 9 \bar{\Delta}^2 \right) \theta_0}{35 \bar{\Delta} (3-\bar{\Delta})}\right]^{1/\bar{\Delta}} \tau^{-1/\bar{\Delta}} &  \mbox{for} \quad  {\frak R}_4 \\
  \left[\frac{2 (6-\bar{\Delta}) \theta_0}{3 \bar{\Delta}}\right]^{1/\bar{\Delta}} \tau^{-1/\bar{\Delta}} & \mbox{for} \quad  {\frak R}_5
\end{dcases}. \label{eq:T_asym_UV}
\ee
Notice that these CPs do not appear when the relaxation time is a constant, $\bar{\Delta} = 0$.
One can numerically make establish the existence of these ISOFs emerging from ${\frak R}_{4,5}$, as it is shown in Fig.\ref{fig:R45}.

Next we consider now the stability around the UV CPs and dimensions of the ISOFs.
This can be easily obtained by the index of the Jacobi matrix for each UV CP.
It is convenient to use the inverse of the temperature $\beta:=1/T$ and redefine the flow time as $s:=1/\tau$ and thus, the UV limit corresponds to the large $s$ limit. After recasting $F_{T,\bar{\pi},\bar{\Pi}}$ in Eqs.(\ref{eq:dTdtau_fm})-(\ref{eq:odediss}) such that $d X/d\tau = F_{X} \rightarrow d X/ds = \tilde{F}_{X}$, we define the Jacobi matrix for ${\frak R}_{1,2,3}$ as
\be
J = \lim_{s \rightarrow +\infty} s \left. \frac{\ppd (\tilde{F}_\beta,\tilde{F}_{\bar{\pi}},\tilde{F}_{\bar{\Pi}})}{\ppd (\beta,\bar{\pi},\bar{\Pi})} \right|_{(\beta,\bar{\pi},\bar{\Pi}) \rightarrow {\frak R}_a}, \label{eq:jacobi}
\ee
where $\tilde{F}_{\beta}:=\tilde{F}_T \cdot (-\beta^2)$.
The factor $s$ is needed to regularize the most dominant term when taking the UV limit.
The eigenvalues of the Jacobi matrix are given by
\be
(\lambda_1,\lambda_2,\lambda_3)=
\begin{dcases}
\left(-\frac{1}{6}, \frac{95 + 3 \sqrt{505}}{105}, \frac{95 - 3 \sqrt{505}}{105} \right) &  \mbox{for} \quad  {\frak R}_1 \\
\left(-\frac{55 - \sqrt{505}}{140}, -\frac{95 - 3 \sqrt{505}}{105}, \frac{2\sqrt{505}}{35} \right) &  \mbox{for} \quad  {\frak R}_2 \\
\left(-\frac{55 + \sqrt{505}}{140}, -\frac{95 + 3 \sqrt{505}}{105},- \frac{2\sqrt{505}}{35} \right) &  \mbox{for} \quad  {\frak R}_3
\end{dcases}, \label{eq:eigen_J}
\ee
and this value gives stability of the projected flows onto the fixed $\tau$ space, denoted by ${\cal F}$, around the associated UV CP.
The definition of the Jacobi matrix given in Eq.(\ref{eq:jacobi}) is not directly applicable to ${\frak R}_{4,5}$ because of the specific temperature behavior~\eqref{eq:T_asym_UV} around those CPs.
Due to this, we slightly modify the Jacobi matrix for ${\frak R}_{4,5}$ by excluding $\beta$ but use it as an input to determine the corresponding eigenvalues in the UV limit, i.e.,
\be
J = \lim_{s \rightarrow +\infty} s \left. \frac{\ppd (\tilde{F}_{\bar{\pi}},\tilde{F}_{\bar{\Pi}})}{\ppd (\bar{\pi},\bar{\Pi})} \right|_{\beta \rightarrow \sigma_\beta s^{-1/\bar{\Delta}}, (\bar{\pi},\bar{\Pi}) \rightarrow {\frak R}_a}, \label{eq:jacobi2}
\ee
where $\sigma_\beta$ is inverse of the overall factor given in Eq.(\ref{eq:T_asym_UV}).
The corresponding eigenvalues read as
\be
(\lambda_1,\lambda_2)=
\begin{dcases}
 \left( - \frac{4(90-23 \bar{\Delta})}{105 (3-\bar{\Delta})}, - \frac{60-8\bar{\Delta}(5-\bar{\Delta})}{5\bar{\Delta}(3-\bar{\Delta})} \right) &  \mbox{for} \quad  {\frak R}_4 \\ 
 \left( -\frac{42-26 \bar{\Delta}+6 \sqrt{49-\bar{\Delta}(70-67 \bar{\Delta})/15}}{21 \bar{\Delta}}, -\frac{42-26 \bar{\Delta}-6 \sqrt{49-\bar{\Delta} (70-67 \bar{\Delta})/15}}{21 \bar{\Delta}}\right) &  \mbox{for} \quad  {\frak R}_5
\end{dcases}. \label{eq:eigen_J2} \nl
\ee
In dynamical systems the importance of the eigenvalues of the Jacobi matrix, such as the ones quoted in Eqs.~\eqref{eq:eigen_J} and~\eqref{eq:eigen_J2}, lies on the the information carried out by these which helps to identify attractors.
For instance, the real part of the eigenvalues determine if the flows diverge or merge along a given direction, and the number of them defines the index, which give us the (upper bound of) dimensions of each ISOF from the number of positive or negative the real parts ~\cite{thompson2002nonlinear,Milnor1963-qq}.
From the eigenvalues given in Eqs.(\ref{eq:eigen_J}) and (\ref{eq:eigen_J2}), the dimensions of the ISOFs corresponding to each ${\frak R}_a$ are obtained by
\be
{\rm dim} \, {\cal S}(\frak{R}_a,{\frak A}) =  \mbox{$\#$ of negative $\lambda_i$} + 1,
\ee
where $+1$ comes from adding the dimension along the $\tau$-axis. We clarify to the reader that the sum in the previous expression is carried over negative eigenvalues since we change the flow time $\tau \to s := 1/\tau$. Hence, one finds
\be
&& {\rm dim} \, {\cal S}(\frak{R}_1,{\frak A}) = 2, \qquad {\rm dim} \, {\cal S}(\frak{R}_2,{\frak A}) = 3, \qquad {\rm dim} \, {\cal S}(\frak{R}_3,{\frak A}) = 4, \nl 
&& {\rm dim} \, {\cal S}(\frak{R}_4,{\frak A}) = 2, \qquad {\rm dim} \, {\cal S}(\frak{R}_5,{\frak A}) = 3. \label{eq:dim_ISOF}
\ee
It is remarkable to mention that ${\rm dim} \, {\cal S}(\frak{R}_a,{\frak A})-1$ provides the dimensions of the space of initial conditions needed to stay in the ISOF.  As we shall later, this particular number plays an important role for the construction of the formal UV transseries since the initial conditions $(T(\tau_0),\bar{\pi}(\tau_0)),\bar{\Pi}(\tau_0))$ are non-trivially related to integration constants in solutions of the ODEs (\ref{eq:dTdtau_fm})-(\ref{eq:odediss}).

When considering objects such as attractor and repeller, it becomes useful to not take into account the temperature to focus only on the viscous effects in the time evolution.
Instead, one can look to the time evolution of the projected phase space $(\bar{\pi},\bar{\Pi},\tau)\in \widehat{\cal M}$ of flows defined by the effective inverse Reynolds numbers of the bulk pressure and shear viscous component~\footnote{This means to introduce projection from the fiber space, $(T,\bar{\pi}, \bar{\Pi})$, onto the subspace, $(\bar{\pi},\bar{\Pi})$; ${\cal F} \ni (T,\bar{\pi},\bar{\Pi}) \mapsto (\bar{\pi},\bar{\Pi}) \in \widehat{\cal F}$. 
The trivial bundle consisting of $\widehat{\cal F}$ and ${\cal T}$ as $\widehat{\cal M}=\widehat{\cal F} \times {\cal T}$ where ${\cal T}$ is the $\tau$-space (see Appendix~\ref{sec:const_nonauto} for the definitions).}.
Since flows on $\widehat{\cal M}$ depend on $T$, the initial condition of the temperature, $T(\tau_0)$, is also necessary to distinguish each flows.
In this sense, $T(\tau_0)$ has the role of a control parameter in the dynamical system on $\widehat{\cal M}$. 
By this procedure, one can find dimensions of the ISOFs projected on $\widehat{\cal M}$ which read as
\be
&& {\rm dim} \, \widehat{\cal S}(\frak{R}_1,{\frak A}) = 1, \qquad {\rm dim} \, \widehat{\cal S}(\frak{R}_2,{\frak A}) = 2, \qquad {\rm dim} \, \widehat{\cal S}(\frak{R}_3,{\frak A}) = 3, \nl 
&& {\rm dim} \, \widehat{\cal S}(\frak{R}_4,{\frak A}) = 2, \qquad {\rm dim} \, \widehat{\cal S}(\frak{R}_5,{\frak A}) = 3. \label{eq:dim_ISOF_pj}
\ee
Notice that the dimensions of $\widehat{\cal S}(\frak{R}_{4,5},{\frak A})$ given in Eq.(\ref{eq:dim_ISOF}) do not change compared with the ones quoted in the previous expression. This is due to the need of fine-tuning for ${\cal S}(\frak{R}_{4,5},{\frak A})$ which does not affect to the change of dimensions. 

From now on, for convenience, we call $\widehat{\cal S}(\frak{R}_1,{\frak A})$ `attractor solution'~\footnote{In Ref.~\cite{Jaiswal:2021uvv} the authors identified $\frak{R}_1$ as the UV CP of the 'attractor solution'. We verify that this UV CP coincide with the one found in our work after matching the slightly different definitions of the inverse Reynolds numbers taken in our work and theirs.}.
The reason for this definition is because the neighboring flows around ${\frak R}_1$ and ${\frak A}$ are attracted to it without fine-tuning of $T$, and thus, it is a one-dimensional subspace in $\widehat{\cal M}$ which is reflected in the value of the index given in Eq.(\ref{eq:dim_ISOF_pj}). 
Our definition of attractor solution does not impose specific behaviors at the intermediate region of the flow time, such as attracting/repelling behavior and its rate of convergence to it\footnote{The meaning of `attractor solution' frequently advocated in non-equilibrium hydrodynamics is extremely ambiguous  when analyzing variables in the phase space of higher dimensional dynamical systems~\cite{kloeden2011nonautonomous,caraballo2016applied}.}.
The behavior in the intermediate flow time region is investigated in Sect.\ref{sec:attractor}.

Let us finally make a few comments on the UV and IR limit for the massive Bjorken flows.
For all the UV CPs, the speed of sound becomes $c_s^2=1/3$ in the UV limit, and whether the UV limit satisfies exactly the traceless condition is determined by the value of $\bar{\Pi}_{\rm UV}$ in Eq.(\ref{eq:UV_cp}).
 ${\frak R}_{2,3,4}$ obey exactly the traceless condition while the other UV CPs do not satisfy it. The speed of sound is zero exactly at the equilibrium point ${\frak A}$ despite the fact that $\bar{\Pi}_{\rm IR}=0$. Therefore, it is the D-symmetry broken equilibrium point.
One should remind that Eq.(\ref{eq:cs2}) gives $c_s^2(z) \sim 1/z = 1/(m \beta)$ when $T$ is sufficiently small relative to the particle mass.
This implies 
that the $\beta$-dependence of the transport coefficients and the speed of sound has to be taken into account to see the IR physics by breaking the D-symmetry and thus, the massless limit of the theory does not commute with the IR limit.


\section{Transseries solution} \label{sec:formal_trans}
In this section we present the new transseries solutions for non-conformal fluids. For pedagogical purposes and in order to illustrate the generic structure of the new transseries solutions, we start our presentation by focusing first on the non-conformal perfect fluid case.

\subsubsection{Massive perfect fluid}
\label{subsect:PF}

\begin{figure}
\centering
\includegraphics[width=0.45\textwidth]{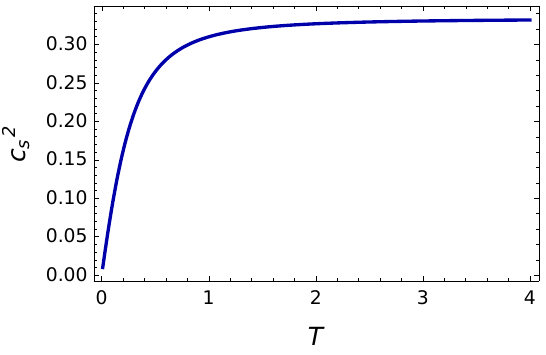} 
\includegraphics[width=0.45\textwidth]{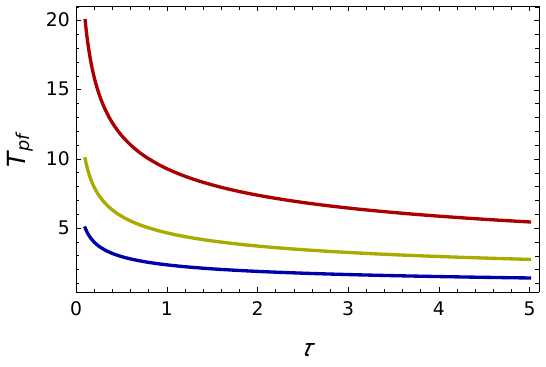}
\caption{Left Panel: The speed of sound as a function of temperature. Right Panel: The numerical solution of Eq.(\ref{eq:dTpfdtau}) for three different initial temperatures, $T_{\rm pf}(\tau_0)=5, 10, 20$ at $\tau_0=0.1$.
}
\label{fig:cs2_PFT}
\end{figure}

 The perfect fluid case is a good example to illustrate how the late-time asymptotic dynamics is changed dramatically by the breaking of conformal symmetry. As a matter of fact, the mass of the particle modifies the speed of sound $c_s^2(T)$~\eqref{eq:cs2} (see Fig.\ref{fig:cs2_PFT}) and determines the asymptotic properties of the IR transseries solution of the temperature $T_{\rm pf}$ compared with the conformal case~\cite{Baier:2007ix}.


The perfect fluid (${\rm PF}$) sector is derived from Eq.(\ref{eq:dTdtau_fm}) under the condition that there are no viscous effects at all. Therefore, the generic solution to the perfect fluid will not have any dependence on the relaxation time~\eqref{eq:tauR}. This results into a single equation that dictates the dynamics of the temperature as a function of proper time $\tau$, i.e., 
\be
\frac{d T_{\rm pf}}{d \tau} = - \frac{T_{\rm pf}}{\tau} c_s^2(1/T_{\rm pf})~. \label{eq:dTpfdtau}
\ee
When $c_s^2$ is a constant, one gets a simple scaling solution $T(\tau)\sim\tau^{-c_s^2}$. In particular, for the conformal case one has $c_s^2=1/3$. However, when mass effects are included Eq.~\eqref{eq:dTpfdtau} it  becomes a highly nonlinear ODE. In Fig.\ref{fig:cs2_PFT} it is shown that the speed of sound is a monotonically increasing function of $T$, approaching $c_s^2=1/3$ for large temperatures. Thus, at early times the power law conformal scaling solution is a good approximation. In contrast at late times, both the temperature and speed of sound $c_s^2$ almost vanish asymptotically. In this IR asymptotic regime we can approximate the speed of sound as follows:
\be
\label{eq:approxcs2}
c_s^2(1/T) \sim T - \frac{T^2}{2} - \frac{33 T^3}{8} + \cdots~.
\ee
By expanding the ODE~\eqref{eq:dTpfdtau} around $T=0$ while keeping only the first leading term of Eq.~\eqref{eq:approxcs2} yields to the dominant term of the solution for the temperature, i.e., $T_{\rm pf}(\tau) \sim (\log \tau)^{-1}$. Since the speed of sound $c_s^2(T)$ admits a Taylor expansion as a function of $T$, the subleading correction at large $\tau$ can be computed by considering a perturbation to the behavior of $T_{\rm pf}(\tau) \sim (\log \tau)^{-1}$ and subsequent linearization of the system. Solution of this problem implies that $T_{\rm pf}(\tau)$ becomes a series expansion in terms of $\logt \tau:= \log \log \tau$ transmonomials as seen from Eq.(\ref{eq:dTpfdtau}) (see Appendix \ref{sec:trans_late} for details).
As a result, the transseries solution for $T_{\rm pf}$ is given by
\be
&& T_{\rm pf}(\tau) = (\log \tau)^{-1}  \sum_{n_2=0,n_3=0}^{\infty,n_2} a_T^{[{\bf 0},(0,n_2,n_3)]} (\log \tau)^{-n_2} (\logt \tau)^{n_3}, \label{eq:trans_Tpf}
\ee
with the real coefficients $a_T^{[{\bf 0},(0,n_2,n_3)]}$ obtained as
\be
&& a_T^{[{\bf 0},(0,0,0)]} = 1, \qquad a_T^{[{\bf 0},(0,1,1)]} = \frac{1}{2}, \qquad a_T^{[{\bf 0},(0,1,0)]} = \sigma_T, \nl
&& a_T^{[{\bf 0},(0,2,2)]}= \frac{1}{4}, \qquad a_T^{[{\bf 0},(0,2,1)]} = \sigma_T-\frac{1}{4}, \qquad a_T^{[{\bf 0},(0,2,0)]} = \sigma_T^2-\frac{\sigma_T}{2}-\frac{35}{8}, \qquad \cdots \label{eq:T_pf} 
\ee
where $\sigma_T$ is the integration constant. From the previous expressions it is clear that some of the transseries coefficients are $\sigma_T$-dependent. It is also worth to mention that the leading order term of the solution does {\it not} depend on $\sigma_T$, and thus all flows converge universally to the equilibrium when $T(\tau)=(\log \tau)^{-1}$ for any integration constant $\sigma_T$. To the best of our knowledge it is the first time that a new ideal non-conformal fluid IR transseries solution~\eqref{eq:trans_Tpf} is found. 

It should not come as a surprise that the leading order behavior appearing in the general solution~\eqref{eq:trans_Tpf} is a logarithm that depends on the particle mass. In the perfect fluid case there is a mass gap $m$ so no excitations can appear if $T<m$. As a result, the mass gap generates the logarithmic behaviour. This fact can be clearly seen by explicitly restoring $m$ in the transseries as $1/\log \tau \rightarrow 1/\log(m \tau)$ and $\logt \tau \rightarrow \logt(m \tau)$, and thus, $T_{\rm pf}(\tau) \sim m/\log (m \tau)$.
This also implies that the IR limit and the massless limit do {\it not} commute with each other because $\lim_{m \rightarrow 0_+} T_{\rm pf}(\tau) = 0$ at fixed $\tau$, which contradicts the conformal temperature behaviour $T(\tau) \sim \sigma_T \tau^{-1/3}$.

An interesting aspect of the transseries solution for the perfect fluid behavior~(\ref{eq:trans_Tpf}) is that it is convergent.
This is understood in simple terms since the Borel transform of Eq.~\ref{eq:trans_Tpf} does not have singularities on the positive real axis \cite{sauzin2014introduction,delabaere2016divergent}.
In order to show it, let us change the variable ${\frak t}:=\log \tau$ in Eq.~(\ref{eq:dTpfdtau}) so we get
\be
\frac{d{T}_{\rm pf}}{d {\frak t}} &=&  - {T}_{\rm pf} c_s^2(1/{T}_{\rm pf}) = - {T}_{\rm pf}^2 + O({T}_{\rm pf}^3)\,. \label{eq:dTpfdtau_mod}
\ee
Then the solution $T_{\rm pf}$ to the previous equation is written as a power series expansion of $1/{\frak t}$ and $\log {\frak t}$.
Using the elementary formula,
\be
{\frak t}^{-n_1} (\log {\frak t})^{n_2} = (-1)^{n_2} \left. \frac{\ppd^{n_2} {\frak t}^{-\sigma}}{\ppd \sigma^{n_2}} \right|_{\sigma=n_1} \quad \mbox{with} \quad \sigma \in {\mathbb C}, \, n_1 \in {\mathbb N}, \, n_2 \in {\mathbb N}_0,
\ee
the Borel transform ${\cal B}$ of the expansion (\ref{eq:T_pf}) is defined as \cite{edgar2010transseries}
\be
{\cal B}[{\frak t}^{-n_1}] = \frac{\xi^{n_1-1}}{\Gamma(n_1)}, \qquad 
{\cal B}[{\frak t}^{-n_1}(\log {\frak t})^{n_2}] = (-1)^{n_2} \left. \frac{\ppd^{n_2}}{\ppd \sigma^{n_2}}\frac{\xi^{\sigma-1}}{\Gamma(\sigma)} \right|_{\sigma=n_1}, \quad \sigma \in {\mathbb C} \setminus {\mathbb Z}_{\le 0}, \, n_1 \in {\mathbb N}, \, n_2 \in {\mathbb N}_0, \label{eq:B_trans}
\ee
where $\xi$ is the Borel parameter.
Thus, the Borel transform for $T_{\rm pf}$ is performed using Eq.(\ref{eq:B_trans}), and 
acting with it on both sides of Eq.(\ref{eq:dTpfdtau_mod}) yields
\be
-\xi \widehat{T}_{\rm pf} = - \widehat{T}_{\rm pf} * \widehat{T}_{\rm pf} + O(\widehat{T}_{\rm pf}^{3*})
\qquad \Rightarrow \qquad  \widehat{T}_{\rm pf} = \xi^{-1} \left[ \widehat{T}_{\rm pf} * \widehat{T}_{\rm pf} + O(\widehat{T}_{\rm pf}^{3*})\right]~ , \label{eq:dTpfdtau_B}
\ee
where $\widehat{T}_{\rm pf}:={\cal B}[T_{\rm pf}]$, $*$ is the convolution product defined as \cite{edgar2010transseries}
\be
&& (F*G)(\xi) := \int_0^{\xi} d \xi_1 \, F(\xi_1) G(\xi-\xi_1), \qquad F^{n*}(\xi) := (\underbrace{F* \cdots *F}_{\mbox{$n$-times}})(\xi)~ ,
\ee
and we used ${\cal B}\left[\frac{d T_{\rm pf}}{d {\frak t}} \right]=-\xi \widehat{T}_{\rm pf}$.
One can find that $\widehat{T}_{\rm pf}^{n*}=O(\xi^{n-1})$ because $\widehat{T}_{\rm pf}^{n*} \sim {\cal B}[{\frak t}^{-n}] = \xi^{n-1}/\Gamma(n)$.
Thus, the r.h.s. in Eq.(\ref{eq:dTpfdtau_B}) is regular for $\xi \in {\mathbb R}_{>0}$. 

Physical implications of this convergent mathematical property is that the transient non-hydrodynamic modes that render the asymptotic series come entirely from the dynamics of the system when the viscous contributions are present. These transient modes are fully absent in the PF case. On the other hand, in the conformal Bjorken case the IR perturbative behavior is written as a power series $\sim 1/\tau^k$ (being $k$ the order of the gradient expansion). The transseries solution~\eqref{eq:trans_Tpf} shows that this specific power law asymptotic behavior does not hold exactly for the non-conformal case even in the PF scenario.

The dissipative corrections are build out of gradients of the perfect fluid quantities and thus, it is often advocated in the literature that the strength of the gradient entirely determines the perturbative hydrodynamic series~\cite{Romatschke:2017ejr,Baier:2007ix,Florkowski:2017olj}. As we shall see, the strength of the gradient is not enough to guarantee us a well defined perturbative series. Our work shows that the value of the transport coefficients together with the breaking of conformality changes completely the way we understand the perturbative series expansion in hydrodynamics. 
As a consequence, our method paves the way to classify all possible transseries structures by their dependence on the strength of the interactions via the $\theta_0$ parameter, namely
\be
\mbox{Perfect Fluid (PF) sector} \quad &:& \quad \mbox{ $\theta_0$ independent} 
\nl
\mbox{Perturbative (PT) sector} \quad &:& \quad \mbox{Power expansion of $\theta_0$} 
\nl
\mbox{Non-perturbative (NP) sector} \quad &:& \quad \mbox{Exponential damping term of $\theta_0^{-1}$ $\times$ power expansion of $\theta_0$} 
\nn
\ee
Further details of this classification are presented in the next section. Our approach does not have an analogy with the standard non-perturbative expansions usually discussed in the literature~\cite{Marino:2015yie}. As we shall see in the next section, this classification allows us to understand physically the meaning of the complicated structure of the transseries solutions for this model.


\begin{figure}[tb]
\begin{center}
\includegraphics[width=0.6\textwidth]{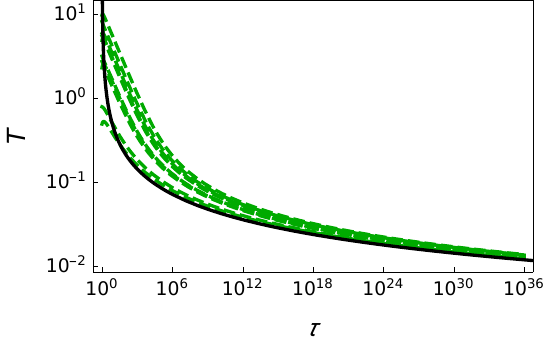}
\includegraphics[width=0.55\textwidth]{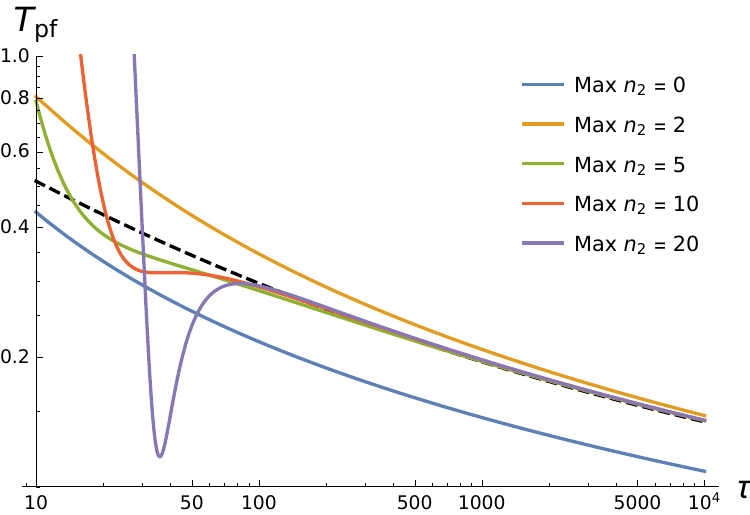}
\caption{Top panel: IR behaviour of the temperature $T$ vs $\tau$ in dimensionless units since $m=1$. The green dashed lines are flows randomly generated initial condition in the range of $T(\tau_0) \in [0.1,1]$ at $\tau_0=1$.
The parameters are chosen $\bar{\Delta}=0.75$, $\theta_0=1$.
The solid red line is the leading term in Eq.(\ref{eq:trans_Tpf}), $T(\tau)=(\log \tau)^{-1}$. 
Bottom panel: comparison of the exact numerical solution of the perfect non-conformal fluid (black dotted line) and the PF transseries solution for a given finite number of terms characterized by the summation index $n_2$ in Eq.~\eqref{eq:trans_Tpf}.}
\label{fig:T_tau}
\end{center}
\end{figure}


We conclude this section by providing some comparisons between the PF transseries solution~\eqref{eq:trans_Tpf} and exact numerical solutions for the temperature in the perfect fluid case. 
In Fig.\ref{fig:T_tau} the IR behavior of the temperature is shown. In the top panel we present numerical solutions of the perfect fluid for the temperature with arbitrary initial conditions (green dashed lines) and the dominant asymptotic term $T(\tau) \sim 1/\log \tau$ (black line). From this figure we conclude that all the flow lines converge to this asymptotic value and thus, the forward attractor of the temperature is described by this term. We also verify that adding the first subleading term $\sim 1/[2(\log \tau)^2] \log_{(2)}\tau$ does not affect the result when $\tau \geq  O(10^{30})$. This subleading term is also insensitive to the initial conditions (see Eq.~\eqref{eq:T_pf}). We strongly emphasize to the reader that the matching between the forward attractor and solutions with arbitrary initial conditions occurs at extreme late times. This happens because the transient terms, which depend on the initial conditions, die out slowly due to their logarithmic dependence with the mass. This is a highly non-trivial effect induced only by the breaking of conformal symmetry via the non-trivial equation of state~\eqref{eq:cs2}. Further consequences of this feature will be discussed in the next section. Now in Fig.~\ref{fig:T_tau}, bottom panel, we compare an exact numerical perfect fluid solution for the temperature and a truncated PF transseries. The truncation is related with the maximum number of terms included in the summation index $n_2$ in Eq.~\eqref{eq:trans_Tpf}. In the bottom panel of Fig.~\ref{fig:T_tau} we consider $n_2=\{0,2,5,10,20\}$. At intermediate times we observe that there is a better agreement between the exact result and the truncated PF transserien when $n_2$ increases. In particular, there is a substantial improvement between the truncated PF transseries and the exact result when $\tau\in [100,10^4]$. 
For $\tau\lesssim 100$,  strong disagreements are found for the summation indexes implemented in our numerical algorithm. This result is not unexpected.
The PF transseries solution~\eqref{eq:trans_Tpf} does not have singularities in the Borel plane, but it only means that the radius of convergence is nonzero. 
So large deviations from the exact solution below $\tau\lesssim 100$ can be interpreted as a regime out of the radius.
In contrast, in $\tau \gtrsim 100$, we could obtain a good agreement with the exact solution by performing numerical fitting of $\sigma_T$ even for a small truncation order such as $n_2=5$.

\subsection{Full transseries for the IR regime} \label{sec:transs_IR}
In this section we consider the transseries for the IR regime.
For pedagogical purposes, we present briefly our results before looking at their derivation details.
The full transseries solutions to the non-conformal hydrodynamical equations~\eqref{eq:dTdtau_fm} and~\eqref{eq:odediss} are
\be
\label{eq:fulltranssols}
T(\tau) = \mu_2 \sum_{{\bf m},{\bf n}} a_{T}^{[{\bf m},{\bf n}]} \bm{\zeta}^{\bf m} \bm{\mu}^{\bf n}, \qquad  \bar{X}(\tau) = \sum_{{\bf m},{\bf n}} a_{\bar{X}}^{[{\bf m},{\bf n}]} \bm{\zeta}^{\bf m} \bm{\mu}^{\bf n}, \qquad \bar{X} \in \{ \bar{\pi}, \bar{\Pi} \}, \label{eq:full_IR_trans}
\ee
where $a^{[{\bf m},{\bf n}]}_{T,\bar{X}}$ are real coefficients while the  transmonomials are given by
\begin{subequations}
\begin{align}
& \mu_1 := \frac{\theta_0 (\log \tau)^{\bar{\Delta}}}{\tau}, \qquad \mu_2 := \frac{1}{\log \tau}, \qquad \mu_3 := \logt \tau = \log \log \tau, \label{eq:mu} \\
& \zeta_{i=1,2} = \sigma_i \frac{\exp \left[ -S(\tau) \right]}{\tau^{\beta_i} (\log \tau)^{\gamma_i}}, \qquad S(\tau) = \mu_1^{-1} \sum_{n_2=0,n_3=0}^{\infty,n_2} a_S^{[(n_2,n_3)]} \, \mu_2^{n_2} \mu_3^{n_3}, \label{eq:zeta}
\end{align}
\end{subequations}
with the integration constants $\sigma_{i=1,2}$ and real constants $\beta_{i=1,2},\gamma_{i=1,2} \in {\mathbb R}$, and the symbolic summation $\sum_{{\bf m},{\bf n}}$ is defined as
\be
 \sum_{{\bf m},{\bf n}} a^{[{\bf m},{\bf n}]} \bm{\zeta}^{\bf m} \bm{\mu}^{\bf n} &:=&  \sum_{n_2=0,n_3=0}^{\infty,n_2} a^{[{\bf 0},(0,n_2,n_3)]} \, \mu_2^{n_2} \mu_3^{n_3} + \sum_{n_1=1,n_2=2,n_3=0}^{\infty,\infty,n_2-2} a^{[{\bf 0},{\bf n}]} \, \mu_1^{n_1} \mu_2^{n_2} \mu_3^{n_3}  \nl
&& + \sum_{\substack{{\bf m}\in {\mathbb N}_0^{2} \\ |{\bf m}|>0}}  \sum_{n_1=|{\bf m}|-1,n_2=-|{\bf m}|+1,n_3=0}^{\infty,\infty,n_1+n_2} a^{[{\bf m},{\bf n}]} \, \zeta_1^{m_1} \zeta_2^{m_2}  \mu_1^{n_1} \mu_2^{n_2} \mu_3^{n_3}, \label{eq:transseries_IR}
\ee
For convention, we decompose the transseries solution in Eq.(\ref{eq:full_IR_trans}) into the perfect fluid (PF) sector, the perturbative (PT) sector, and the non-perturbative (NP) sector as
\be
&& T(\tau) = T_{\rm pf}(\tau) + T_{\rm pt}(\tau) + T_{\rm np}(\tau), \qquad \bar{X}(\tau) =  \bar{X}_{\rm pt}(\tau) + \bar{X}_{\rm np}(\tau).
\ee
These are classified by their dependence on $\theta_0$ as follows:
\be
\mbox{PF sector} \quad &:& \quad \mbox{No dependence of $\theta_0$} \  [\mu_2, \mu_3] \nl
\mbox{PT sector} \quad &:& \quad \mbox{Power expansion of $\theta_0$} \ [\mu_1, \mu_2, \mu_3] \nl
\mbox{NP sector} \quad &:& \quad \mbox{Exponentally damping term of $\theta_0^{-1}$ $\times$ power expansion of $\theta_0$} \ [\zeta_i, \mu_1, \mu_2, \mu_3] \nn
\ee
where $[\cdots]$ denote the respective transmonomials of the associated sector.
The full transseries are obtained from the lower sectors by providing the speed of sound as an input. As we shown in the previous section, the PF sector of the temperature is determined entirely by the speed of sound, Eq.(\ref{eq:cs2}). Once $T_{\rm pf}$ is known, the leading order asymptotic terms for $(\bar{\pi}_{\rm pt},\bar{\Pi}_{\rm pt})$ are computed from Eqs.(\ref{eq:dpidtau_fm}) and (\ref{eq:dPIdtau_fm}), and then, the leading order of $T_{\rm pt}$ and higher orders terms for all variables can be computed recursively from the ODEs. This same strategy holds when calculating the NP sector after calculating the PT sector. The outline of this procedure is shown in Fig.\ref{fig:det_sectors}. The technical derivation of the calculations carried out to compute the transmonomials and transseries is explained in appendix \ref{sec:trans_late}.
\subsubsection{Perturbative sector}
\begin{figure}[tb]
\begin{center}
\includegraphics[width=.42\textwidth]{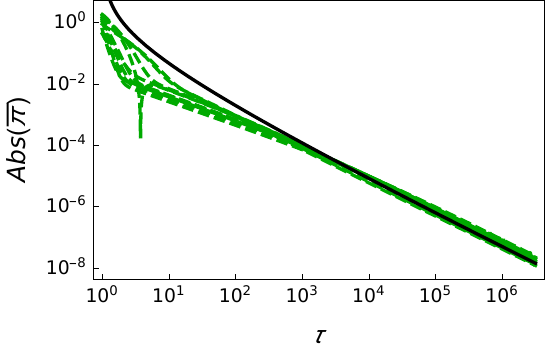} \ \ \
\includegraphics[width=.42\textwidth]{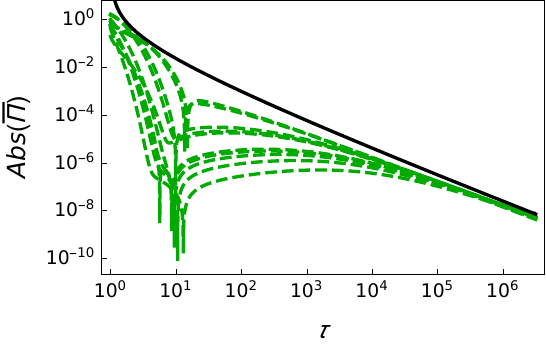}
\caption{
The IR behaviour of the viscosities.
The green dashed lines are randomly generated initial condition in the range of $\bar{X}(\tau_0) \in [-2.0,2.0]$ at $\tau_0=1.0$ for $\bar{X} \in \{ \bar{\pi}, \bar{\Pi}\}$.
The parameters are chosen as $\bar{\Delta}=0.75$ and $\theta_0=1.0$.
The solid black line is the leading term in Eq.(\ref{eq:X_pt}).
}
\label{fig:pi_tau}
\end{center}
\end{figure}

Next, let us look more closely at the PT sector. The leading order dominant term of the solutions for the viscous fluid variables is determined by  asymptotic form of the RHS of Eqs.(\ref{eq:dpidtau_fm}) and (\ref{eq:dPIdtau_fm}), which read as
\be
&& - \frac{T_{\rm pf}^{\bar{\Delta}}}{\theta_0} \bar{\pi}_{\rm pt} + \frac{4}{3\tau} \bar{\beta}_\pi(1/T_{\rm pf}) \sim 0, \qquad - \frac{T_{\rm pf}^{\bar{\Delta}}}{\theta_0} \bar{\Pi}_{\rm pt} - \frac{1}{\tau} \bar{\beta}_\pi(1/T_{\rm pf}) \sim 0. \label{eq:lead_Xpt}
\ee
Clearly, the dependence of $T_{\rm pf}$ enters explicitly in the asymptotic form of the ODEs, such that the PT sectors of $\bar{\pi}_{\rm pt}$ and $\bar{\Pi}_{\rm pt}$ are affected and carry information of the PF sector. 
Hence, the PT sector inherits an expansion in terms of $(\log \tau)^{-1}$ and $\logt  \tau$. As a result, the new transmonomial generated from the PF sector reads  Eq.(\ref{eq:lead_Xpt}) as
\be
\label{eq:transm_nonconf}
\left( \frac{\theta_0}{T_{\rm pf}^{\bar{\Delta}}} \right) \frac{1}{\tau} = \frac{\theta_0(\log \tau)^{\bar{\Delta}}}{\tau} \left( 1 + \mathcal{O}(\logt \tau/\log \tau) \right).
\ee
The LHS of this expression is nothing else than the late-time value of the  Knudsen number which measures effectively the competition between the collisions, determined mostly by the strength of the relaxation time scale~\eqref{eq:tauR}, i.e. $\theta_0$, and the expansion rate $\sim 1/\tau$. This term also appears in conformal systems~\cite{Behtash:2019txb}. However, in the non-conformal case and due to the PF sector of the temperature~\eqref{eq:trans_Tpf}, the late-time Knudsen number is modified and receive logarithmic contributions which effectively increase its size at late times compared with the conformal case. In the conformal case the Knudsen number drops faster due to its power law decay~\cite{Behtash:2019txb}. Thus, the PT sector is given by
\begin{subequations}
\begin{align}
& T_{\rm pt}(\tau) = (\log \tau)^{-1} \sum_{n_1=1,n_2=2,n_3=0}^{\infty,\infty,n_2-2} a_T^{[{\bf 0},{\bf n}]} \, \left[ \frac{ \theta_0(\log \tau)^{\bar{\Delta}}}{\tau} \right]^{n_1} (\log \tau)^{-n_2} (\logt \tau)^{n_3}, \label{eq:T_pt} \\
&  \bar{X}_{\rm pt}(\tau) =  \sum_{n_1=1,n_2=2,n_3=0}^{\infty,\infty,n_2-2} a_{\bar{X}}^{[{\bf 0},{\bf n}]} \, \left[ \frac{ \theta_0(\log \tau)^{\bar{\Delta}}}{\tau} \right]^{n_1} (\log \tau)^{-n_2} (\logt \tau)^{n_3}. \label{eq:X_pt}
\end{align}
\end{subequations}
$\theta_0$ appears explicitly only in the term $\mu_1$~\eqref{eq:mu} while the $\bar{\Delta}$-dependence of the relaxation time scale~\eqref{eq:tauR} propagates into the coefficients of the PT expansion. Some of the coefficients appearing in the previous expression are given explicitly as follows:
\be
&& a_T^{[{\bf 0},(1,2,0)]} = -2, \qquad a_T^{[{\bf 0},(1,3,1)]} = \bar{\Delta} -3, \qquad a_T^{[{\bf 0},(1,3,0)]} = 2 \bar{\Delta} \sigma_T-2 \bar{\Delta}-6 \sigma_T+13, \qquad \cdots \nl
&& a_T^{[{\bf 0},(2,2,0)]} = 1, \qquad a_T^{[{\bf 0},(2,3,1)]} = \frac{1}{2} (3-2 \bar{\Delta}), \qquad a_T^{[{\bf 0},(2,3,0)]} = -2 \bar{\Delta} \sigma_T+2 \bar{\Delta}+3 \sigma_T-15, \qquad \cdots  \nl
&& a_{\bar{\pi}}^{[{\bf 0},(1,2,0)]} = \frac{4}{3}, \qquad a_{\bar{\pi}}^{[{\bf 0},(1,3,1)]} = -\frac{2}{3}  (\bar{\Delta}-2), \qquad a_{\bar{\pi}}^{[{\bf 0},(1,3,0)]} = -\frac{4}{3} (\bar{\Delta} \sigma_T-2 \sigma_T+4), \qquad \cdots \nl
&& a_{\bar{\pi}}^{[{\bf 0},(2,2,0)]} = -\frac{4}{3}, \qquad a_{\bar{\pi}}^{[{\bf 0},(2,3,1)]} = \frac{4}{3} (\bar{\Delta}-1), \qquad a_{\bar{\pi}}^{[{\bf 0},(2,3,0)]} = \frac{4}{3} (2 \bar{\Delta} \sigma_T-\bar{\Delta}-2 \sigma_T+14), \qquad \cdots \label{eq:pi_pt} \nl
&& a_{\bar{\Pi}}^{[{\bf 0},(1,2,0)]} = -\frac{2}{3}, \qquad a_{\bar{\Pi}}^{[{\bf 0},(1,3,1)]} = \frac{1}{3}(\bar{\Delta}-2), \qquad a_{\bar{\Pi}}^{[{\bf 0},(1,3,0)]} = \frac{1}{3} (2 \bar{\Delta} \sigma_T-4 \sigma_T+17),  \qquad \cdots \nl
&& a_{\bar{\Pi}}^{[{\bf 0},(2,2,0)]} = \frac{2}{3}, \qquad a_{\bar{\Pi}}^{[{\bf 0},(2,3,1)]} = -\frac{2}{3} (\bar{\Delta}-1), \qquad a_{\bar{\Pi}}^{[{\bf 0},(2,3,0)]} = -\frac{1}{3} (4 \bar{\Delta} \sigma_T-2 \bar{\Delta}-4 \sigma_T + 31),  \qquad \cdots \label{eq:PI_pt}
\ee
Fig.\ref{fig:pi_tau} shows the asymptotic behavior of $\bar{X}_{\rm pt}$.
Similarly to the PF sector, the leading terms of the PT sector do not depend on $\sigma_T$ and universally converge to the local equilibrium. It is important to emphasize that the convergence occurs at long times and thus, hydrodynamization gets delayed due to the breaking of conformal symmetry. The explanation for the delay of hydrodynamization in the non-conformal case is explained in the next section.

\subsubsection{Non-perturbative sector}
Finally, we consider the NP sector. This sector correspond to the transient non-hydrodynamic modes which are characterized by the exponentially damping term in $\zeta_i$ in Eq.(\ref{eq:mu}). These modes die off prior to its arrival at the IR regime. In other words, the system hydrodynamizes when these transient modes of the transseries solution have effectively vanished. In this sense, the NP sector has almost no influence in the IR dynamics but interestingly, the terms entering in $\zeta_i$~\eqref{eq:mu} are determined entirely by the PF sector as we shall see below. In the conformal case, the transient exponential modes are uniquely determined by the PT sector of the transseries~\cite{Heller:2015dha,Basar:2015ava,Heller:2016rtz,Aniceto:2018uik,Behtash:2018moe,Behtash:2019txb,Behtash:2020vqk,Blaizot:2020gql,Blaizot:2021cdv} which corresponds to the hydrodynamic gradient expansion. On the other hand, far from the IR regime the NP sector becomes the dominant mechanism due to the intrinsic non-linear nature of the non-conformal fluid equations. 

The term $S(\tau)$ in the exponentially decaying damping rate~\eqref{eq:zeta} is determined by the PF sector in the collision part of the ODEs for the dissipative variables and it is computed by simply solving the following equation
\be
\frac{d S}{d \tau} = \frac{T_{\rm pf}^{\bar{\Delta}}}{\theta_0}, \label{eq:dSdtau}
\ee
whose general solution is obtained given the PF sector for the temperature, i.e.,
\be
\label{eq:timedep_lyap}
S(\tau) = \frac{\tau}{\theta_0 (\log \tau)^{\bar{\Delta}}} \sum_{n_2=0,n_3=0}^{\infty,n_2} a_{S}^{[(n_2,n_3)]} (\log \tau)^{-n_2} (\logt \tau)^{n_3}\,.
\ee
Some of the coefficients entering into the previous expression are quoted below
\be
\label{eq:listcoeff}
&& a_S^{[(0,0)]} = 1, \qquad  a_S^{[(1,1)]} = \frac{\bar{\Delta}}{2}, \qquad a_S^{[(1,0)]} = \bar{\Delta}(\sigma_T + 1), \nl
&& a_S^{[(2,2)]}= \frac{\bar{\Delta}}{8}(\bar{\Delta}+1), \qquad a_S^{[(2,1)]}= \frac{\bar{\Delta}}{2} \left[ (\bar{\Delta} +1) \sigma_T +  \bar{\Delta} + \frac{1}{2} \right], \nl
&& a_S^{[(2,0)]} = \frac{\bar{\Delta}}{2} \left[ (\bar{\Delta} +1) \sigma_T^2 + (2 \bar{\Delta}+1) \sigma_T + 2 \bar{\Delta} - \frac{31}{4}\right], \cdots
\ee
From the previous expressions, the leading dominant term in the solution~\eqref{eq:timedep_lyap} is nothing else than the late-time inverse Knudsen number $Kn^{-1}$. As we pointed out in the previous section, the logarithms of the mass enhance the value of the late-time Knudsen number and thus, the decay rate of the non-hydrodynamic modes $S(\tau)$ decreases which effectively slows down the arrival towards the asymptotic forward attractor. This naturally explains why in the non-conformal model studied here hydrodynamization happens at extreme long time scales. Now, strictly speaking $S(\tau)$~\eqref{eq:timedep_lyap} is not a monomial but a expansion in terms of $(\log \tau)^{-1}$ and $\logt \tau$ multiplied by the asymptotic form of the inverse Knudsen number $Kn^{-1}=\tau/[\theta_0(\log \tau)^{\bar{\Delta}}]$. This is one of the most remarkable differences with the conformal case and again, these non-trivial modifications are a consequence of the breaking of the conformal symmetry and the changes induced in $T_{\rm pf}$~\eqref{eq:trans_Tpf}.
In addition, $\exp[-S(\tau)]$ can not be expanded around $\tau=+\infty$ and becomes zero by taking the IR limit~\footnote{An interesting case arises when the relaxation time scale is a constant, i.e. $\bar{\Delta}=0$ in Eq.~\eqref{eq:tauR}, then only the term $a_S^{[(0,0)]} = 1$ survives in Eq.~\eqref{eq:listcoeff} while the other ones vanish exactly. For this particular case this does not mean that the perfect fluid sector vanishes identically in the full transseries solution of the temperature $T$. Instead, the precise form of $S(\tau)$ emerges from the non-linear couplings of the temperature and the viscous dissipative variables.}. The other constants appearing in $\zeta_i$~\eqref{eq:zeta}, $\beta_i$ and $\gamma_i$ are given by respectively (see appendix \ref{sec:trans_late})
\be
(\beta_1,\beta_2)=(2,0), \qquad (\gamma_1,\gamma_2)=(-7,-1).
\ee
The previous coefficients are determined by the PF sector of the temperature $T_{\rm pf}$ and the asymptotic form of the thermodynamic transport coefficients which couple linearly with $\bar{X}$ at the level of the ODEs.
The first NP sector generates higher exponentially damping parts through the nonlinear terms of $F_{\bar{\pi}}$ and $F_{\bar{\Pi}}$ in Eqs.(\ref{eq:dpidtau_fm}) and (\ref{eq:dPIdtau_fm}).
As a result, one can write down the NP sector of the transseries solutions as follows:
\begin{subequations}
\begin{align}
& T_{\rm np}(\tau) = (\log \tau)^{-1}  \sum_{\substack{{\bf m}\in {\mathbb N}_0^{2} \\ |{\bf m}|>0}}  \left[\prod_{i=1}^2  \frac{\sigma_i e^{-S(\tau)}}{\tau^{\beta_i}(\log \tau)^{\gamma_i}} \right] \sum_{n_1=|{\bf m}|-1,n_2=-|{\bf m}|+1,n_3=0}^{\infty,\infty,n_1+n_2} \, a_T^{[{\bf m},{\bf n}]}   \left[ \frac{ \theta_0(\log \tau)^{\bar{\Delta}}}{\tau} \right]^{n_1} (\log \tau)^{-n_2} (\logt \tau)^{n_3}, \nl \\ 
& \bar{X}_{\rm np}(\tau) =  \sum_{\substack{{\bf m}\in {\mathbb N}_0^{2} \\ |{\bf m}|>0}}  \left[\prod_{i=1}^2  \frac{\sigma_i e^{-S(\tau)}}{\tau^{\beta_i}(\log \tau)^{\gamma_i}} \right] \sum_{n_1=|{\bf m}|-1,n_2=-|{\bf m}|+1,n_3=0}^{\infty,\infty,n_1+n_2} \, a_{\bar{X}}^{[{\bf m},{\bf n}]} \left[ \frac{ \theta_0(\log \tau)^{\bar{\Delta}}}{\tau} \right]^{n_1} (\log \tau)^{-n_2} (\logt \tau)^{n_3}, \nl
\end{align}
\end{subequations}
where $|{\bf m}|:=m_1+m_2$, $\sigma_{i=1,2}$ is the integration constant.
The transeries has two ambiguous coefficients corresponding to the normalization of the integration constants $\sigma_{1,2}$.
By fixing the normalization, e.g. $a_{\bar{\pi}}^{[(1,0),(0,0,0)]} = a_{\bar{\Pi}}^{[(0,1),(0,0,0)]} = 1$, the remaining coefficients can be uniquely determined. When using this normalization, the coefficients of the non-perturbative sector of the temperature $T_{\rm np}$ are obtained as $a_{T}^{[(1,0),(0,n_2,n_3)]} = a_{T}^{[(0,1),(0,n_2,n_3)]} = 0$  for any $n_2, n_3 \in {\mathbb N_0}$, and the non-zero coefficients start when $n_1=1$ as follows:
\be
&& a_{T}^{[(1,0),(1,0,0)]} = -\frac{3}{2}, \qquad a_{T}^{[(1,0),(1,1,1)]} = \frac{3}{4}(\bar{\Delta}+6), \qquad a_{T}^{[(1,0),(1,1,0)]} = \frac{3}{4}(2 \bar{\Delta} \sigma_T + 4 \bar{\Delta} + 12 \sigma_T - 71), \qquad \cdots \nl
&& a_{T}^{[(0,1),(1,0,0)]} = 0, \ \ \ \qquad a_{T}^{[(0,1),(1,1,1)]} = 0, \qquad \qquad \quad \ \  a_{T}^{[(0,1),(1,1,0)]} = 0, \qquad \cdots
\ee
With the same normalization and for completeness, we quote below some of coefficients entering into the transseries for $\bar{X}_{\rm np}$ beginning with $n_1=0$ as follows:
\be
    && a_{\bar{\pi}}^{[(1,0),(0,0,0)]} = 1, \qquad a_{\bar{\pi}}^{[(1,0),(0,1,1)]} = - \frac{7}{2}, \qquad a_{\bar{\pi}}^{[(1,0),(0,1,0)]} = -2 \bar{\Delta} - 7\sigma_T + \frac{73}{2}, \qquad \cdots \nl
&& a_{\bar{\pi}}^{[(0,1),(0,0,0)]} = 1, \qquad a_{\bar{\pi}}^{[(0,1),(0,1,1)]} = - \frac{1}{2}, \qquad a_{\bar{\pi}}^{[(0,1),(0,1,0)]} = -2 \bar{\Delta} -\sigma_T + \frac{11}{2}, \qquad \cdots 
\nl
&& a_{\bar{\Pi}}^{[(1,0),(0,0,0)]} = -\frac{1}{2}, \qquad a_{\bar{\Pi}}^{[(1,0),(0,1,1)]} =  \frac{7}{4}, \ \ \, \qquad a_{\bar{\Pi}}^{[(1,0),(0,1,0)]} =  \bar{\Delta} + \frac{7}{2}\sigma_T - \frac{67}{4}, \qquad \cdots \nl
&& a_{\bar{\Pi}}^{[(0,1),(0,0,0)]} = 1, \ \ \ \qquad a_{\bar{\Pi}}^{[(0,1),(0,1,1)]} = - \frac{1}{2}, \qquad a_{\bar{\Pi}}^{[(0,1),(0,1,0)]} = -2 \bar{\Delta} -\sigma_T + \frac{11}{2}, \qquad \cdots
\ee

\subsubsection{IR Transseries of the non-conformal energy-momentum tensor}

The dissipative fluid variables $\bar{X} = \left(\bar{\pi},\bar{\Pi}\right)$ and the temperature $T$ are the fundamental building blocks for other observables that depend explicitly on those variables. In the remaining part of this section we show how the components of the energy momentum tensor of our non-conformal fluid model~\eqref{eq:EMtensor}  can be straightforwardly reconstructed from the transseries solutions~(\ref{eq:full_IR_trans}). One can start by writing down the energy density and thermodynamic pressure in Eqs.~\eqref{eq:def_epsP} and speed of sound~(\ref{eq:cs2}) by substituting the full transseries of the temperature $T$~\eqref{eq:fulltranssols}  into these expressions after expanding ${\rm K}_n(z)$ and $\alpha_n(z)$ around $z=\infty$. This yields into the following expressions
\be
&& \varepsilon(\tau) =  \frac{e^{\sigma_T}}{\sqrt{8 \pi^3}} \frac{\mu_2}{\tau} \sideset{}{^\prime}\sum_{{\bf m},{\bf n}} a_{\varepsilon}^{[{\bf m},{\bf n}]} \bm{\zeta}^{\bf m} \bm{\mu}^{\bf n}, \qquad P(\tau) = \frac{e^{\sigma_T}}{\sqrt{8 \pi^3}} \frac{\mu_2^{2}}{\tau} \sideset{}{^\prime} \sum_{{\bf m},{\bf n}} a_{P}^{[{\bf m},{\bf n}]} \bm{\zeta}^{\bf m} \bm{\mu}^{\bf n}, \qquad c_s^2(\tau) = \mu_2 \sum_{{\bf m},{\bf n}} a_{c_s^2}^{[{\bf m},{\bf n}]} \bm{\zeta}^{\bf m} \bm{\mu}^{\bf n},  \label{eq:trans_epsP} \nl
\ee
where the symbolic summation $\sideset{}{^\prime} \sum_{{\bf m},{\bf n}}$ in $\varepsilon(\tau)$ and $P(\tau)$ is defined as 
\be
\sideset{}{^\prime} \sum_{{\bf m},{\bf n}} a^{[{\bf m},{\bf n}]} \bm{\zeta}^{\bf m} \bm{\mu}^{\bf n} 
&:=&  \sum_{n_2=0,n_3=0}^{\infty,n_2} a^{[{\bf 0},(0,n_2,n_3)]} \mu_2^{n_2} \mu_3^{n_3}  + \sum_{n_1=1,n_2=1,n_3=0}^{\infty,\infty,n_2-1} a^{[{\bf 0},{\bf n}]} \mu_{1}^{n_1} \mu_2^{n_2} \mu_3^{n_3}  \nl
&& + \sum_{\substack{{\bf m}\in {\mathbb N}_0^{2} \\ |{\bf m}|>0}}  \sum_{n_1=|{\bf m}|-1,n_2=-|{\bf m}|,n_3=0}^{\infty,\infty,n_1+n_2+1} a^{[{\bf m},{\bf n}]} \, \zeta_1^{m_1} \zeta_2^{m_2}  \mu_1^{n_1} \mu_2^{n_2} \mu_3^{n_3}. \label{eq:transseries_IR_prime} 
\ee
The first terms of Eqs.~\eqref{eq:trans_epsP} look like
\begin{subequations}
\begin{align}
{\varepsilon}(\tau) &\sim \frac{e^{\sigma_T}}{\sqrt{8 \pi^3} \tau \log \tau}  \left[ 1  + \frac{\logt \tau}{2 \log \tau} - \frac{1-\sigma_T}{\log \tau} + \cdots  \right], \\ \nl
P(\tau) &\sim \frac{e^{\sigma_T}}{\sqrt{8 \pi^3} \tau (\log \tau)^2} \left[1  +  \frac{\logt \tau}{\log \tau} - \frac{5-4\sigma_T}{2 \log \tau} + \cdots \right], \\ \nl
c_s^2(\tau) &\sim \frac{1}{\log \tau} + \frac{\logt \tau}{2(\log \tau)^2} - \frac{1-2\sigma_T}{2(\log \tau)^2} + \cdots \nl
& - \frac{\theta_0}{\tau (\log \tau)^{3-\bar{\Delta}}} \left[ 2 + \frac{(3-\bar{\Delta})\logt \tau}{\log \tau} - \frac{15 - 2 \bar{\Delta} - 2 \sigma_T(3-\bar{\Delta})}{\log \tau} + \cdots \right] + \cdots.
\end{align}
\end{subequations}
The normalized transverse and longitudinal pressures, $P_T/\left(\varepsilon(\tau) + P(\tau)\right)$ and $P_L/\left(\varepsilon(\tau) + P(\tau)\right)$ respectively, can be computed by using their definitions Eqs.(\ref{eq:PTPL2}). The substitution of the transseries solutions~\eqref{eq:fulltranssols} into these expressions yields
\be
\frac{P_T(\tau)}{\varepsilon(\tau) + P(\tau)} &\sim& \frac{1}{\log \tau} + \frac{\logt \tau}{2(\log \tau)^2} - \frac{5 - 2 \sigma_T}{2(\log \tau)^2} + \cdots \nl
&& + \frac{\theta_0}{\tau (\log \tau)^{2-\bar{\Delta}}} \left[  3 + \frac{3(2-\bar{\Delta})\logt \tau}{2\log \tau} - \frac{35 - 6 \sigma_T (2-\bar{\Delta})}{2 \log \tau} + \cdots \right] + \cdots, \\ \nl
\frac{P_L(\tau)}{\varepsilon(\tau) + P(\tau)} &\sim& \frac{1}{\log \tau} + \frac{\logt \tau}{2(\log \tau)^2} - \frac{5 - 2 \sigma_T}{2(\log \tau)^2} + \cdots \nl
&& - \frac{\theta_0}{\tau (\log \tau)^{1-\bar{\Delta}}} \left[ 2 + \frac{(1-\bar{\Delta})\logt \tau}{\log \tau} - \frac{10 - 2 \sigma_T (1-\bar{\Delta})}{ \log \tau} + \cdots \right] + \cdots.
\ee
Interestingly, the leading order asymptotic terms for the normalized transverse and longitudinal pressures is independent of the initial conditions and thus, these are universal in the deep IR. These asymptotic behavior of these quantities depends on the mass which is clearly seen when rewriting $\log\tau\to \log(m\tau)$ and $\log_{(2)}\tau\to \log_{(2)}(m\tau)$. 

From these expressions we conclude that the speed of sound has a crucial role in the IR transseries. What is truly remarkable of the results derived in this section is that the transmonomials of both PT and NP sectors are determined by the speed of sound via the PF sector of the temperature $T_{\rm pf}$. When a given theory provides a power expansion of the PT sector, e.g. the hydrodynamic gradient expansion of the massless Bjorken flow, the coefficients in the PT and NP sectors are normally related via a resurgent relation~\cite{Dorigoni:2014hea}.
As a matter of fact, this mathematical property strongly suggest that in the non-conformal fluid model studied here there exists a sort of correlation mechanism among transmonomials such that information of the local equilibrium propagates to the PT and NP sectors. How to determine the resurgent relation of this non-conformal model goes beyond the scope of this work and we left it for future research works.

In order to clarify further our previous paragraph let us recall to the reader the logical steps behind the derivation of the full transseries solutions~\eqref{eq:fulltranssols}. Our starting point was the full complete knowledge of the PF sector for the temperature. In this particular regime, the transmonomials of the perfect fluid sector are determined exclusively by the asymptotic form of the speed of sound $c_s^2(T)$. If $c_s^2$ is a positive constant, one obtains a power law scaling solution for the temperature, $T(\tau) \sim \tau^{-c_s^2}$, such as in the massless case when $c_s^2=1/3$. In contrast, for the non-conformal model the leading dominant term at early times for $c_s^2 \sim T$, and this asymptotic form generates a series of terms $\mu_2=\log\tau$ and $\mu_3=\log_{(2)}\tau$, which we call the PF monomials.
The transmonomial of the PF sector $\mu_1$~\eqref{eq:mu}, which we call the perturbative monomial, emerges from collisions quantified by the relaxation time scale $\tau_R$ expanded around $T=0$ and from $\beta_{\pi,\Pi}$ whose nature is purely non-collisional (see Eq.(\ref{eq:lead_Xpt})). Thus, the PF sector $T_{\rm pf}$ becomes the keystone upon which all the PT monomials are determined.
The NP monomial, $\zeta_i$~\eqref{eq:zeta}, is given in terms $S(\tau)$ and $(\beta_i,\gamma_i)$, and both of these terms have a different origin. $S(\tau)$ arises from the functional dependence of $\tau_R(T)$,
whereas $(\beta_i,\gamma_i)$ arise from the linear couplings of the thermodynamic transport coefficients~\eqref{eq:transportcoeff} with  $\bar{\pi}$ and $\bar{\Pi}$ in the non-collision part of the ODEs~\ref{eq:odediss}. Similar to the PT monomial, $T_{\rm pf}$ affects both $S(\tau)$ and $(\beta_i,\gamma_i)$ (see Eqs.(\ref{eq:dSdtau}) and (\ref{eq:dx_vector3}) respectively).
However, there exists no direct correlations between the PT and NP monomials. The schematic figure of the correlation mechanism is summarized in Fig.\ref{fig:corr_mono}. A very similar mechanism, albeit with different origin, can be also seen in the massless case when using $\tau$ as flow time~\cite{Behtash:2019txb}. We shall discuss the comparison between the massive and massless Bjorken expanding fluids in Sec.\ref{sec:massless_B}.

\subsection{Full transseries for the UV regime} \label{sec:transs_UV}

Let us now focus on the transseries in the UV regime. For convenience, we choose as dynamical variable the inverse temperature $\beta$ instead of the temperature $T$. The essential difference from the IR transseries studied in the previous section is that the UV transseries is obtained under the condition that the inverse temperature is zero in the UV limit. From Eq.~\eqref{eq:dTdtau_ml} one finds the evolution equation for $\beta$, i.e.,
\begin{equation}
    \label{eq:betaeqn}
    \frac{d\beta}{d\tau}=\frac{\beta}{\tau}\left[\bar{\Pi}-\bar{\pi}+c_s^2(\beta)\right]\,.
\end{equation}
Now the UV asymptotic form of the speed of sound is
\be
c_s^2(\beta) \sim \frac{1}{3} - \frac{\beta^2}{36} + \frac{5 \beta^4}{864} + \cdots,
\ee
so in the UV limit $\lim_{\beta \rightarrow 0_+}c_s^2(\beta)=1/3$ regardless of the values of the UV CPs.  
At a given finite value of the flow time $\tau$, the RHS of  Eq.(\ref{eq:betaeqn}) does not vanish identically and thus, the leading term of $\beta(\tau) \sim \sigma_\beta \tau^{\delta}$, where $\delta = \bar{\Pi}_{\rm UV}-\bar{\pi}_{\rm UV}+1/3$ which depends on the value of the associated UV CP, and being $\sigma_\beta \in {\mathbb R}_{>0}$ the integration constant of $\beta$.
The asymptotic expansion of the thermodynamic transport coefficients~\eqref{eq:transportcoeff} around $\beta=0$ involves $\log \beta$ (see appendix \ref{sec:tpcoeff_alpha}). These type of terms arise from the non-collisional part of the ODEs and correspond to the shear-bulk couplings and thus, it is expected to find non-trivial $\log$-type transmonomial terms in the UV transseries. As a matter of fact, the UV transseries can be written in a compact manner as follows (see appendix \ref{sec:trans_early} for details of the derivation):
\be
&& \beta(\tau) = \nu_2 \sum_{{\bf m},{\bf n}} b_{\beta}^{[{\bf m},{\bf n}]} \bm{\eta}^{\bf m} \bm{\nu}^{\bf n}, \qquad  \bar{X}(\tau) = \sum_{{\bf m},{\bf n}} b_{\bar{X}}^{[{\bf m},{\bf n}]} \bm{\eta}^{\bf m} \bm{\nu}^{\bf n}, \qquad \bar{X} \in \{ \bar{\pi}, \bar{\Pi} \}, \label{eq:trans_UV}
\ee
where the definition of the symbolic summation $\sum_{{\bf m},{\bf n}}$ depends explicitly on the location of the UV CPs discussed in Sect.~\ref{sec:find_ISOF}, Eqs.~\eqref{eq:CPs}. $\sum_{{\bf m},{\bf n}}$ is defined as
\be
\sum_{{\bf m},{\bf n}} b^{[{\bf m},{\bf n}]} \bm{\eta}^{\bf m} \bm{\nu}^{\bf n} &:=&
\begin{dcases}
\sum_{{\bf m}\in {\mathbb N}_0^{2}}   \sum_{n_1=0,n_2=0,n_3=0}^{\infty,\infty,\lfloor n_2/2\rfloor} b^{[{\bf m},{\bf n}]} \, \eta_1^{m_1} \eta_2^{m_2} \nu_1^{n_1} \nu_2^{n_2} \nu_3^{n_3} & \quad  \mbox{for} \quad {\frak R}_{1,2,3} \\
\sum_{{\bf m} \in {\mathbb N}_0^2} \sum_{n_2=0,n_3=0}^{\infty,\lfloor n_2/2\rfloor} b^{[{\bf m},{\bf n}]} \,  \eta_1^{m_1} \eta_2^{m_2} \nu_2^{n_2} \nu_3^{n_3} & \quad  \mbox{for} \quad {\frak R}_{4,5}
\end{dcases}, \label{eq:sum_betanug1}
\ee
with the floor function $\lfloor x \rfloor$, and the transmonomials are given by
\be
&& \nu_1:= \frac{\tau^{1-\bar{\Delta}\delta}}{\theta_0}, \qquad \nu_2 := \tau^\delta, \qquad \nu_3 := \log \tau,  \label{eq:transmono_UV} \\
&& \eta_{i=1,2}:= \sigma_i \tau^{\rho_i}, \label{eq:eta_i}
\ee
where $\sigma_{i}$ is the integration constant associated to $\bar{\pi}$ and $\bar{\Pi}$.
The values found for $\delta$ and $\rho_i$ (see appendix \ref{sec:trans_early}) are summarized in Table \ref{table:para_UV}.
Since the UV limit gives $\lim_{\tau \rightarrow 0_+} \bar{X}(\tau)=\bar{X}_{\rm UV}$, the constant part of $\bar{X}$ should be chosen as $b_{\bar{X}}^{[{\bf 0},{\bf 0}]}= \bar{X}_{\rm UV}$.
The integration constant of $\beta$, $\sigma_\beta$, explicitly enters into coefficients $b_{\beta,\bar{X}}^{[{\bf m},{\bf n}]}$, and for the case of ${\frak R}_{4,5}$ it should be chosen as
\be
\sigma_\beta =
\begin{cases}
  \left[\frac{6 \left(70 -55 \bar{\Delta} + 9 \bar{\Delta}^2 \right) \theta_0}{35 \bar{\Delta} (3-\bar{\Delta})}\right]^{-1/\bar{\Delta}} &  \mbox{for} \quad  {\frak R}_4 \\
  \left[\frac{2 (6-\bar{\Delta}) \theta_0}{3 \bar{\Delta}}\right]^{-1/\bar{\Delta}} & \mbox{for} \quad  {\frak R}_5
\end{cases}. \label{eq:b_beta00}
\ee
These transseries involves two ambiguous coefficients due to the normalization of $\sigma_i$ ($i =(\bar{\pi},\bar{\Pi})$). By choosing the normalization conditions $b_{\bar{\pi}}^{[(0,1),{\bf 0}]}=b_{\bar{\Pi}}^{[(1,0),{\bf 0}]}=1$, the other coefficients are uniquely determined.
In order to obtain consistency with the dimensions of the associated ISOFs given in Eq.(\ref{eq:dim_ISOF}), one has to eliminate some of the integration constants $\sigma_i$. This can be achieved by taking $\sigma_i=0$ which has $\rho_i<0$, such that all flows converge to the respective UV CPs in the UV limit.
Hence, $\sigma_{1}=0$ for ${\frak R}_{1}$ and $\sigma_{2}=0$ for ${\frak R}_{1,2,5}$.
The number of free integration constants indeed corresponds to ${\rm dim}\,{\cal S}({\frak R}_a,{\frak A})-1$.
As a result, one can obtain $\bar{X}_{\rm UV}$ + (leading term) as
\be
\bar{\pi}(\tau) &\sim& 
\begin{dcases}
  -\frac{7}{54}  -\frac{448 \pi \sigma_\beta}{13905} \tau^{1/6} & \quad \mbox{for} \quad {\frak R}_1 \\ 
  -\frac{25-3\sqrt{505}}{420} + \frac{265+27 \sqrt{505}}{1120} \sigma_1 \tau^{\frac{95-3 \sqrt{505}}{105}} 
  & \quad \mbox{for} \quad {\frak R}_2 \\ 
  -\frac{25+3\sqrt{505}}{420} + \frac{25 + 3 \sqrt{505}}{3 \sigma_\beta \left[ 140-8 \sqrt{505} - \left(55+\sqrt{505}\right)\bar{\Delta}  \right]\theta_0}  \tau^{\frac{140-(55+\sqrt{505})\bar{\Delta}}{140}}  & \quad \mbox{for} \quad {\frak R}_3 \\
  \frac{1}{3} -\frac{1}{\bar{\Delta}} + \frac{7(15-8\bar{\Delta})}{40+65 (1-\bar{\Delta})} \sigma_1 \tau^{\frac{4(90-23\bar{\Delta})}{105(3-\bar{\Delta})}}
  & \quad \mbox{for} \quad {\frak R}_4 \\
 -\frac{14}{27} + \frac{7}{3\bar{\Delta}} +  \frac{7 \pi (90-23\bar{\Delta})}{81(175+60\bar{\Delta})} \left[\frac{ 3 \bar{\Delta} }{2^{1-\bar{\Delta}} (6-\bar{\Delta})} \right]^{1/\bar{\Delta}} \tau^{1/\bar{\Delta}} & \quad \mbox{for} \quad {\frak R}_5 
\end{dcases}, \\ \nl
\bar{\Pi}(\tau) &\sim&
\begin{dcases}
  -\frac{8}{27} - \frac{88 \pi \sigma_\beta}{13905} \tau^{1/6} & \quad \mbox{for} \quad {\frak R}_1 \\
  \sigma_1 \tau^{\frac{95-3 \sqrt{505}}{105}} 
  & \quad \mbox{for} \quad {\frak R}_2 \\
  \frac{\sigma_\beta^2}{30} \tau^{\frac{55+\sqrt{505}}{70}}
 &\quad \mbox{for} \quad {\frak R}_3 \\
  \sigma_1 \tau^{\frac{4(90-23\bar{\Delta})}{105(3-\bar{\Delta})}}
  & \quad \mbox{for} \quad {\frak R}_4 \\
 - \frac{23}{27} + \frac{10}{3 \bar{\Delta}} + \frac{7 \pi (90-23\bar{\Delta})}{81(175+60\bar{\Delta})} \left[\frac{ 3 \bar{\Delta}}{2^{1-\bar{\Delta}} (6-\bar{\Delta})} \right]^{1/\bar{\Delta}} \tau^{1/\bar{\Delta}} & \quad \mbox{for} \quad {\frak R}_5 
\end{dcases}. 
\ee
All together and including the temperature, all of these leading terms are expansions of fractional powers in time. The UV transseries is a convergent series for all ${\frak R}_{1,\cdots,5}$, and this fact can be made sure in the similar way to the Borel analysis using Eq.(\ref{eq:B_trans}).

\begin{table}[t]
  \centering
\renewcommand{\arraystretch}{1.4}
\begin{tabular}{||c||c c||} 
\hline 
 \ UV CP \ & \ $\delta$ \ & \ $(\rho_1,\rho_2)$ \  \\ [0.5ex] 
 \hline\hline
 ${\frak R}_1$ & $\frac{1}{6}$ & $(-\frac{95+3 \sqrt{505}}{105},-\frac{95-3 \sqrt{505}}{105})$  \\ [0.5ex] 
 ${\frak R}_2$ & $\frac{55-\sqrt{505}}{140}$ & $(\frac{95-3 \sqrt{505}}{105},-\frac{2 \sqrt{505}}{35}) $  \\ [0.5ex] 
 ${\frak R}_3$ & $\frac{55+\sqrt{505}}{140}$ & $(\frac{95+3 \sqrt{505}}{105},\frac{2 \sqrt{505}}{35})$ \\ [0.5ex] 
 ${\frak R}_4$ & $1/\bar{\Delta}$ & $\left( \frac{4(90-23 \bar{\Delta})}{105 (3-\bar{\Delta})}, \frac{60-8\bar{\Delta}(5-\bar{\Delta})}{5\bar{\Delta}(3-\bar{\Delta})} \right)$  \\ [0.5ex] 
 ${\frak R}_5$ & $1/\bar{\Delta}$ & $\left( \frac{42-26 \bar{\Delta}+6 \sqrt{49-\bar{\Delta}(70-67 \bar{\Delta})/15}}{21 \bar{\Delta}}, \frac{42-26 \bar{\Delta}-6 \sqrt{49-\bar{\Delta} (70-67 \bar{\Delta})/15}}{21 \bar{\Delta}}\right)$  \\ [0.5ex] 
 \hline
\end{tabular}
\renewcommand{\arraystretch}{1}
\caption{
  The value of $\delta$ and $(\rho_1,\rho_2)$.
}
\label{table:para_UV}
\end{table}

The observables in the EM tensor can be computed from $T$ and $\bar{X}$, and they provide the physical information around the UV CPs.
The leading orders of the energy density and the thermodynamic pressure are given as
\be
\varepsilon(\tau) \sim 3P(\tau), \qquad \varepsilon(\tau) \sim \frac{3}{\pi^2} T(\tau)^4 \sim \frac{3}{\pi^2} \sigma_T^4 \tau^{-4\delta}, 
\ee
where $\sigma_T=1/\sigma_\beta$.
Notice that $\sigma_T$ for ${\frak R}_{4,5}$ has a value given in  Eq.(\ref{eq:b_beta00}).
Furthermore, the transverse and longitudinal pressures are given in Eq.(\ref{eq:PTPL2}), 
and the UV limit of the normalized transverse and longitudinal pressures is evaluated as follows:
\be
\lim_{\tau \rightarrow 0_+} \left( \frac{P_T(\tau)}{\varepsilon(\tau)+P(\tau)}, \frac{P_L(\tau)}{\varepsilon(\tau)+P(\tau)} \right) &=& \left( \frac{1}{4} + 3 \left(\bar{\Pi}_{\rm UV}+\frac{\bar{\pi}_{\rm UV}}{2}\right), \frac{1}{4} + 3 \left(\bar{\Pi}_{\rm UV}-\bar{\pi}_{\rm UV} \right) \right) \nl
&=&
\begin{dcases}
\left(  - \frac{5}{6}, - \frac{1}{4} \right)  & \quad \mbox{for} \quad {\frak R}_1 \\
\left( \frac{3(15 + \sqrt{505})}{280}, \frac{3(20-\sqrt{505})}{140} \right)  & \quad \mbox{for} \quad {\frak R}_2 \\
\left( \frac{3(15 - \sqrt{505})}{280}, \frac{3(20+\sqrt{505})}{140} \right)  & \quad \mbox{for} \quad {\frak R}_3 \\
\left( \frac{3}{4} - \frac{4}{2\bar{\Delta}}, - \frac{3}{4} + \frac{3}{\bar{\Delta}} \right)  & \quad \mbox{for} \quad {\frak R}_4 \\
\left( - \frac{37}{12} + \frac{27}{2\bar{\Delta}}, - \frac{3}{4} + \frac{3}{\bar{\Delta}} \right)  & \quad \mbox{for} \quad {\frak R}_5 \\
\end{dcases}.
\ee
It is important to notice that this specific limit does not depend on the values of the mass of the particle and thus, in the deep UV their values are uniquely determined by the UV CPs of $\bar{\pi}$ and $\bar{\Pi}$ respectively. 
As we expected, the traceless condition is indeed satisfied for ${\frak R}_{2,3,4}$.

\subsection{Transseries for $w=\tau/\tau_R$} \label{sec:transs_w}
The transseries analysis of conformal dynamical systems has shown that the system is dimensionally reduced when analyzed in terms of the asymptotic value of the inverse Knudsen number which for this particular case is simply $w = T\tau$~\cite{Heller:2015dha}. Thus, it becomes relevant to present the transasymptotic analysis of the non-conformal fluid model in terms of the inverse Knudsen number $w=\tau/\tau_R$. For our convenience, we keep $\theta_0$ explicitly in the transseries in order to account for the order of the PT expansion. Thus, we use a slightly modified definition for the $w$ flow time as $w:= \theta_0 \tau/\tau_R = \tau T^{\bar{\Delta}}$ with $\tau_R$ given by Eq.~\eqref{eq:tauR}.
In terms of this flow time, the ODEs~(\ref{eq:dTdtau_fm})-(\ref{eq:dPIdtau_fm}) are rewritten as
\begin{subequations}
\label{eq:ODE_wtime}
\begin{align}
\frac{dT}{dw} &= -\frac{T}{w} \cdot \frac{\bar{\Pi} - \bar{\pi} + c_s^2(\beta)}{1 - \bar{\Delta} \left( \bar{\Pi} - \bar{\pi} + c_s^2(\beta)  \right)}, \label{eq:dTdw_fm}  \\
 \frac{d \bar{\pi}}{d w} 
 &=   - \frac{1}{1 - \bar{\Delta} \left( \bar{\Pi} - \bar{\pi} + c_s^2(\beta)  \right)} \left[ \frac{\bar{\pi}}{\theta_0} + \frac{1}{w} \left( C_\pi(\beta) \bar{\pi} -  \frac{2 \lambda_{\pi \Pi}(\beta)}{3}  \bar{\Pi} -  \frac{4 \bar{\beta}_{\pi}(\beta)}{3} - D (\beta) \left( \bar{\Pi} - \bar{\pi} \right) \bar{\pi}   \right) \right], \label{eq:dpidw_fm} \\  
 \frac{d \bar{\Pi}}{d w} &= - \frac{1}{1 - \bar{\Delta} \left( \bar{\Pi} - \bar{\pi} + c_s^2(\beta)  \right)} \left[ \frac{\bar{\Pi}}{\theta_0} + \frac{1}{w} \left( C_{\Pi}(\beta)  \bar{\Pi} -  \lambda_{\Pi \pi}(\beta)  \bar{\pi} +  \bar{\beta}_{\Pi}(\beta)  - D (\beta) \left( \bar{\Pi} - \bar{\pi} \right) \bar{\Pi} \right) \right]. \label{eq:dPIdw_fm}
\end{align}
\end{subequations}
The reader clearly notices that there is no dimensional reduction in this case.
Construction of the transseries can be carried out in the similar way to the case of the $\tau$-time based on the technical methods outlined explicitly in appendices \ref{sec:trans_late} and \ref{sec:trans_early}. 
Instead of directly solving Eqs.~\eqref{eq:ODE_wtime}, we can construct the transseries solutions in $w$ flow time from the transseries solutions written in terms of the $\tau$-time. This is achieved by using the identity
\be
\frac{d \tau}{d w} = \frac{\tau}{w} \cdot \frac{1}{1-\bar{\Delta}(\bar{\Pi}-\bar{\pi}+c_s^2(\beta))}, \label{eq:dtau_dw}
\ee
which follows directly from the conservation law~\eqref{eq:dTdtau_fm} and the definition of the $w$ flow time. 
Then, by solving Eq.~\eqref{eq:dtau_dw} and plugging its solution $\tau(w)$ into Eq.(\ref{eq:full_IR_trans}) one obtains the transseries in the $w$-time.
In consequence, and after a bit of algebra, one finds that the transseries with the $w$-time look similar to the ones given in Eqs.(\ref{eq:full_IR_trans}) and (\ref{eq:trans_UV}). Nonetheless, the transmonomials are slightly different in terms of the $w$-time. 

For the IR transseries, Eq.(\ref{eq:dtau_dw}) gives $\tau(w) \approx w (\log w)^{\bar{\Delta}}$, and from Eqs.(\ref{eq:mu}) and (\ref{eq:zeta}) one can immediately obtain the transmonomials for $w$ as
\be
&& \widehat{\mu}_1 := \frac{\theta_0}{w}, \qquad \widehat{\mu}_2 := \frac{1}{\log w}, \qquad \widehat{\mu}_3 := \logt w = \log \log w. \label{eq:mu_w} \\
&& \widehat{\zeta}_{i=1,2} := \sigma_i \frac{\exp \left[ -S(w) \right]}{w^{\widehat{\beta}_i} (\log w)^{\widehat{\gamma}_i}}, \qquad S(w) = \widehat{\mu}_1^{-1} \sum_{n_2=0,n_3=0}^{\infty,n_2} a_S^{[(n_2,n_3)]} \, \widehat{\mu}_2^{n_2} \widehat{\mu}_3^{n_3}, \label{eq:zeta_w}
\ee
with
\be
(\widehat{\beta}_1,\widehat{\beta}_2)=(2,0), \qquad (\widehat{\gamma}_1,\widehat{\gamma}_2)=(-7+2\bar{\Delta},-1).
\ee
The exponent of the exponentially damping function, $S(w)$, is determined from the asymptotic form of $c_s^2$ and the PF sector in $T$ as
\be
\frac{d S}{d w} = \frac{1}{[1-\bar{\Delta}c_s^2(1/T_{\rm pf})]\theta_0}.
\ee
After all, the asymptotic expansion takes the form as
\be
T(w) &\sim&  \frac{1}{\log w} + \frac{(1-2\bar{\Delta})\logt w}{2(\log w)^2} +  \frac{\sigma_T}{(\log w)^2} + \cdots  \nl
&& - \frac{\theta_0}{w} \left[ \frac{2}{(\log w)^3} + \frac{3(1-2 \bar{\Delta}) \logt w}{(\log w)^4} + \frac{4 \bar{\Delta}+6 \sigma_T - 13}{(\log w)^4} + \cdots  \right]  + \cdots \\ \nl
\bar{\pi}(w) &\sim& \frac{\theta_0}{w} \left[ \frac{4}{3 (\log w)^2} + \frac{4 (1-2 \bar{\Delta}) \logt w}{3 (\log w)^3} - \frac{8(2-\sigma_T)}{3 (\log w)^3} + \cdots \right]  + \cdots, \\ \nl
\bar{\Pi}(w) &\sim& \frac{\theta_0}{w} \left[ - \frac{2}{3 (\log w)^2} - \frac{2 (1-2 \bar{\Delta}) \logt w}{3 (\log w)^3} + \frac{17-4\sigma_T}{3 (\log w)^3} + \cdots \right]  + \cdots.
\ee
The transseries solution for the energy density and the thermodynamic pressure are also slightly changed from Eq.(\ref{eq:trans_epsP}). These read as respectively
\be
&& \varepsilon(w) =  \frac{e^{\sigma_T}}{\sqrt{8 \pi^3}} \frac{\widehat{\mu}_2^{1+\bar{\Delta}}}{w} \sideset{}{^\prime}\sum_{{\bf m},{\bf n}} a_{\varepsilon}^{[{\bf m},{\bf n}]} \widehat{\bm{\zeta}}^{\bf m} \widehat{\bm{\mu}}^{\bf n}, \qquad P(w) = \frac{e^{\sigma_T}}{\sqrt{8 \pi^3}} \frac{\widehat{\mu}_2^{2+\bar{\Delta}}}{w} \sideset{}{^\prime} \sum_{{\bf m},{\bf n}} a_{P}^{[{\bf m},{\bf n}]} \widehat{\bm{\zeta}}^{\bf m} \widehat{\bm{\mu}}^{\bf n}, \label{eq:trans_epsPw}
\ee
and some of the first terms can be explicitly evaluated as follows:
\begin{subequations}
\begin{align}
{\varepsilon}(w) &\sim \frac{e^{\sigma_T}}{\sqrt{8 \pi^3} w (\log w)^{1+\bar{\Delta}}}  \left[ 1  + \frac{(1-\bar{\Delta} - 2\bar{\Delta}^2)\logt w}{2\log w} - \frac{1-\sigma_T(1+\bar{\Delta})}{\log w} + \cdots  \right], \\ \nl
P(w) &\sim \frac{e^{\sigma_T}}{\sqrt{8 \pi^3} w (\log w)^{2+\bar{\Delta}}} \left[ 1  +  \frac{(2-3 \bar{\Delta}-2 \bar{\Delta}^2)\logt w}{2\log w} - \frac{35-8\sigma_T(2+\bar{\Delta})}{8 \log w} + \cdots \right], \\ \nl
c_s^2(w) &\sim \frac{1}{\log w} + \frac{(1-2\bar{\Delta})\logt  w}{2(\log w)^2} - \frac{1-2\sigma_T}{2(\log w)^2} + \cdots \nl
& - \frac{\theta_0}{w (\log w)^3} \left[ 2 + \frac{3(1-2\bar{\Delta})\logt w}{\log w} - \frac{15 - 4 \bar{\Delta} - 6 \sigma_T}{\log w} + \cdots \right] + \cdots.
\end{align}
\end{subequations}
In addition, the normalized transverse and  longitudinal pressures are written down as
\be
\frac{P_T(w)}{\varepsilon(w) + P(w)} &\sim& \frac{1}{\log w} + \frac{(1-2 \bar{\Delta})\logt w}{2(\log w)^2} - \frac{35 - 8 \sigma_T}{8(\log w)^2} + \cdots \nl
&& + \frac{\theta_0}{w (\log w)^{2}} \left[  3 + \frac{3(1-2\bar{\Delta})\logt w}{\log w} - \frac{1+ 6 \bar{\Delta}-12 \sigma_T}{2 \log w} + \cdots \right] + \cdots, \\ \nl
\frac{P_L(w)}{\varepsilon(w) + P(w)} &\sim& \frac{1}{\log w} + \frac{(1-2 \bar{\Delta})\logt w}{2(\log w)^2} - \frac{35 - 8 \sigma_T}{8(\log w)^2} + \cdots \nl
&& - \frac{\theta_0}{w \log w} \left[ 2 + \frac{(1-2\bar{\Delta})\logt w}{\log w} - \frac{10-2 \sigma_T}{ \log w} + \cdots \right] + \cdots.
\ee
For the UV transseries expanded around ${\frak R}_{1,2,3}$ (see Eqs.~\eqref{eq:CPs}), one can obtain $\tau(w) \sim w^{1/[(1-\bar{\Delta}) \delta]}$ from Eq.(\ref{eq:dtau_dw}), where the $\delta$ values are quoted in Table \ref{table:para_UV}.
Thus, one can define the formal transseries by the change of transmonomials as keeping the same summed form in Eq.(\ref{eq:sum_betanug1}).
From Eqs.(\ref{eq:transmono_UV}) and (\ref{eq:eta_i}), the UV transmonomials are modifed as
\be
&& \widehat{\nu}_1:= \frac{w}{\theta_0}, \qquad \widehat{\nu}_2 := w^{\widehat{\delta}}, \qquad \widehat{\nu}_3 := \log w, \\ \label{eq:transmono_UV_w}
&& \widehat{\eta}_{i=1,2}:= \sigma_i w^{\widehat{\rho}_i}, \label{eq:eta_i_w}
\ee
where $\widehat{\delta}:=\delta/(1-\bar{\Delta}\delta)$ and $\widehat{\rho}_i:=\rho_i/(1-\bar{\Delta}\delta)$. The values for $\rho_i$ are listed in Table \ref{table:para_UV},
The other CPs, ${\frak R}_{4,5}$ (see Eq.~\eqref{eq:CPs}), do not appear in the $w$-time because the ODEs~\eqref{eq:ODE_wtime} becomes singular at these UV CPs. $\bar{X}_{\rm UV}$ + (leading term) is evaluated as
\be
\bar{\pi}(w) &\sim& 
\begin{dcases}
  -\frac{7}{54}  -\frac{448 \pi \sigma_\beta}{13905} w^{1/(6-\bar{\Delta})} & \quad \mbox{for} \quad {\frak R}_1 \\ 
  -\frac{25-3\sqrt{505}}{420} + \frac{265+27 \sqrt{505}}{1120} \sigma_1 w^{\frac{4(95-3 \sqrt{505})}{3[140-(55-\sqrt{505})\bar{\Delta}]}} 
  & \quad \mbox{for} \quad {\frak R}_2 \\ 
  -\frac{25+3\sqrt{505}}{420} + \frac{25 + 3 \sqrt{505}}{3 \sigma_\beta \left[ 140-8 \sqrt{505} - \left(55+\sqrt{505}\right)\bar{\Delta} \right] \theta_0} w  & \quad \mbox{for} \quad {\frak R}_3 
\end{dcases}, \\ \nl
\bar{\Pi}(w) &\sim&
\begin{dcases}
  -\frac{8}{27} - \frac{88 \pi \sigma_\beta}{13905} w^{1/(6-\bar{\Delta})} & \quad \mbox{for} \quad {\frak R}_1 \\
  \sigma_1 w^{\frac{4(95-3 \sqrt{505})}{3[140-(55-\sqrt{505})\bar{\Delta}]}} 
  & \quad \mbox{for} \quad {\frak R}_2 \\
  \frac{\sigma_\beta^2}{30} w^{\frac{2(55+\sqrt{505})}{140-(55+\sqrt{505})\bar{\Delta}}}
 &\quad \mbox{for} \quad {\frak R}_3 
\end{dcases}. 
\ee
Similar to the $\tau$-time case, the $\log$- and $\log \log$-types transmonomials entirely arise in the all the sectors in the transseries, which is the consequence of the nontrivial PF sector and functional forms of the transport coefficients due to the particle mass.
The $\sigma_T$-dependence in the coefficients also appears explicitly and carries information about the initial data to the IR regime. In other words, in both $w$ and $\tau$ flow times, there is a memory effect in the late time dynamics. 
It is important to mention that in the $w$-time the lower sectors of the transseries,  such as the PF sector, includes information of the higher order sectors of the $\tau$-time such as the NP sector. The reason is the non-trivial relation between $\tau$ and $w$ via the ODE~\eqref{eq:dtau_dw}. In addition, as we will see in Sec.\ref{sec:NS_lim}, the perturbative expansion in the $w$-time does not match exactly with the results derived from the Chapman-Enskog expansion. This aspect is one of the main differences compared with the massless Bjorken flow where such an asymptotic match can be made~\cite{Behtash:2018moe,Behtash:2019txb,Blaizot:2017ucy,Blaizot:2017lht}.

\subsection{Navier-Stokes regime} \label{sec:NS_lim}
\begin{figure}[tb]
\begin{center}
\includegraphics[width=.42\textwidth]{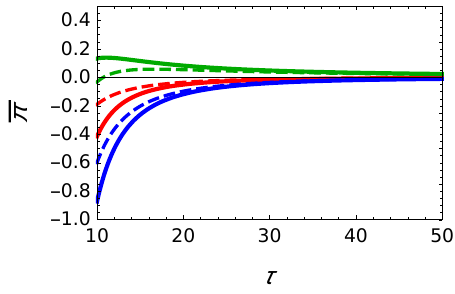} \ \ \
\includegraphics[width=.42\textwidth]{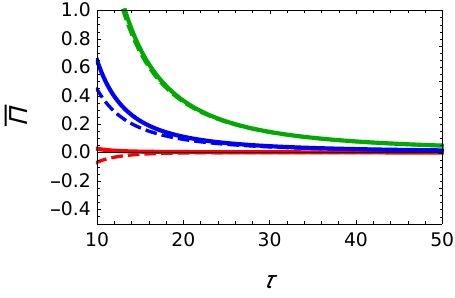}
\caption{Comparison of the PT sector of IR transseries and the NS theory with different integration constants $\sigma_T$.
We took $\theta_0=1$ and $\bar{\Delta}=1$ in this figure.
The solid and dashed lines denote $(\bar{\pi}_{\rm pt},\bar{\Pi}_{\rm pt})$ and $(\bar{\pi}_{NS},\bar{\Pi}_{NS})$, respectively. 
The integration constant is chosen as $\sigma_T=2.5$ (Red), $\sigma_T=5$ (Blue), and $\sigma_T=10$ (Green).
$(\bar{\pi}_{NS},\bar{\Pi}_{NS})$ is computed by substituting the PF and PT sectors of temperature, $T=T_{\rm pf}+T_{\rm pt}$, into Eq.(\ref{eq:X(1)}).
The order of expansion is taken into account up to $O(\mu_2^3)$ in the PF sector and $O(\mu_1^3\mu_2^3)$ in the PT sector.
}
\label{fig:NS}
\end{center}
\end{figure}

In this section we consider the Navier-Stokes (NS) regime of the fluid equations from the transasymptotics perspective. The  Chapman-Enskog (CE) method in kinetic theory provides a systematic method to calculate the constitutive relations of the fluid dissipative variables~\cite{DeGroot:1980dk}. For the boost invariant case and within the relaxation time approximation the Navier Stokes constitutive relations read as~\cite{Jaiswal:2014lsa}
\be
\pi_{NS}(T) = -\tau \pi^{\zeta \zeta}_{NS}(T) = \frac{4 \tau_R(T)}{3 \tau} \beta_{\pi}(1/T), 
\qquad  \Pi_{NS}(T) = - \frac{\tau_R(T)}{\tau} \beta_{\Pi}(1/T).
\label{eq:constrel}
\ee
In terms of the effective inverse Reynolds numbers~\eqref{eq:effRey} the previous NS constitutive relations can be rewritten as
\be
\bar{\pi}_{NS}(T) = \frac{4 \theta_0}{3 T^{\bar{\Delta}} \tau} \bar{\beta}_\pi(1/T), \qquad \bar{\Pi}_{NS}(T) = -\frac{\theta_0}{T^{\bar{\Delta}} \tau} \bar{\beta}_\Pi(1/T), \label{eq:X(1)}
\ee
where we used explicitly Eq.~\eqref{eq:tauR}. Both constitutive relations are solely functions of $T(\tau)$~\footnote{The reader must bear in mind that we set $m=1$.} and $\tau$ and thus, their values can be evaluated by substituting the transseries solution of the temperature~\eqref{eq:fulltranssols} into Eq.(\ref{eq:X(1)}). 
By doing so it brings all of the sectors, perturbative and non-perturbative ones, in the transseries of $T$ and coincides with $\bar{X}_{\rm pt}$ when $n_1=1$ in Eq.~\eqref{eq:X_pt} when using only the PF sector, i.e. $T = T_{\rm pf}$. This means that the IR transseries of $(\bar{\pi},\bar{\Pi})$ reproduces the NS limit only in the region that is asymptotically equal up to $\mathcal{O}(\theta_0/\tau)$. In addition, both $\bar{X}_{\rm pt}$ and $\bar{X}_{NS}$ have information about the initial data dependence which is encoded in the integration constant $\sigma_T$. In Fig.\ref{fig:NS} we show the comparison of the IR transseries and the NS regime for the effective inverse Reynolds numbers~\eqref{eq:effRey}. In this figure we observe that $(\bar{\pi}_{\rm pt},\bar{\Pi}_{\rm pt})$ and $(\bar{\pi}_{NS},\bar{\Pi}_{NS})$ have good agreement with each other only at extreme long times $\tau \sim 35$.

Let us briefly discuss the NS regime in terms of the $w$ flow time. The transseries of the temperature in terms of $w$ has the same structure than its counterpart in $\tau$-time. However, when replacing the full transseries for the temperature into the constitutive relations~\eqref{eq:constrel}, all the sectors of the solution enter there together with the dependence on the initial condition. Thus, the standard asymptotic Chapman-Enskog expansion differs from the asymptotic transseries solutions since the former does not carry any information of neither initial conditions nor non-perturbative phenomena at long times. This can be easily understood mathematically as a certain no-go theorem as follows: It is known that in conformal Bjorken flows the asymptotic gradient hydrodynamic expansion can be written as a power series of the inverse Knudsen number $w=\tau T$~\cite{Heller:2015dha}. For instance, the inverse Reynolds number $\bar{\pi}=\sum_{k=1}^\infty f^{(k)}w^{-k}$ and thus its Navier-Stokes value $\pi_{NS}=f^{(0)}/w$ where $f^{(0)} \sim \eta/s$. In contrast, the terms $\bar{\beta}_{\pi}$ and $\bar{\beta}_{\Pi}$ entering into Eqs.~\eqref{eq:constrel} are independent functions of $T$ from each other, so that one can easily find that it is impossible to formulate a flow time $\tau \mapsto {\frak w}(\tau)$ to give a functional dependence $\sim \frak w^{-1}$ for both $\bar{\pi}_{NS}$ and $\bar{\Pi}_{NS}$ simultaneously. Therefore, such a flow time can not be formulated. This can be also understood from the view point of the D-symmetry. Suppose that the dissipative fluid variable can be expanded in terms of the ${\frak w}=\tau/\tau_R$ variable as a power series, i.e., up to the $n$-th order this series expansion the dissipative variables can be written as $\bar{X}_{(n)}(T)=\sum_{m=1}^n {\frak w}^{-m} F_{\bar{X}}^{(m)}(T)$. 
If the D-symmetry  holds at the level of the ODEs and transforms $(\tau,\tau_R) \rightarrow (\lambda \tau, \lambda \tau_R)$, then the temperature dependence can be eliminated from $F_{\bar{X}}^{(m)}(T)$ for any $m \in {\mathbb N}$, and $\frak{w}$.
If it does not exist, then $\frak{w}$ has to take the form as $\frak{w}=\tau/\tau_R \cdot f(T)^{-1}$, and $f(T)$ has to be given such that $ F^{(m)}_{\bar{X}}(T)=c_{\bar{X}}^{(m)}f(T)^m$ with a constant $c_{\bar{X}}^{(m)}$ for all $m$ and $\bar{X} \in \{\bar{\pi}, \bar{\Pi} \}$, which is impossible.
Therefore, the D-symmetry is required  to formulate such a $\frak{w}$.
The previous discussion follows for any $0 < \bar{\Delta} \le 1$ with $w \propto \tau T^{\bar{\Delta}}$.

We conclude this section by generalizing previous findings~\cite{Behtash:2018moe,Behtash:2019txb,Behtash:2020vqk} related with transient rheology and renormalized transport coefficients. One can go further and extract the shear and bulk viscosities from the constitutive relations~\eqref{eq:constrel} as follows~\cite{Jaiswal:2014lsa}:
\begin{equation}
    \label{eq:renortransport}
    \eta = 2\tau_R(T)\beta_\pi(1/T)=\frac{\theta_0}{T^{\bar{\Delta}}}\beta_\pi(1/T) \,,\quad\quad 
    \zeta =\tau_R(T)\beta_\Pi = \frac{\theta_0}{T^{\bar{\Delta}}}\beta_\Pi(1/T)\,.
\end{equation}
Each of these terms depend on the initial data as well as non-perturbative physics given the temperature dependence of the functions $\beta_{\pi,\Pi}$. This behavior has a neat physical interpretation in terms of physics of non-newtonian fluids~\cite{Behtash:2018moe,Behtash:2019txb,Behtash:2020vqk}: 

\textit{Non-equilibrium constitutive relations of the macroscopic dissipative quantities emerge in far-from-equilibrium fluids and thus, the meaning of transport coefficients is extended to non-equilibrium setups.  While the system hydrodynamizes transport coefficients change their values and get effectively renormalized due to the summation over all the transient non-hydrodynamical modes. Furthermore, transport coefficients carry information on the initial conditions and thus, their values depend on the deformation history of the fluid, i.e. its transient rheology.} 

It is important to emphasize this picture of hydrodynamization holds provided that the system is described in terms of a few degrees of freedom in the long wavelength limit, i.e., it depends on the existence of a slow invariant manifold~\cite{kunihiro2022geometrical}. In other words if most of the fastest modes survive in the hydrodynamic limit one cannot integrate them out so no constitutive relations emerges in non-equilibrium situations as it happens in the Gubser flow~\cite{Behtash:2019qtk}. 

\section{Attractors and the meaning of attraction in phase space} \label{sec:attractor}
\begin{figure}[tp]
\begin{minipage}[b]{1.0\textwidth}
\centering
\includegraphics[width=0.4\textwidth]{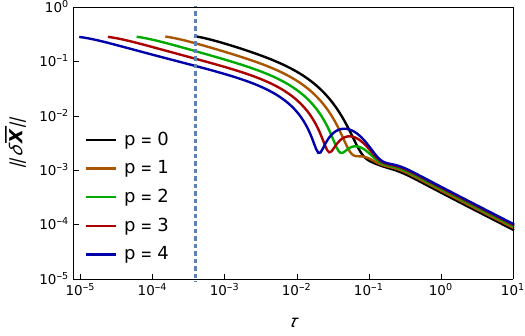} \ \ \
\includegraphics[width=0.4\textwidth]{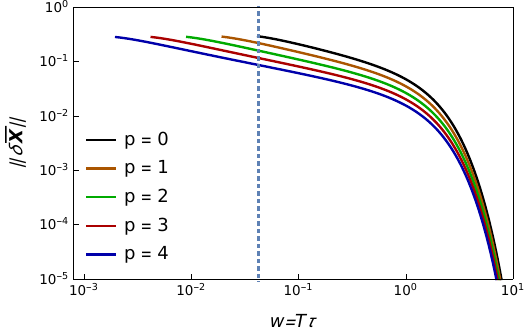}
\subcaption{$\bar{\pi}_{\rm ini}= \bar{\pi}_{\rm UV}-0.2$, $\bar{\Pi}_{\rm ini}=\bar{\Pi}_{\rm UV}-0.2$}
\label{fig:pullback_a}
\end{minipage} \\
\hspace{10mm}
\begin{minipage}[b]{1.0\textwidth}
\centering
\includegraphics[width=0.4\textwidth]{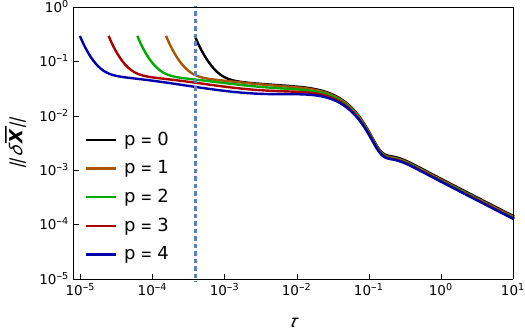} \ \ \
\includegraphics[width=0.4\textwidth]{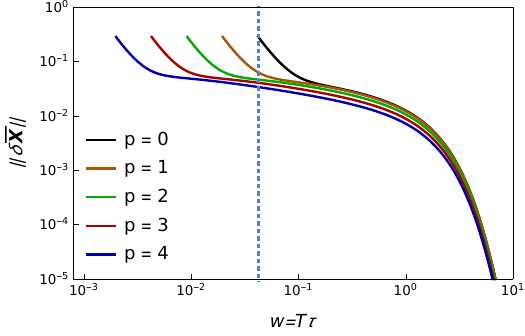}
\subcaption{$\bar{\pi}_{\rm ini}= \bar{\pi}_{\rm UV}+0.2$, $\bar{\Pi}_{\rm ini}=\bar{\Pi}_{\rm UV}-0.2$}
\label{fig:pullback_b}
\end{minipage}
\caption{
The pullback attraction for the $\tau$ and $w$.
The conformal dimensions and the relaxation scale are chosen as $\bar{\Delta}=1$ and $\theta_0=1$, respectively.
The vertical axis $||\delta \bar{\bf X}||$ is the distance between $(\bar{\pi},\bar{\Pi})$ and a flow on ${\cal S}({\frak R}_1,{\frak A})$ with the initial temperature, $T_0 \approx 1.08 \times 10^{2}$ for $\tau_0 \approx 3.91 \times 10^{-4}$.
The colored plots are drawn by the fixed initial condition $(\bar{\pi}_{\rm ini},\bar{\Pi}_{\rm ini})$ with different initial times $\tau_{\rm ini} (\le \tau_0)$, $\tau_{\rm ini}=\tau_0 \times 2.5^{-p}$ with $p=0,\cdots,4$.
$(\bar{\pi}_{\rm UV},\bar{\Pi}_{\rm UV})=(-7/54,-8/27)$ is the CP associated with ${\frak R}_1$.
The vertical gray dotted lines are $\tau=\tau_0$ and 
$w=w_0 (=T_0 \tau_0$).
}
\label{fig:pullback}
\end{figure}

In this section, we investigate the attracting behavior in the phase space when using the proper time $\tau$ and $w$ as a flow time variable. As described in Sect.\ref{sec:find_ISOF}, we identified $\widehat{\cal S}({\frak R}_1, {\frak A})$ as attractor solution from the stability analysis around the UV and IR CPs and its dimensions.
However, the attracting behavior at the intermediate region can not be captured by local analyses such as transseries analysis.
Consequently, in order to investigate the attraction strength of the attractor in arbitrary regions of the phase space in a quantitative way, we introduce the \textit{distance} between a particular flow in the phase space and the attractor solution and analyse its time-dependence by numerical computation~\footnote{See Refs.~\cite{kloeden2011nonautonomous,caraballo2016applied} for more precise mathematical definitions of atractors in non-autonomous dynamical systems.}. Recent works have also focused on introducing metrics that quantify the attraction of rapidly expanding plasmas~\cite{Du:2022bel,Ambrus:2021sjg}.

We remind to the reader that the attractor solution is defined in the projected subspace, $(\bar{\pi},\bar{\Pi},\tau)\in  \widehat{\mathcal{M}}$, and it depends on the initial condition of the temperature. This fact implies that specifying the value of $T(\tau_0)=T_0$ at a fixed time $\tau=\tau_0$ uniquely determines $\widehat{\cal S}({\frak R}_1,{\frak A})$ in $\widehat{\cal M}$.
We measure the distance between an arbitrary flow and the attractor solution $\bar{\bf X}_{\rm att}(\tau;T_0)$ as follows:
\be
|| \delta \bar{\bf X} || (\tau;Y_{\rm ini}) := \sqrt{\left( \bar{\pi}(\tau;Y_{\rm ini}) - \bar{\pi}_{\rm att}(\tau;T_0) \right)^2 + \left( \bar{\Pi}(\tau;Y_{\rm ini}) - \bar{\Pi}_{\rm att}(\tau;T_0) \right)^2}, \label{eq:distance}
\ee
where $Y_{\rm ini}=(T_{\rm att}(\tau_{\rm ini};T_0),\bar{\bf X}_{\rm ini},\tau_{\rm ini})$ is a set of the initial conditions.
Here, the initial condition for the temperature in $Y_{\rm ini}$ was taken the solution of $T_{\rm att}(\tau;T_0)$
\footnote{
The precise way to obtain $T_{\rm att}(\tau;T_0)$ is as follows:
We set the initial condition of temperature as $T(\tau_0)=T_0$.
Because of the stability of ${\frak R}_1$, $\bar{\bf X}(\tau_0)$ that converges to ${\frak R}_1$ in the UV limit is uniquely determined, which we denote it as $\bar{\bf X}_{\rm att}(\tau_0;T_0)$.
$(T_{\rm att}(\tau;T_0),\bar{\bf X}_{\rm att}(\tau;T_0))$ is the solution of the ODEs by taking the initial condition, $(T_0,\bar{\bf X}_{\rm att}(\tau_0;T_0))$ at $\tau = \tau_0$.
}.
This definition can be generalized to any observable ${\cal O}$ that is a function of $(T,\bar{\bf X})$ like the transverse and longitudinal pressures. In that case, the distance is defined as
\be
&& || \delta {\cal O} || (\tau;Y_{\rm ini}) := \sqrt{\left( {\cal O}(\tau;Y_{\rm ini}) - {\cal O}_{\rm att}(\tau;T_0) \right)^2}, \qquad 
\label{eq:distance_obs}
\ee
where ${\cal O}(\tau;Y_{\rm ini}):={\cal O}(T(\tau;Y_{\rm ini}),\bar{\pi}(\tau;Y_{\rm ini}),\bar{\Pi}(\tau;Y_{\rm ini}))$ and ${\cal O}_{\rm att}(\tau;T_0):={\cal O}(T_{\rm att}(\tau;T_0),\bar{\pi}_{\rm att}(\tau;T_0),\bar{\Pi}_{\rm att}(\tau;T_0))$.
Similar definitions follow in the flow $w$-time which in this case are obtained by simply replacing $\tau \mapsto w$ in Eqs.(\ref{eq:distance}) and (\ref{eq:distance_obs}). For compactness in our notation we use ${\frak w}$ as a unified symbol of $\tau$ and $w$.

We present our results by analyzing first the behavior of the distance in the UV region. Fig.~\ref{fig:pullback} shows the ${\frak w}_{\rm ini}$-dependence of the distance for fixed values of the initial conditions ${\bf X}_{\rm ini}$. In order to generate the numerical results shown in Fig.~\ref{fig:pullback} the initial conditions are fixed while pushing backward the initial value of the flow time ${\frak w}_{\rm ini}$. The initial flow time satisfies ${\frak w}_{\rm ini}\leq {\frak w}_{0}$ with ${\frak w}_{0}$ being a constant (horizontal dashed blue line in Fig.~\ref{fig:pullback}) but very small number. Fig.~\ref{fig:pullback} shows clearly that there is indeed early-time attraction since all the flow lines merge towards the attractor solution below $\frak{w}_0$. 
Thus, there exists a well located region in the UV projected phase space, i.e. the domain $\widehat{\cal D} \times (0,{\frak w}_0)  \subset \widehat{\cal M}$, such that $\lim_{{\frak w}_{\rm ini} \rightarrow 0_+}|| \delta \bar{\bf X} || ({\frak w}_0;Y_{\rm ini})=0$ with a fixed albeit small ${\frak w}_0$. 
This intuitively means that, when decreasing ${\frak w}$, there exist flows around ${\frak R}_1$ which deviate from the attractor solution but can be trapped in an extremely small positive ${\frak w}$. When one considers the limit where ${\frak w}\to 0_+$,  flows around ${\frak R}_1$ go to another UV CP or move out of the basin of attraction. 
When increasing ${\frak w}$, flows neighboring to ${\cal S}({\frak R}_1,{\frak A})$ go forward as being attracted to the ISOF until a certain ${\frak w}_0$ which depends on the choice of flow time. ${\frak w}_0$ is a parameter to describe how long the attractor solution keeps attracting behavior. For instance, in the top left panel of Fig.~\ref{fig:pullback_a} we observe that the distance is a monotonically decreasing function of $\tau_{\rm ini}$ by looking at $\tau_0=3.91\times 10^{-4}$ or smaller. This means that $\widehat{\cal S}({\frak R}_1,{\frak A})$ has the property of being a local pullback attractor~\cite{thompson2002nonlinear,caraballo2016applied}. Nonetheless, as $\tau$ increases $||\delta\bar{\bf X}||$ develops a small convex around $\tau \approx (10^{-2},10^{-1})$. This means that in that interval of time the distance can increase depending on the UV region of the phase space. 

Being a local pullback attractor means that all initial flows defined in a domain converge to it while increasing the flow time $\frak{w}_{\rm ini}\leq {\frak w}_0$.
In Fig.~\ref{fig:pullback_a} we  observe that the distance between different flows and the pullback attractor depends on the choice of the initial condition before  $\tau \lsim 5.0 \times 10^{-3}$ ($w \lsim 5.0 \times 10^{-1}$). In this flow time region the distance decreases by following a power law decay until  $||\delta \bar{\bf X}|| \approx 3.0 \times 10^{-2}$ -- $1.0 \times 10^{-1}$. Nonetheless we observe a rather distinct behavior in Fig.\ref{fig:pullback_b} where the distances decrease until $||\delta \bar{\bf X}|| \approx 5.0 \times 10^{-2}$ by following a power law decay, and then it curves in such a way that continues decaying with a steeper secondary  power law. The difference of these behaviors can be explained in terms of the UV transseries explained in the previous section.
Now, $(\bar{\pi}_{\rm att},\bar{\Pi}_{\rm att})$ in Eq.(\ref{eq:distance}) is nothing else but the transseries solution associated to the UV CP ${\frak R}_1$ which does not depend on $\sigma_{1,2}$, see Eq.(\ref{eq:trans_UV}).
If one sets the initial condition slightly deviating from the attractor around the UV regime, then both $\sigma_{1,2}$ have a small nonzero value, and the modes $\eta_i$ in Eq.(\ref{eq:eta_i}) contribute (see also Eq.(\ref{eq:eta_i_w}) for the $w$-time). Due to these two modes, the convergence to the pullback attractor in the UV regime behaves as the power law obeying $||\delta \bar{\bf X}|| \sim \tau^{\rho_i} \, (w^{\widehat{\rho}_i})$, and as increasing ${\frak w}$ the transition to the secondary weak power law  can generally happen depending on the prepared initial condition.
This fact also means that the pullback attractor in the UV regime can be characterized by the value of $(\rho_1,\rho_2)$ for the $\tau$-time ($(\widehat{\rho}_1,\widehat{\rho}_2)$ for the $w$-time).

\begin{figure}[tp]
\begin{minipage}[b]{1.0\textwidth}
\centering
\includegraphics[width=0.4\textwidth]{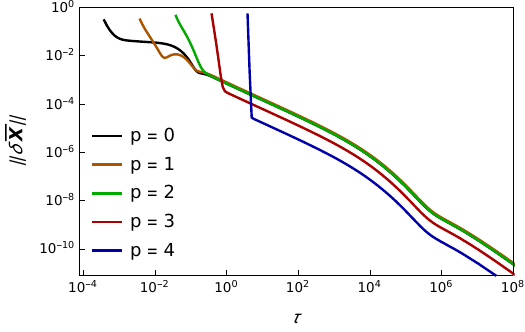} \ \ \
\includegraphics[width=0.4\textwidth]{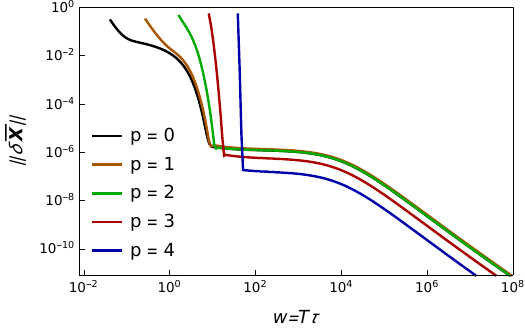}
\end{minipage}
\caption{Time dependence of the distance.
The colored plots are drawn with different initial times $\tau_{\rm ini} (\ge \tau_0)$, $\tau_{\rm ini}=\tau_0 \times 10^{p}$ with $p=0,\cdots,4$ with a fixed initial condition $(\bar{\pi}_{\rm ini},\bar{\Pi}_{\rm ini})$ .
The values of the parameters were the same as in Fig.\ref{fig:pullback_b}.
}
\label{fig:pullback_tot}
\end{figure}

\begin{figure}[tp]
\begin{minipage}[b]{1.0\textwidth}
\centering
\includegraphics[width=0.4\textwidth]{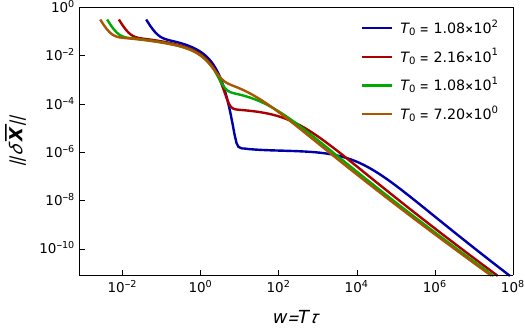} \ \ \
\includegraphics[width=0.4\textwidth]{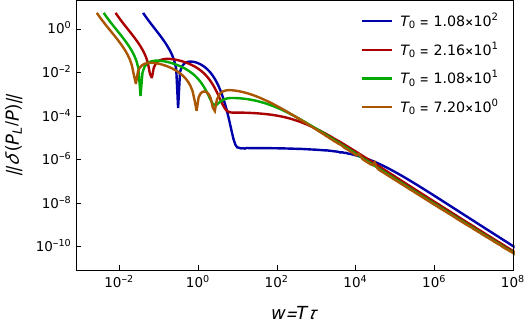}
\end{minipage}
\caption{
$T_0$-dependence of the distance of $\bar{\bf X}$ (Left) and $P_L/P$ (Right) in the $w$-time.
The plots denote the distance in Eq.(\ref{eq:distance}) with several $T_0$.
$\bar{\bf X}_{\rm ini}$ and the other parameters are taken the same values in Fig.\ref{fig:pullback_a} with $\tau_{\rm ini}=\tau_0$ ($p=0$).
The right panel is generated from the same flow in the left.
}
\label{fig:pullback_w_b}
\end{figure}

Let us look now at the distance in the entire time evolution including the IR regime, which is shown in Fig.\ref{fig:pullback_tot}. 
The panels of this figure show that the distance presents several phases in the flow time evolution which depends on choice of the flow time, $\tau$ or $w$. 
After the power law behavior dominating regime in the UV regime which we explained above, an exponentially damping phase emerges. The exponential damping effect in the $w$-time is stronger than in $\tau$-time, and in this phase or later,  initial flows which are far from the attractor approach it later with an extremely sharp gradient.
In the case of the $\tau$-time, once the flows are reaching close to the attactor and depending on the initial condition, these can develop a transient repelling and attracting concavity and then again start to obey a power law decay.
In the $w$-time, in contrast, the flows behave in the plateau phase for a while, and then enter to the final smooth power law phase.

Here, we focus on the plateau flat phase in the $w$-time.
The left panel in Fig.\ref{fig:pullback_w_b} shows the $T_0$-dependence of the distance of $||\delta {\bf X}||$.  Clearly this figure shows that the plateau phase depends on the temperature. When the initial values of the temperature are large the plateau phase of the distance extends over a large period of time until it drops out as a power law in the long time. However, as the initial temperature decreases the plateau phase lasts for shorter times compared with the initial higher temperature values while developing a smooth slope. Eventually the plateau phase disappears in the low temperature regime. Similarly, the exponential damping phase gets weaker as soon as the plateau phase starts to develop a weak gradient. The $T_0$-dependence of the plateau phase can be partially explained from the fact that the high temperature limit corresponds to the small $z$ limit, and the D-symmetry is restored, as we saw in Sec.\ref{sec:setup}.
When the D-symmetry holds, the situation is similar to the massless Bjorken flow, and changing the initial data of the the dissipative variables in the UV regime  does not affect much the IR regime. We also notice that as the initial temperature decreases, the particle mass effect becomes more relevant, and the initial data in the UV regime propagates faster to the IR regime.
In the low temperature regime, the plateau phase eventually vanishes and shifts to the final power law phase.
Similar observations can be generally obtained in other observables which are functions of $(T,\bar{\bf X})$ such as $P_L/P$, as shown in the right panel of Fig.\ref{fig:pullback_w_b}.

When the flows are sufficiently close to $\widehat{\cal S}({\frak R}_1,{\frak A})$, the asymptotic behavior of the distance can be evaluated from the transseries solution of $\bar{X}$ as
\be
&& || \delta \bar{\bf X} || (\tau;Y_{\rm ini}) \sim 
\begin{dcases}
C_1 \tau^{\rho_1} + C_2 \tau^{\rho_2} & \quad \mbox{as} \quad \tau \rightarrow 0_+ \\
D \frac{\theta_0}{\tau (\log \tau)^{3-\bar{\Delta}}}  & \quad \mbox{as} \quad \tau \rightarrow + \infty 
\end{dcases},  \label{eq:dis_tau_X} \\ \nl
&& || \delta \bar{\bf X} || (w;Y_{\rm ini}) \sim 
\begin{dcases}
C_1 w^{\widehat{\rho}_1} + C_2 w^{\widehat{\rho}_2} & \quad \mbox{as} \quad w \rightarrow 0_+ \\
D \frac{\theta_0}{w (\log w)^3}  & \quad \mbox{as} \quad w \rightarrow + \infty 
\end{dcases}, \label{eq:dis_w_X} 
\ee
with real constants $C_{1,2}$ and $D$, where $\rho_i$ and $\widehat{\rho}_i$ are defined in Eqs.(\ref{eq:transmono_UV}) and (\ref{eq:transmono_UV_w}) as
where
\be
&& \rho_1 = - \frac{95+3 \sqrt{505}}{105} \approx -0.2627, \qquad \rho_2 = - \frac{95-3 \sqrt{505}}{105} \approx -1.5468, \\
&& \widehat{\rho}_1 = - \frac{95+3 \sqrt{505}}{105 (1-\bar{\Delta}/6 )} \approx - \frac{0.2627}{(1-\bar{\Delta}/6 )}, \qquad \widehat{\rho}_2 = - \frac{95-3 \sqrt{505}}{105 (1-\bar{\Delta}/6 )} \approx - \frac{1.5468}{(1-\bar{\Delta}/6)},
\ee
respectively.
In this derivation, we used the fact that change of the initial condition of $(\bar{\pi},\bar{\Pi})$ affects \textit{all} the integration constants in the transseries, $(\bar{\pi}(\tau_{\rm ini})+ \delta \bar{\pi}(\tau_{\rm ini}),\bar{\Pi}(\tau_{\rm ini}) + \delta \bar{\Pi}(\tau_{\rm ini})) \mapsto (\sigma_T+\delta \sigma_T,\sigma_1+\delta \sigma_1,\sigma_2+\delta \sigma_2)$\footnote{The initial condition and the integration constant are related to each other through a nontrivial relationship.
Thus, the change of initial condition, $(\bar{\pi}(\tau_{\rm ini}),\bar{\Pi}(\tau_{\rm ini}))$, affects to all of integration constants, $(\sigma_T, \sigma_{1}, \sigma_{2})$.}.
Similarly, one finds for $P_L/P$ that 
\be
&& || \delta (P_L/P) || (\tau;Y_{\rm ini}) \sim 
\begin{dcases}
C_1 \tau^{\rho_1} + C_2 \tau^{\rho_2} & \quad \mbox{as} \quad \tau \rightarrow 0_+ \\
D \frac{\theta_0}{\tau (\log \tau)^{2-\bar{\Delta}}}  & \quad \mbox{as} \quad \tau \rightarrow + \infty 
\end{dcases}, \label{eq:dis_tau_P} \\ \nl
&& || \delta (P_L/P)|| (w;Y_{\rm ini}) \sim 
\begin{dcases}
C_1 w^{\widehat{\rho}_1} + C_2 w^{\widehat{\rho}_2} & \quad \mbox{as} \quad w \rightarrow 0_+ \\
D \frac{\theta_0}{w (\log w)^2}  & \quad \mbox{as} \quad w \rightarrow + \infty 
\end{dcases}. \label{eq:dis_w_P}
\ee
By considering the high temperature limit, which is up to some extent the massless regime, provides further understanding of the particle mass contribution to the asymptotics of the attracting behavior.
The ODEs in the small $z$ limit are given in Eqs.(\ref{eq:dTdtau_ml})-(\ref{eq:dPIdtau_ml}) are invariant under the D-scale transform with $(\Delta_T,\Delta_{\bar{\pi}},\Delta_{\bar{\Pi}})=(1/\bar{\Delta},0,0)$.
This limit does not affect the UV structure such as the UV CPs and their stabilities, and thus the attractor $\widehat{\cal S}({\frak R}_1,{\frak A})$ still exists.
Since the IR transseries solution for $\bar{X} \in \{\bar{\pi},\bar{\Pi} \}$ can be written down as
\be
&& \bar{X}(\tau)= \sum_{{\bf m} \in {\mathbb N}_0^2} \left[ \prod_{i=1} \sigma_i \frac{e^{-\frac{\sigma_T^{2\bar{\Delta}/3}}{(1-\bar{\Delta}/3)\theta_0}  \tau^{1-\bar{\Delta}/3}}}{\tau^{\beta_i}}\right] \sum_{n \in {\mathbb N}_0} a_{\bar{X}}^{[{\bf m},n]} \left[  \frac{\theta_0}{\sigma_T^{2\bar{\Delta}/3} \tau^{1-\bar{\Delta}/3}} \right]^n \quad \mbox{with} \quad (\beta_1,\beta_2)=\left( \frac{10}{21},-\frac{2}{3} \right), \label{eq:Xtau_massless} \\
&& \bar{X}(w)= \sum_{{\bf m} \in {\mathbb N}_0^2} \left[ \prod_{i=1} \sigma_i \frac{e^{-\frac{w}{\theta_0}}}{w^{\widehat{\beta}_i}}\right] \sum_{n \in {\mathbb N}_0} a_{\bar{X}}^{[{\bf m},n]}  \cdot \left[  \frac{\theta_0}{w} \right]^n \quad \mbox{with} \quad (\widehat{\beta}_1,\widehat{\beta}_2)=\left( \frac{10}{7(3-\bar{\Delta})},-\frac{2}{3-\bar{\Delta}} \right),
\ee
where all the coefficients do not have the $\sigma_T$-dependence, the distance in the IR regime can be evaluated as
\be
&& || \delta \bar{\bf X} || (\tau;Y_{\rm ini}) \sim 
D \frac{\theta_0}{\tau^{1-\bar{\Delta}/3}}  \qquad \qquad \quad \ \, \mbox{as} \quad \tau \rightarrow + \infty, \label{eq:dis_tau_ml} \\ 
&& || \delta \bar{\bf X} || (w;Y_{\rm ini}) \sim 
C_1 \frac{e^{-\frac{w}{\theta_0}}}{w^{\widehat{\beta}_1}} + C_2 \frac{e^{-\frac{w}{\theta_0}}}{w^{\widehat{\beta}_2}}   \quad \mbox{as} \quad w \rightarrow + \infty. \label{eq:dis_w_ml}
\ee
One can immediately find that these results, in particular the $w$-time case, are drastically changed from Eqs.(\ref{eq:dis_tau_X}) and (\ref{eq:dis_w_X}).
Thanks to the D-symmetry, the attracting behavior of the $w$-time always has the plateau phase after the exponentially damping phase, but the final power law phase does not appear.
As a matter of fact, this result is consistent with the attracting behavior in the massive case which gives the plateau phase when the temperature is high. Nonetheless, it also implies that the constant approximation of the speed of sound and the transport coefficients generally does not provide physical information of the IR regime in the massive Bjorken flow. Moreover, the D-symmetry is crucial for the attracting behavior with the $w$-time in the intermediate and later regimes.

This discussion can be generalized to other observables, ${\cal O}(T,\bar{\pi},\bar{\Pi})$.
In order to obtain the similar global structure to $(\bar{\pi}, \bar{\Pi})$, 
they satisfy $\lim_{T \rightarrow +\infty}{\cal O}(T,\bar{\pi},\bar{\Pi})=O(\bar{\pi}^{a_1}\bar{\Pi}^{a_2})$ with $a_1+a_2>0$ and $\lim_{\bar{\pi},\bar{\Pi} \rightarrow 0}{\cal O}(T,\bar{\pi},\bar{\Pi})=O(T^b)$ with $b \ge 0$.
In such a case, ${\cal O}(T,\bar{\pi},{\Pi})$ has the same number of UV CPs to the case of  $(\bar{\pi},\bar{\Pi})$ and converges to the local equilibrium in the IR limit. The ratio between the longitudinal and equilibrium pressure
$P_L/P$ satisfies this condition and thus, it the behavior of the flow time dependence of the distance of this observable has similar features like $||\delta \bar{\bf X}||$, such as the plateau phase in the $w$-time which is observed in Fig.\ref{fig:pullback_w_b}.

It is important to mention that the pullback attractor we discussed through this section is of local nature but it is not a global early-time (or pullback) attractor due to the existence of multiple UV CPs. When moving backwards in time, one has three possibilities:
(1) some of the flows near to the attractor solution move away towards another UV CP, (2) these diverge to infinity, or (3) reach exactly to the UV CP associated to the pullback attractor. In other words, for the non-conformal case a three dimensional basin of attraction in $\widehat{\cal M}$ exists and is restricted to a particular subset of initial values in the phase space. Additionally, flows are more sensitive to choice of the initial conditions than the scale invariant case since the dimensions of the phase space increase in the non-conformal case. 
Conversely, if the pullback attractor is global, then flows always diverge when moving backwards in the time except the ones located exactly on the attractor solution. 

We conclude this section by emphasizing that  attractor is a two dimensional ISOF in the $(\bar{\pi}, T, w)$-system in the scale invariant case after projection onto ${\cal M} \rightarrow \widehat{\cal M}$. In this sense, the numerical findings studied in the literature where the words ``single line" and ``universality" are frequently advocated are a very specific example which works only in the scale invariance case. These concepts do not always help when understanding and quantifying attractors in more general higher dimensional systems like the non-conformal one studied here or the Gubser flow~\cite{Behtash:2017wqg,Behtash:2019qtk}. It is also notable that the particle mass always couples with $\sigma_T$ in $(\bar{\pi},\bar{\Pi})$ when using the $w$-time. This fact means that the IR asymptotics has the mass dependence through $\sigma_T$ in the PT and NP sectors\footnote{More precisely, the energy dimensions of parameters in the transsries are \be
[w]=[\theta_0]=\bar{\Delta}-1, \qquad [\sigma_T]=1, \qquad [\sigma_{1,2}]=[\bar{\Delta}]=0.
\ee
Therefore, $\log w$, $\log_{(2)} w$, and $\sigma_T$ in the transseries of $(\bar{\pi}, \bar{\Pi})$ vary under the change of mass  in a generic value of $\bar{\Delta}$.
In Eq.(\ref{eq:Xtau_massless}), the power of $\sigma_T$ is determined to be $[\sigma_T]=1$ but can be different by redefining the energy dimensions of $\sigma_T$.
}. This feature does not change in \textit{any} observables such as $P_{L, T}/(\varepsilon + P)$.

\section{Comparing massive vs. massless Bjorken expanding fluids} \label{sec:massless_B}
In this section, we comment on the differences of the massive and massless dissipative fluids undergoing Bjorken flow. Our discussion addresses the differences at the level of the global flow structure, the IR transseries solutions, and the attractor solution.
The massless Bjorken flow can be directly obtained by setting exactly the traceless condition to Eqs.(\ref{eq:dTdtau_fm})-(\ref{eq:odediss}). This yields to the following reduced set of ODEs
\be 
&& \frac{d T}{d \tau} =- \frac{T}{\tau} \left( - \bar{\pi}  + c_s^2 \right), \qquad  \frac{d \bar{\pi}}{d \tau} =  - \frac{T^{\bar{\Delta}}}{\theta_0} \bar{\pi} - \frac{1}{\tau} \left( \frac{10}{21} \bar{\pi}  - \frac{4}{45} + 4 \bar{\pi}^2  \right), \label{eq:ODE_m0}
\ee
where $c_s^2=1/3$.
Because of the traceless condition, the ODEs~\eqref{eq:ODE_m0} possesses exact D-symmetry, i.e. $(T,\bar{\pi},\tau) \rightarrow (\lambda^{-\Delta_T} T, \lambda^{-\Delta_{\bar{\pi}}} \bar{\pi},\lambda \tau)$ with  $(\Delta_T,\Delta_{\bar{\pi}})=(1/\bar{\Delta},0)$. Therefore, Eqs.~\eqref{eq:ODE_m0} can be reduced dimensionally in terms of the $w =\tau T$ variable.   

\begin{figure}[tb]
\begin{center}
\includegraphics[width=0.80\textwidth]{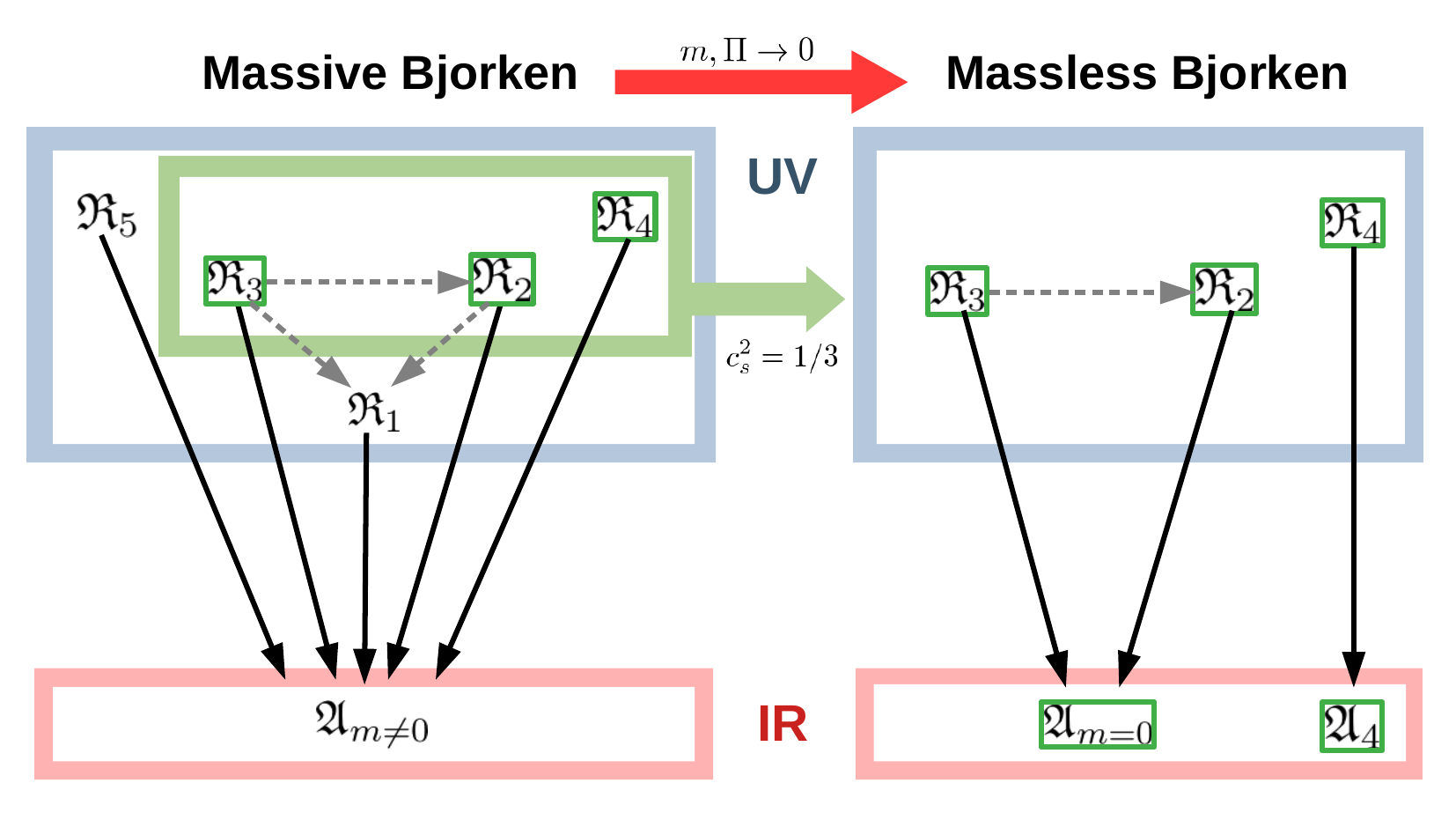}
\caption{
Flow structure of the massive and massless Bjorken flow.
The UV CPs with the green box are the scale invariant points.
The gray dashed lines are projection of flows emerging from UV CPs onto ${\cal F} \times \{ \tau_0 \}$  with $\tau_0 \ll 1$.
The black solid lines denote the ISOFs.
}
\label{fig:flow_fig}
\end{center}
\end{figure}

Let us firstly address the global flow structure.
For the comparison with the massive Bjorken, we consider the massless Bjorken using the $\tau$ as the flow time variable. 
In the IR, there exists two IR points $\Sigma_{\rm IR} = \{{\frak A}_{m=0}, {\frak A}_4\}$.
${\frak A}_{m=0}$ is the equilibrium point, and the asymptotic behavior of temperature is given by  $T(\tau) \sim \sigma_T \tau^{-1/3}$, where $\sigma_T$ is the integration constant associated with the temperature. 
At the UV CP, ${\frak A}_4$, the temperature $T(\tau) \sim \tau^{-1/\bar{\Delta}}$ which is unphysical as we explain below. 
In the UV, there exists three of the UV points in the massive case $\Sigma_{\rm UV} = \{ {\frak R}_2,{\frak R}_3,{\frak R}_4 \}$.
${\frak R}_2$ is the free-streaming point and gives $T(\tau) \sim  \sigma_T \tau^{-(55-\sqrt{505})/140}$, and ${\frak R}_3$ gives $T \sim \sigma_T \tau^{-(55+\sqrt{505})/140}$.
In order to find ${\frak R}_4$ fine-tuning of the initial conditions for both, $T$ and $\bar{\pi}$, is needed which gives $(T(\tau_0),\bar{\pi}(\tau_0))=(\frac{6 (9 \bar{\Delta}^2 - 55 \bar{\Delta} + 70) \theta_0}{35 \bar{\Delta}(3-\bar{\Delta})} \tau_0^{-1/\bar{\Delta}}, 1/3-1/\bar{\Delta})$. In this ISOF the flow arrives at ${\frak A}_4$ without the change of the value $\bar{\pi}$, and the temperature keeps the form $T(\tau)= \frac{6 (9 \bar{\Delta}^2 - 55 \bar{\Delta} + 70) \theta_0}{35 \bar{\Delta}(3-\bar{\Delta})} \tau^{-1/\bar{\Delta}}$ during the entire $\tau$-evolution. 
As a result, one finds in total three ISOFs, ${\cal S}({\frak R}_2,{\frak A}_{m=0})$, ${\cal S}({\frak R}_3,{\frak A}_{m=0})$ and ${\cal S}({\frak R}_4,{\frak A}_{4})$ respectively.
It is remarkable that the massive and massless Bjorken expanding fluids have in common the UV CPs ${\frak R}_{2,3,4}$  because the speed of sound is $c_s^2=1/3$ in these points while ${\frak A}_{m=0,4}$ is different.
Therefore, the traceless condition does not continuously maps the non-conformal ISOFs onto the conformal ones by taking the massless limit. Notice that the ${\cal S}({\frak R}_{4},{\frak A}_4)$ is unphysical because $\tau/\tau_R(\tau)$=${\rm const}$ for all $\tau \in {\mathbb R}_{\ge 0}$, meaning that under fine tuning certain initial conditions will never hydrodynamize. Moreover, this ISOF can not be observed when using $w=\tau/\tau_R(\tau)$.
When the constant relaxation time $\tau_R={\rm const}$, both of ${\frak A}_{4}$ and ${\frak R}_{4}$ do not appear, and thus ${\cal S}({\frak R}_4,{\frak A}_4)$ does not exist.
The global flow structure is summarized in Fig.\ref{fig:flow_fig}.

Now we consider the IR transseries expanding around ${\frak A}_{m=0}$.
These solutions take the following form~\cite{Behtash:2019txb}
\be
T(\tau) = \sigma_T \tau^{-1/3} \sum_{m \in {\mathbb N_0}} \sum_{n \in {\mathbb N_0}} a_{T}^{[m,n]} \zeta^m \mu^{n}, \qquad \bar{\pi}(\tau) =  \sum_{m \in {\mathbb N_0}} \sum_{n \in {\mathbb N}_0} a_{\bar{\pi}}^{[m,n]} \zeta^m \mu^{n},
\ee
where the NP transmonomial associated to the transient non-hydrodynamic mode $\zeta$ is
\begin{equation}
    \label{eq:conf_transm}
    \zeta = \sigma_{\bar \pi} \frac{e^{-S(\tau)}}{\tau^{\beta}}
\end{equation}
while $\mu$ is the PT transmonomial related with the inverse Knudsen number as shown below. As we argued in Sect.~\ref{sec:transs_IR}, the transseries structure of the massive fluid case arises from the asymptotic form of the speed of sound in terms of the temperature in the absence of viscous corrections. The asymptotic information of the PF sector propagates to the structure of transseries entirely when viscous corrections are turned on. One can outline a similar analogy for the massless Bjorken case as follows: Near to equilibrium $|\bar{\pi}(\tau)| \ll T(\tau)$ around the IR CP, ${\frak A}_{m=0}$.
The PF sector is computed from Eq.(\ref{eq:ODE_m0}) with $\bar{\pi}=0$ which yields the following ODE
\be
\frac{d T_{\rm pf}}{d \tau} = - \frac{T_{\rm pf}}{3\tau}\,,
\ee
whose solution reads easily as $T_{\rm pf}(\tau) = \sigma_T \tau^{-1/3}$.
$T_{\rm pf}$ is a monomial whose power law decay is uniquely determined by the equation of state $c_s^2=1/3$. In addition, the integration constant $\sigma_T$ arises as an overall factor.
The PF sector gives a transmonomial in the PT sector under the condition that $d \bar{\pi}/d \tau \sim 0$.
From the collision and non-collision parts of the ODE~\eqref{eq:ODE_m0}, one finds that
\be
&& 0 \sim -\frac{T_{\rm pf}^{\bar{\Delta}}}{\theta_0} + \frac{4}{45 \tau} \qquad \Rightarrow \qquad  \mu := \left( \frac{\theta_0}{T_{\rm pf}^{\bar{\Delta}}} \right)\,\frac{1}{\tau}
 = \frac{\theta_0}{ \sigma_T^{\bar{\Delta}}\tau^{1-{\bar{\Delta}}/3}}, \label{eq:mu_m0}
\ee
and thus, $\mu$ is identified with the Knudsen number. The transseries solution associated with the PT sector is expanded in power laws of $\mu$ because $T_{\rm pf}(\tau)$ is a monomial. This sector corresponds precisely to the asymptotic gradient hydrodynamic expansion whose coefficients are constants. In the non-conformal case this is not the case since the speed of sound depends non-trivially on the temperature. In the transmonomial (\ref{eq:mu_m0}) it is seen that $\sigma_T^{\bar{\Delta}}$ always couples with $\theta_0$, and thus, roughly speaking, the change of $\sigma_T$ can be regarded as rescaling of $\theta_0$ in the transseries. This does not happen in the massive Bjorken fluid case. In order to obtain the transmonomial of the NP sector, one needs to obtain the decay rate $S$ of the transmonomial $\zeta$~\eqref{eq:conf_transm}.
This term can be obtained by knowing only the PF sector and it is given by
\be
\frac{d S}{d \tau} = \frac{T_{\rm pf}^{\bar{\Delta}}}{\theta_0} \qquad \Rightarrow \qquad S(\tau) = \frac{1}{1-\bar{\Delta}/3} \cdot \frac{\sigma_T^{\bar{\Delta}}}{\theta_0} \tau^{1-\bar{\Delta}/3}+ C = \frac{\mu^{-1}}{1-\bar{\Delta}/3} + C,
\label{eq:dec_massless}
\ee
where $C$ is the integration constant which can vanish exactly without the loss of generality by absorbing it into $\sigma_{\bar{\pi}}$. The constant
$\beta$ in Eq.~\eqref{eq:conf_transm} is obtained from the non-collision part in Eq.(\ref{eq:ODE_m0}) and is $\beta=10/21$. Therefore, the precise form of the transmonomials are different from the massive case, but the correlation mechanism among each sector is indeed the same; they originate from $c_s^2$ in the IR limit, and $\mu$ and $S(\tau)$ are indeed determined only by the PF sector as well as the the strength of the relaxation time $\theta_0$. Furthermore, the decay rate of the non-hydrodynamic mode~\eqref{eq:dec_massless} grows as a power law thus, hydrodynamization happens very rapidly in the massless Bjorken fluid. In contrast, the decay of the non-conformal non-hydrodynamic mode is slowed down due to the PF sector of the temperature which results into the system hydrodynamizing at extremely long time scales.  

The definition of the distance in Eq.(\ref{eq:distance}) is applicable to the massless case by taking $\bar{\Pi}=0$.
$\widehat{\cal S}({\frak R}_2,{\frak A}_{m=0})$ in Fig.\ref{fig:flow_fig} is a stable ISOF and satisfies the condition of the pullback attractor. When comparing with the massive case, the distance in terms of the $\tau$ flow time behaves similarly . However, it is important to clarify that the terms $\propto C_2$ in Eq.~\eqref{eq:dis_tau_X} vanish exactly in the UV regime of the massless fluid case due to the absence of bulk pressure in the conformal case. For the same reasons, these terms also vanish exactly in Eq.~\eqref{eq:dis_w_X} when describing the dynamics in the $w$ time. In the same flow time $w$ the IR power law in Eq.~\eqref{eq:dis_w_X} cannot appear due to the D-symmetry.

\section{Summary and outlook} \label{sec:summary} 
In this paper, we have derived new transseries solutions of the non-conformal fluid equations for a fluid undergoing Bjorken expansion. We obtained the complete flow structure of these equations while providing important insights on the attractor behavior in the phase space of macroscopic fluid variables.
We list below the summary of our main findings:
\begin{itemize}
\item 
We have classified the flow structure of the non-conformal fluid equations. In total, We found five ISOFs. All flows on the ISOFs reach out asymptotically to the unique local equilibrium point, which is D-scale broken point and incompatible with the massless limit.
Three of the ISOFs ${\frak R}_{2,3,4}$ in Eq.(\ref{eq:UV_cp}) fulfill the traceless condition in the UV limit while ${\frak R}_{1,5}$ do not satisfy it.
In order to converge to ${\frak R}_{4,5}$ in the UV limit, it is needed to fine-tune in the phase space of initial conditions for the temperature.
Around these specific CPs, the temperature asymptotically behaves $T \propto \tau^{-1/\bar{\Delta}}$.
The longitudinal pressure is negative for both ${\frak R}_{4,5}$, and the transverse pressure is negative for ${\frak R}_4$ but positive for ${\frak R}_5$.
\item
We have found exact transseries solutions to the non-conformal fluid equations given in terms of $\tau$ and $w=\theta_0 \tau/\tau_R$ for the UV and the IR CPs.
The transseries drastically changes as compared with the massless Bjorken due to the particle mass effect and involves the $\log$- and $\log \log$-types transmonomials. In particular, the speed of sound plays a key role to understand and determine the physics of the IR transseries since the breaking of the conformal symmetry, given in terms of the value for the mass of the particle, propagates to the PT and the NP sectors via the PF sector as a form of transmonomials. The mass of the particle also effects the speed of sound, and thus, it gives an extremely slow decay for the temperature, $T \sim m/\log (m \tau) \sim m/\log(m^{1-\bar{\Delta}} w)$, around the local equilibrium. Furthermore, the Knudsen number $Kn$ increases its value in the IR regime and thus slows down the decay rate of the non-hydrodynamic modes since the later is inversely proportional to $Kn$. For non-conformal fluids hydrodynamization is affected by the equation of state and in our model it happens at extremely late times. This is in contrast with conformal systems where the emergence of fluid behavior occurs at very short time scales.  
\item The initial data including temperature carries to the IR regime as the $\sigma_T$-dependence in the transseries, which can not be eliminated from the transseries for both the $\tau$-time and the $w$-time. Thus, transport coefficients preserve the information of the transient rheology of the fluid, and thus its transient behavior while hydrodynamizing at extremely long times. Moreover, transport coefficients change their values as the non-hydrodynamic modes decay towards the attractor so the non-conformal fluid features a neat transient non-newtonian behavior. 
\item We have investigated the time dependence of the attracting behavior of the attractor solution, $\widehat{\cal S}({\frak R}_1,{\frak A})$, by introducing the distance on the (projected) phase space, $(\bar{\pi},\bar{\Pi}) $. 
We have found that the attractor exhibits several phases for the convergence rate of neighboring flows in the time evolution;
\begin{itemize}
    \item \underline{\bf $\tau$-time} \quad \,$\mbox{1st. power law} \ \Rightarrow \ (\mbox{2nd. power law}) \ \Rightarrow \ \mbox{exp damping (+ convex)}\ \ \Rightarrow \ \mbox{final power law}$
    \item \underline{\bf $w$-time} \quad $\mbox{1st. power law} \ \Rightarrow \ (\mbox{2nd. power law}) \ \Rightarrow \ \mbox{exp damping} \ \Rightarrow \ 
    \begin{pmatrix}
      \mbox{plateau} \\
      \mbox{in high $T$}
    \end{pmatrix} \ \Rightarrow \ \mbox{final power law}$
\end{itemize}
\begin{figure}[tb]
\begin{center}
\includegraphics[width=1.0\textwidth]{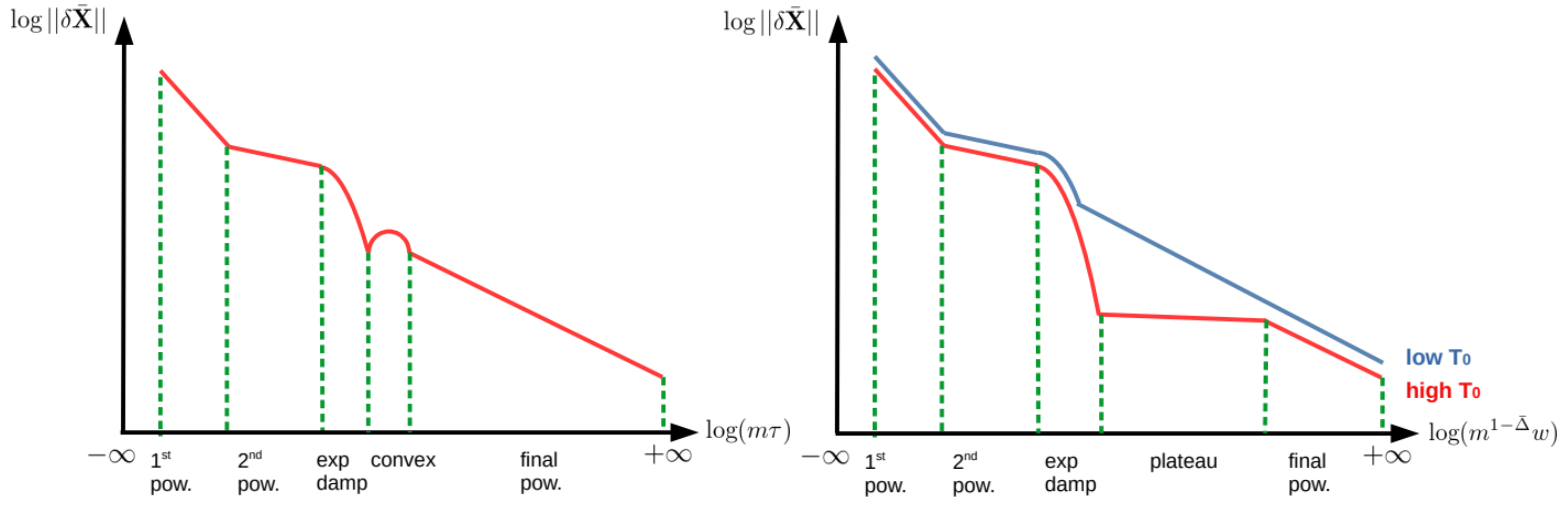}
\caption{Phases of the convergence rate $||\delta \bar{\bm X}||$ of neighboring flows to the attractor in different flow times, $\tau$ and $w$ (left and right panels respectively).
}
\label{fig:phase_fig}
\end{center}
\end{figure}
The schematic figure is shown in Fig.\ref{fig:phase_fig}.
The attractor satisfies the condition for a local pullback attractor in  the UV regime.
The second power law phase comes from existence of the bulk viscosity, and the final power law phase in the $w$-time is a direct consequence of the mass of the particle constituents of the fluid.
Each the power law phases can be characterized from the transseries solution.
In the $w$-time, the plateau phase appears when the temperature is high relative to the particle mass but vanishes as the temperature is getting lower values. The exponential damping phase gets weaker when decreasing the temperature as illustrated in Fig.\ref{fig:pullback_w_b}.
This means that, in the $w$-time, flows in low temperature (or with a heavy mass) can easily carry information about the UV initial data to the IR regime than in high temperature (or with a light mass).
The same feature can be observed for other obaservables such as $P_L/P$ and $P_T/P$.
\item We have made a comment on the difference of the massive and the massless Bjorken flows from the perspectives of the global flow structure, the exact transseries solution, and the phase of the attractor solution.
\end{itemize}
\vspace{3mm}

The non-conformal Bjorken fluid equations studied in this work is one of the simplest models with broken conformal symmetry. The transseries analysis together with the global flow structure of this model provides a better understanding of the hydrodynamization and non-linear transport processes fluids under extreme far-from-equilibrium conditions. On the mathematical side, one of the most important questions is the construction of the resurgent relation.
Even though the PT sectors look like divergent series because of existence of the NP sectors, the Borel transform and the alien calculus also have to be generalized from a standard definition. Furthermore, it would be relevant to determine the RG flow structure of the dynamical system by generalizing the ideas of transasymptotic matching~\cite{costin2008asymptotics} to the new IR transseries solutions found in this work. We left these interesting topics as a future work.

\begin{acknowledgments}
It is a pleasure to thank to M.~Spali\'nski for initial collaboration, insightful conversations about this project and encouragement to conclude it. We also thank to A.~Jaiswal, S.~Jaiswal, A.~Behtash and C.~Chattopadhyay for useful correspondence and discussions. M.~M. thanks Dr. Bhawana Bariar and Churro Jr. for making useful grammatical corrections in an earlier version of this article. The authors thank to the anonimous referees for their constructive feedback which helped to improve the presentation of this article. S.~K. is supported by the Polish National Science Centre grant
2018/29/B/ST2/02457. 
M.~M. was supported in part by the US Department
of Energy Grant No. DE-FG02-03ER41260 and BEST (Beam Energy Scan Theory) DOE
Topical Collaboration. 

\end{acknowledgments}

\appendix

\section{Formulation of nonautonomous dynamical systems}  \label{sec:const_nonauto}
We briefly review of nonautonomous systems \cite{kloeden2011nonautonomous,caraballo2016applied}.
The difference from autonomous systems is that a given ODE breaks the constant shift of a flow time.
This fact implies that the time influences the flow structure of the configuration space.
For taking into account this time dependence, the phase space is formulated by adding the time axis.
Nonautonomous systems are defined by introducing a trivial bundle ${\cal M}={\cal F} \times {\cal T}$ with the fiber space $(T,\bar{\pi},\bar{\Pi})\in {\cal F} = \overline{\mathbb R}_{\ge 0} \times {\mathbb R}^2$ and the base space $\tau \in {\cal T} = {\mathbb R}_{> 0}$, where $\overline{\mathbb R}_{\ge 0}:={\mathbb R}_{\ge 0} \cup \{ + \infty \}$.
By defining a tangent bundle of ${\cal M}$ as $T {\cal M} = \bigcup_{x \in {\cal M}} T_x {\cal M}$, the flows can be expressed by
\be
\frac{d}{d \tau} = F_T \frac{\ppd}{\ppd T} + F_{\bar{\pi}} \frac{\ppd }{\ppd \bar{\pi}} + F_{\bar{\Pi}} \frac{\ppd }{\ppd \bar{\Pi}} + \frac{\ppd}{\ppd \tau},
\ee
where $F_{\cal O} = \frac{d {\cal O}}{d \tau}$.
Each of flows can be distinguished by the initial condition, $(T(\tau_0),\bar{\pi}(\tau_0),\bar{\Pi}(\tau_0),\tau_0)$.
It is notable that due to breaking the constant time-shift invariance, $\tau_0$ is also required in the initial condition.
Since $(T,\bar{\pi},\bar{\Pi})$ are functions of $\tau$, a flow can be regarded as a section on ${\cal M}$ obeying Eqs.(\ref{eq:dTdtau_fm})-(\ref{eq:dPIdtau_fm}) by introducing the projection, $\phi:\, {\cal M} \rightarrow {\cal T}$~\cite{2020gub}.
Instead of thinking of such a fixed point, one considers \textit{invariant subspace of flows} (ISOF) in ${\cal M}$ and its attraction/repulsion as objects controlling flows in the time evolution. 

When considering attractor and repeller, it is important to find points to which flows converge in the early and late time limit.
In this paper, we call them \textit{convergent points} (CPs).
They are end-points of flows and can be formally defined by
\be
\Sigma_{\rm UV} &:=& \{\, {\frak R} \in {\cal F} \, |\,  \lim_{\tau \rightarrow 0_+} {\bf X}(\tau) = {\frak R}, \  \exists ({\bf X}(\tau), \tau) \in {\cal M} \,\}, \\
\Sigma_{\rm IR} &:=& \{\, {\frak A} \in {\cal F} \, |\, 
\lim_{\tau \rightarrow +\infty} {\bf X}(\tau) = {\frak A}, \ \exists ({\bf X}(\tau), \tau) \in {\cal M}  \,\},
\ee
where ${\bf X}(\tau)$ is a flow of $(T(\tau),\bar{\pi}(\tau),\bar{\Pi}(\tau))$ on ${\cal M}$. 
An ISOF on ${\cal M}$ is defined as the collection of all flows in the same UV and IR CPs as
\be
{\cal S}({\frak R}_a,{\frak A}_b) := \{\, \bigcup_{\tau \in {\cal T}} ({\bf X}, \tau) \subset {\cal M}  \, | \,  \lim_{\tau \rightarrow 0_+} {\bf X}(\tau)= {\frak R}_a \in \Sigma_{\rm UV} \ \mbox{and} \  \lim_{\tau \rightarrow +\infty} {\bf X}(\tau)= {\frak A}_b \in \Sigma_{\rm IR} \,\}, \label{eq:ISOF}
\ee
which are time independent subspaces in ${\cal M}$.
Notice that ${\cal S}({\frak R_{a}},{\frak A}_{b}) \cap {\cal S}({\frak R_{a^\prime}},{\frak A}_{b^\prime}) = \emptyset$ if $a \ne a^\prime$ or $b \ne b^\prime$\footnote{Remind that
the base space ${\cal T}$ is taken as an open domain in this definition.
Thus, the CPs are not included in the domain of ISOFs and are given by the limit of flows.
}.
Therefore, in our problem, the union of ISOFs provides the basin of attraction,
\be
\bigcup_{\substack{{\frak R}_a \in \Sigma_{\rm UV}\\{\frak A}_b \in \Sigma_{\rm IR}}} {\cal S}({\frak R}_a,{\frak A}_b) \subset {\cal M}. 
\ee
In order to focus only on the physics of viscous effects, we define a projected phase space, $\widehat{\cal M} \subset {\cal M}$, by dropping the $T$-axis, i.e. $(\bar{\pi}, \bar{\Pi}, \tau) \in \widehat{\cal M}$.
The flows on $T\widehat{\cal M}$ is expressed by
\be
\frac{d}{d \tau} =  F_{\bar{\pi}} \frac{\ppd }{\ppd \bar{\pi}} + F_{\bar{\Pi}} \frac{\ppd }{\ppd \bar{\Pi}} + \frac{\ppd}{\ppd \tau}.
\ee
It is important to note that the initial condition of the temperature propagates into $F_{\bar{\pi},\bar{\Pi}}$, i.e. $F_{\bar{\pi},\bar{\Pi}}$ has the $T_0$-dependence and $T_0$ changes the flow structure in $\widehat{\cal M}$.
In this reason, we offer the definition of the distance in Eq.(\ref{eq:distance_obs}) by keeping the information of the temperature, $T_0$.
It has the similar role to a control parameter in the context of bifurcation theory.

\section{Transport coefficients} \label{sec:tpcoeff_alpha}
The transport coefficients are written down as \cite{2014ARM}
\be 
\delta_{\pi \pi}(z) &=& \frac{5}{3} + \frac{7}{3}\frac{\alpha_3(z)}{\alpha_5(z)}, \\
\tau_{\pi \pi}(z) &=& 2 +  4 \frac{\alpha_3(z)}{\alpha_5(z)} ,\\
\lambda_{\pi \Pi}(z) &=& - \frac{2}{3} \chi(z), \\
\beta_{\pi}(z) &=& m^4 z \alpha_5(z), \\
\lambda_{\Pi \pi}(z) &=& \frac{2}{3} + \frac{7}{3} \frac{\alpha_3(z)}{\alpha_{5}(z)}  - c_s^2(z),\\
\delta_{\Pi \Pi}(z) &=& - \frac{5}{9} \chi(z) - c_s^2(z),\\
\beta_{\Pi}(z) &=& \frac{5}{3} m^4 z \alpha_{5}(z) - \frac{m^4 \alpha_1(z)}{z^2} c_s^2(\beta), \\
\chi(z) &=& \frac{m^4 z}{\beta_\Pi(z)} \left[ \left(1-3 c_s^2(z) \right)\left( \alpha_5(z) -  \frac{\alpha_1(z)}{z^3} \right)-\alpha_4(z) \right],
\ee
where $z:=m \beta$ with the inverse temperture $\beta$, and $\alpha_{n}(z)$ is expressed by the modified bessel function ${\rm K}_n(x)$ and the Bickley-Naylor function ${\rm Ki}_n(x)$ as
\be
 \alpha_1(z) &:=& \frac{1}{2 \pi^2}  \left[ 4 {\rm K}_2(z) + z {\rm K}_1(z) \right],\\
 \alpha_2(z) &:=& \frac{1}{2 \pi^2} \left[ (12+z^2){\rm K}_2(z)+ 3z {\rm K}_1(z) \right],\\
 \alpha_3(z) &:=& -\frac{1}{3360 \pi^2} \left[ {\rm K}_5(z) - 11 {\rm K}_3(z) + 58 {\rm K}_1(z) + 16 {\rm Ki}_3(z) - 64 {\rm Ki}_1(z) \right], \\
 \alpha_4(z)&:=& -\frac{1}{30 \pi^2} \left[ {\rm K}_3(z) -4 {\rm K}_1(z)  + {\rm Ki}_3(z) + 2 {\rm Ki}_1(z)  \right], \\
 \alpha_5(z) &:=& \frac{1}{480 \pi^2} \left[ {\rm K}_5(z) -7 {\rm K}_3(z) + 22 {\rm K}_1(z) - 16 {\rm Ki}_1(z) \right], \\
 \alpha_6(z) &:=& \frac{1}{2 \pi^2} \left[  (48 + 5 z^2) {\rm K}_2(z) + z (12 + z^2) {\rm K}_1(z) \right].
\ee

The asymptotic form for the coefficients in the small $z$ limit in Eqs.(\ref{eq:dTdtau_fm})-(\ref{eq:dPIdtau_fm}) is given by the following expressions
\be
&& C_\pi(z) = \frac{4}{3}- \frac{11}{3 z} + \frac{70}{3z} + O(z^{-3}), \\
&& \lambda_{\pi \Pi}(z) = 2- \frac{4}{z^2} + \frac{26}{z^2} + O(z^{-3}), \\
&& \bar{\beta}_\pi(z) = \frac{1}{z^2} + O(z^{-3}), \\
&& C_\Pi(z) = \frac{2}{3}- \frac{13}{3 z} + \frac{65}{3 z^2} + O(z^{-3}), \\
&& \lambda_{\Pi \pi}(z) = \frac{2}{3}- \frac{10}{3 z} + \frac{47}{3 z^2} + O(z^{-3}), \\
&& \bar{\beta}_\Pi(z) = \frac{2}{3 z^2}  + O(z^{-3}), \\
&& D(z) = z + \frac{1}{2} + \frac{39}{8z} + \frac{33}{8 z^2}  + O(z^{-3}), \\
&& c_s^2(z) = \frac{1}{z} - \frac{1}{2 z^2}  + O(z^{-3}).
\ee
On the other hand the large $z$ limit of Eqs.(\ref{eq:dTdtau_fm})-(\ref{eq:dPIdtau_fm}) yields the following asymptotic expressions
\be
&& C_\pi(z) = \frac{10}{21} + \frac{8}{63}z^2  + O(z^4)  \label{eq:Cpiz0}, \\
&& \lambda_{\pi \Pi}(z) = \frac{6}{5} +\frac{3 \pi}{25}z + \frac{1}{125} z^2 \left( 9 \pi^2 + 60 \gamma -5 +60 \log  \frac{z}{2}  \right) + O(z^3), \\
&& \bar{\beta}_\pi(z) = \frac{1}{15} - \frac{1}{90}z^2  + O(z^4), \\
&& C_\Pi(z) =  -\frac{2}{3} + \frac{\pi}{10}z + \frac{1}{450} z^2 \left( 27 \pi^2 + 180 \gamma + 35 + 180 \log  \frac{z}{2}  \right) + O(z^3), \\
&& \lambda_{\Pi \pi}(z) = \frac{1}{18} z^2+ O(z^4), \\
&& \bar{\beta}_\Pi(z) =  O(z^4), \\
&& D(z) =  4 + \frac{1}{12}z^2 + O(z^4), \label{eq:Dz0} \\
&& c_s^2(z) = \frac{1}{3} - \frac{1}{36} z^2 + O(z^4),
\ee
for the small $z$ limit, where $\gamma$ is the Euler-Mascheroni constant.

\section{Construction of transmonomials for the IR regime} \label{sec:trans_late}
We consider the transmonomial for the IR regime.
Assume that $|\bar{X}(\tau)| \ll T(\tau)$ as $\tau \rightarrow \infty$ where $X(\tau) \in \{ \bar{\pi}(\tau), \bar{\Pi}(\tau) \}$.
In this case, the approximated equation of Eq.(\ref{eq:dTdtau_fm}) is given by
\be
\frac{dT}{d\tau} \sim -\frac{T^2}{\tau}.
\ee
The solution is given by
\be
T(\tau) \sim \frac{1}{- \sigma_T + \log \tau}  \sim \frac{1}{\log \tau} + O((\log \tau)^{-2}).
\ee
Then, we define $T(\tau) = (\log \tau)^{-1} + \delta T(\tau)$.
By taking up to $O(\delta T,\tau^{-1} (\log \tau)^{-3})$ in Eq.(\ref{eq:dTdtau_fm}), one obtains
\be
\frac{d \delta T}{d \tau}  \sim -\frac{\delta T}{\tau \log \tau} \left[ 2 - \frac{3}{2 \log \tau} \right]  + \frac{1}{2 \tau (\log \tau)^3},
\ee
and the solution is given by
\be
\delta T(\tau) \sim \frac{e^{-\frac{3}{2 \log \tau}} \left[2 \sigma_T - {\rm Ei}\left( \frac{3}{2 \log \tau}\right) \right]}{2 (\log \tau)^2} \sim \frac{\sigma_T^\prime}{(\log \tau)^2} + \frac{\logt \tau }{2(\log \tau)^2},
\ee
where $\sigma_T^\prime:= \sigma_T   - (\gamma +\log \frac{3}{2})/2$, ${\rm Ei}(x)$ is the exponential integral, and  $\gamma$ is the Euler constant.
In order to obtain the leading order of $\bar{X}(\tau)$, we consider Eqs.(\ref{eq:dpidtau_fm}) and (\ref{eq:dPIdtau_fm}) by setting $T(\tau) \sim (\log \tau)^{-1}$ and $d \bar{X}/d \tau \sim 0$:
\be
&& -\frac{\bar{\pi}}{(\log \tau)^{\bar{\Delta}} \theta_0} + \frac{4}{3 \tau (\log \tau)^2} \sim 0, \qquad  -\frac{\bar{\Pi}}{(\log \tau)^{\bar{\Delta}} \theta_0} - \frac{2}{3 \tau (\log \tau)^2} \sim 0,
\ee
and the solution is given by
\be
&& \bar{\pi} (\tau)  \sim \frac{4 \theta_0}{3\tau (\log \tau)^{2-\bar{\Delta}}}, \qquad \bar{\Pi} (\tau) \sim -  \frac{2 \theta_0}{3 \tau (\log \tau)^{2-\bar{\Delta}}}.
\ee

Then, in order to obtain the transmonomial corresponding to non-perturbative modes, we consider the linearized equation for $\bar{X}(\tau)$ with $T(\tau)\sim (\log \tau)^{-1}$.
It is given by
\be 
&& \frac{d \delta \bar{\pi}}{d \tau} \sim -\frac{\delta \bar{\pi}}{ (\log \tau)^{\bar{\Delta}} \theta_0}- \frac{1}{3\tau} \left[ \left( 4 - \frac{11}{ \log \tau} \right)\delta \bar{\pi} + \left( - 4 + \frac{8}{\log \tau} \right) \delta \bar{\Pi} \right], \\
&& \frac{d \delta \bar{\Pi}}{d \tau} \sim -\frac{\delta \bar{\Pi}}{ (\log \tau)^{\bar{\Delta}} \theta_0}- \frac{1}{3\tau} \left[ \left( -2 + \frac{10}{\log \tau} \right) \delta \bar{\pi} + \left( 2 - \frac{13}{\log \tau}  \right) \delta \bar{\Pi} \right], 
\ee
and it can be expressed by the matrix form as
\be
\frac{d \delta \bar{\bf X}}{d \tau} \sim - \left[ {\mathbb I}_2 \frac{(\log \tau)^{-\bar{\Delta}}}{\theta_0} + \frac{1}{\tau} \left( {\frak B}^{(1,0)} + {\frak B}^{(1,1)} (\log \tau)^{-1} \right) \right] \delta \bar{\bf X}, \label{eq:dx_vector}
\ee
where
\be
&& \delta \bar{\bf X} := (\delta \bar{\pi},\delta \bar{\Pi})^{\top}, \qquad {\frak B}^{(1,0)} := \frac{1}{3}
\begin{pmatrix}
  4 && -4 \\
  -2  && 2
\end{pmatrix}, \qquad
{\frak B}^{(1,1)} := \frac{1}{3}
\begin{pmatrix}
  -11 && 8 \\
  10  && -13
\end{pmatrix}.
\ee
By diagonalizing ${\frak B}^{(1,0)}$ using the invartible matrix defined as
\be
U = \frac{1}{3}
\begin{pmatrix}
  -1 && 1 \\
  1 && 2
\end{pmatrix}, \qquad
U^{-1} =
\begin{pmatrix}
  -2 && 1 \\
  1 && 1
\end{pmatrix},
\ee
Eq.(\ref{eq:dx_vector}) gives
\be
\frac{d \delta \tilde{\bf X}}{d \tau} \sim - \left[ {\mathbb I}_2 \frac{(\log \tau)^{-\bar{\Delta}}}{\theta_0} + \frac{1}{\tau} \left( \tilde{\frak B}^{(1,0)} + \tilde{\frak B}^{(1,1)} (\log \tau)^{-1} \right) \right] \delta \tilde{\bf X}, \label{eq:dx_vector2}
\ee
where
\be
&& \delta \tilde{\bf X}:=U \delta \bar{\bf X}, \nl
&& \tilde{\frak B}^{(1,0)} := U {\frak B}^{(1,0)} U^{-1} = {\rm diag}(2,0), \qquad  \tilde{\frak B}^{(1,1)} := U {\frak B}^{(1,1)} U^{-1} = -
\begin{pmatrix}
  7 && 0 \\
  4 && 1
\end{pmatrix}.
\ee
For diagonalization of $\tilde{\frak B}^{(1,1)}$, we define $\delta \check{\bf X}= V(\tau) \delta \tilde{\bf X}$ where $V(\tau):= {\mathbb I}_2 + V_{\rm c} (\log \tau)^{-1}$ with a 2-by-2 constant matrix $V_{\rm c}$\cite{costin2008asymptotics}. 
Thus, by taking into account up to $O(\tau^{-1}(\log \tau)^{-1})$,  Eq.(\ref{eq:dx_vector2}) becomes 
\be
\frac{d \delta \check{\bf X}}{d \tau} &\sim&  - V(t) \left[ {\mathbb I}_2 \frac{(\log \tau)^{-\bar{\Delta}}}{\theta_0} + \frac{1}{\tau} \left( \tilde{\frak B}^{(1,0)} + \tilde{\frak B}^{(1,1)} (\log \tau)^{-1} \right) \right] V(\tau)^{-1} \delta \check{\bf X} \nl
&\sim&  -  \left[ {\mathbb I}_2 \frac{(\log \tau)^{-\bar{\Delta}}}{\theta_0} + \frac{1}{\tau} \left\{ \tilde{\frak B}^{(1,0)} + \left( \tilde{\frak B}^{(1,1)} + \left[ V_{\rm c},\tilde{\frak B}^{(1,0)} \right] \right)  (\log \tau)^{-1} \right\} \right]  \delta \check{\bf X}, \label{eq:dx_vector3}
\ee
where $[A,B]:=A B-B A$.
We determine $V_{\rm c}$ such that $\tilde{\frak B}^{(1,1)}+ \left[ V_{\rm c},\tilde{\frak B}^{(1,0)} \right]$ gives the diagonal matrix whose elements are the diagonal parts of $\tilde{\frak B}^{(1,1)}$, i.e.
\be
V_{\rm c} =
\begin{pmatrix}
  0 && 0 \\
  2 && 0
\end{pmatrix}.
\ee
From the diagonalized linearized equation, the solution is obtained by
\be
\delta \bar{\pi}(\tau) &\sim& -2 \sigma_1 e^{ - \frac{\tau}{ (\log \tau)^{\bar{\Delta}} \theta_0} \left[ 1 + O((\log \tau)^{-1})\right]}\frac{(\log \tau)^7}{\tau^2}\left[1+O((\log \tau)^{-1})\right] \nl
&& + \sigma_2 e^{ - \frac{\tau}{ (\log \tau)^{\bar{\Delta}} \theta_0} \left[ 1 + O(\log \tau)^{-1})\right]} \log \tau \left[1+O((\log \tau)^{-1})\right], \\ \nl
\delta \bar{\Pi}(\tau) &\sim&
\sigma_1 e^{ - \frac{\tau}{ (\log \tau)^{\bar{\Delta}} \theta_0} \left[ 1 + O((\log \tau)^{-1})\right]}\frac{(\log \tau)^7}{\tau^2}\left[1+O((\log \tau)^{-1})\right] \nl
&& + \sigma_2 e^{ - \frac{\tau}{ (\log \tau)^{\bar{\Delta}} \theta_0} \left[ 1 + O(\log \tau)^{-1})\right]} \log \tau \left[1+O((\log \tau)^{-1})\right],
\ee
where $\sigma_{i=1,2}$ is the integration constant, which couples with the non-perturbative part.

\section{Construction of transmonomials for the UV regime} \label{sec:trans_early}
We consider the transmonomial for the UV regime.
For technical convenience, we rewrite Eqs.(\ref{eq:dTdtau_fm})-(\ref{eq:dPIdtau_fm}) as
\be
 \frac{d \beta}{d s} &=& -\frac{\beta}{s} \left[ \bar{\Pi} - \bar{\pi}  + c_s^2(\beta) \right], \label{eq:dTdtau_fm_UV} \\
 \frac{d \bar{\pi}}{d s} 
 &=&     \frac{\bar{\pi}}{\beta^{\bar{\Delta}} \theta_0 s^2}  +\frac{1}{s} \left[ C_\pi(\beta)
   \bar{\pi} -  \frac{2 \lambda_{\pi \Pi}(\beta)}{3}  \bar{\Pi} -  \frac{4 \bar{\beta}_{\pi}(\beta)}{3}  - D (\beta) \left( \bar{\Pi} - \bar{\pi} \right) \bar{\pi}   \right],  \label{eq:dpidtau_fm_UV} \\
 \frac{d \bar{\Pi}}{d s} &=&   \frac{\bar{\Pi}}{\beta^{\bar{\Delta}} \theta_0 s^2} + \frac{1}{s} \left[ C_{\Pi}(\beta)  \bar{\Pi} -  \lambda_{\Pi \pi}(\beta)  \bar{\pi} +  \bar{\beta}_{\Pi}(\beta)  - D (\beta) \left( \bar{\Pi} - \bar{\pi} \right) \bar{\Pi} \right],  \label{eq:dPIdtau_fm_UV} 
\ee
where $s=\tau^{-1}$, and we employ the inverse temperature $\beta$ instead of thinking of $T$.

Firstly, we find the UV CPs by considering that $\beta \rightarrow 0$ and $d \bar{X}/ds \rightarrow 0$ as $s \rightarrow +\infty$.
We define 
$\bar{X}_{\rm UV} :=  \lim_{s \rightarrow \infty} \bar{X}(s)$.
Since $c_s^2 \rightarrow 1/3$, the leading order of temperature is obtained as $\beta \sim \sigma_\beta s^{-\delta}$, where
\be
\delta:={\bar \Pi}_{\rm UV}-{\bar \pi}_{\rm UV}+c_s^2(0). \qquad \left(c_s^2(0)=1/3 \right) \label{eq:delta}
\ee
Substituting the asymptotic forms of the transport coefficients, Eqs.(\ref{eq:Cpiz0})-(\ref{eq:Dz0}), into the ODEs of $\bar{X}$ gives
\be
&&  \frac{\bar{\pi}_{\rm UV}}{\sigma_\beta^{\bar{\Delta}} \theta_0 s^{2-\bar{\Delta} \delta}}  +\frac{1}{s} \left[ \frac{10}{21} \bar{\pi}_{\rm UV} -  \frac{4}{5}  \bar{\Pi}_{\rm UV} - \frac{4}{45}  - 4 \left( \bar{\Pi}_{\rm UV} - \bar{\pi}_{\rm UV} \right) \bar{\pi}_{\rm UV}   \right] \sim 0,  \label{eq:dpidtau_fm_UV3} \\
&& \frac{\bar{\Pi}_{\rm UV}}{\sigma_\beta^{\bar{\Delta}} \theta_0 s^{2-\bar{\Delta} \delta}} + \frac{1}{s} \left[ - \frac{2}{3}  \bar{\Pi}_{\rm UV}   - 4 \left( \bar{\Pi}_{\rm UV} - \bar{\pi}_{\rm UV} \right) \bar{\Pi}_{\rm UV} \right] \sim 0,  \label{eq:dPIdtau_fm_UV3} 
\ee

One can consider the two cases: 
\begin{enumerate}
\item The non-collision part is dominant. 
In this case, one only has to consider the part of $\frac{1}{s}[\cdots] = 0$ in Eqs.(\ref{eq:dpidtau_fm_UV3})(\ref{eq:dPIdtau_fm_UV3}), which would be ``Class 1" shown in (\ref{eq:class1}).
Notice that $\sigma_\beta$ can be arbitrary.
\item The collision and non-collision parts have the same order to each others.
In this case, one has to set the constraint that $2 - \bar{\Delta} \delta = 1$ and solve Eqs.(\ref{eq:dpidtau_fm_UV3})(\ref{eq:dPIdtau_fm_UV3}).
As a result, $\sigma_\beta$ can not be an arbitrary constant because of the constraint as
\be
2 - \bar{\Delta} \delta = 1 \quad \Rightarrow \quad 
  {\bar \Pi}_{\rm UV} = \frac{1}{\bar{\Delta}} - \frac{1}{3} + {\bar \pi}_{\rm UV},
\ee
which would be ``Class 2" shown in (\ref{eq:class2}).
\end{enumerate}

For the technical convenience, we shift $\bar{X} \rightarrow \bar{X} + \bar{X}_{\rm UV}$, and thus the redefined $\bar{X}$ satisfies $\bar{X}(\tau) \rightarrow 0$ in the UV limit (i.e. the large $s$ limit).
From Eqs.(\ref{eq:dTdtau_fm_UV})-(\ref{eq:dPIdtau_fm_UV}), one obtains   
\be
 \frac{d \beta}{d s} &=& -\frac{\beta}{s} \left[ \delta  + \bar{\Pi} - \bar{\pi} + \tilde{c}_s^2(\beta) \right], \label{eq:dTdtau_fm_UV2} \\
 \frac{d \bar{\pi}}{d s}  &=&     \frac{\bar{\pi}_{\rm UV} + \bar{\pi}}{\beta^{\bar{\Delta}} \theta_0 s^2}  +\frac{1}{s} \left[ \tilde{G}_{\pi}(\beta)  + C_\pi(\beta) \bar{\pi}   -  \frac{2 \lambda_{\pi \Pi}(\beta)}{3} \bar{\Pi} \right. \nl
   && \left.  - {D}(\beta) \left( \bar{\Pi}_{\rm UV} - 2\bar{\pi}_{\rm UV} \right) \bar{\pi} - {D}(\beta) \bar{\pi}_{\rm UV} \bar{\Pi} - D(\beta) \left( \bar{\Pi} - \bar{\pi} \right) \bar{\pi}  \right],  \label{eq:dpidtau_fm_UV2} \\
 \frac{d \bar{\Pi}}{d s} &=&   \frac{\bar{\Pi}_{\rm UV} + \bar{\Pi}}{\beta^{\bar{\Delta}} \theta_0 s^2} + \frac{1}{s} \left[ \tilde{G}_{\Pi}(\beta) + C_{\Pi}(\beta)  \bar{\Pi} - \lambda_{\Pi \pi}(\beta)  \bar{\pi}   \right. \nl
   && \left.  - D(\beta) \left( 2\bar{\Pi}_{\rm UV} - \bar{\pi}_{\rm UV} \right) \bar{\Pi}  + D (\beta)  \bar{\Pi}_{\rm UV} \bar{\pi}  - D (\beta) \left( \bar{\Pi} - \bar{\pi} \right) \bar{\Pi} \right],  \label{eq:dPIdtau_fm_UV2} 
\ee
where $\tilde{A}(\beta):=A(\beta)-A(0)$,
and 
\be
\tilde{G}_{\pi}(\beta) &:=& \tilde{C}_\pi(\beta) \bar{\pi}_{\rm UV}-  \frac{2 \tilde{\lambda}_{\pi \Pi}(\beta)}{3} \bar{\Pi}_{\rm UV} - \frac{4 \tilde{\bar{\beta}}_{\pi}(\beta)}{3}- \tilde{D}(\beta) \left( \bar{\Pi}_{\rm UV} - \bar{\pi}_{\rm UV} \right) \bar{\pi}_{\rm UV}, \\
\tilde{G}_{\Pi}(\beta) &:=& \tilde{C}_{\Pi}(\beta) \bar{\Pi}_{\rm UV} - \lambda_{\Pi \pi}(\beta)  \bar{\pi}_{\rm UV} +  \bar{\beta}_{\Pi}(\beta)   - \tilde{D}(\beta) \left( \bar{\Pi}_{\rm UV} - \bar{\pi}_{\rm UV} \right) \bar{\Pi}_{\rm UV}.
\ee
Below, we separately consider two classes,
\be
\mbox{Class I} && \nl
 {\frak R}_1 &:& \ (\beta_{\rm UV},\bar{\pi}_{\rm UV}, \bar{\Pi}_{\rm UV}) =
\left(0, - \frac{7}{54}, -\frac{8}{27} \right), \nl 
{\frak R}_2 &:& \ (\beta_{\rm UV},\bar{\pi}_{\rm UV}, \bar{\Pi}_{\rm UV}) = \left(0, - \frac{25 - 3 \sqrt{505}}{420}, 0 \right), \label{eq:class1} \\
{\frak R}_3 &:& \ (\beta_{\rm UV},\bar{\pi}_{\rm UV}, \bar{\Pi}_{\rm UV}) = \left(0, - \frac{25 + 3 \sqrt{505}}{420}, 0 \right), \nl \nl
\mbox{Class II} && \nl
 {\frak R}_4 &:& \ (\beta_{\rm UV},\bar{\pi}_{\rm UV}, \bar{\Pi}_{\rm UV}) = \left(0,\frac{1}{3} - \frac{1}{\bar{\Delta}},0 \right), \nl
 {\frak R}_5 &:& \ (\beta_{\rm UV},\bar{\pi}_{\rm UV}, \bar{\Pi}_{\rm UV}) =
 \left(0, -\frac{14}{27} + \frac{7}{3 \bar{\Delta}}, - \frac{23}{27} + \frac{10}{3\bar{\Delta}}  \right). \label{eq:class2}
\ee
\subsection{Class I $({\frak R}_{1,2,3})$} \label{sec:trans_early1}
In the case of Class I, the non-collision part is dominant in the UV limit.
From Eq.(\ref{eq:dTdtau_fm_UV2}), one can immediately find the leading order of temperature as
\be
\beta(s) \sim \frac{\sigma_\beta}{s^{\delta}}, \qquad \delta = {\bar \Pi}_{\rm UV}-{\bar \pi}_{\rm UV}+c_s^2(0), \quad (c_s^2(0)=1/3) \label{eq:beta_leading_I}
\ee
where $\sigma_\beta$ is the integration constant.
Substituting Eq.(\ref{eq:beta_leading_I}) into Eqs.(\ref{eq:dpidtau_fm_UV2}) and (\ref{eq:dPIdtau_fm_UV2}) yields
\be 
\frac{d \bar{\pi}}{ds} &\sim& \frac{\bar{\pi}_{\rm UV}}{\sigma_\beta^{\bar{\Delta}} \theta_0 s^{2-{\bar{\Delta}}\delta}} + \frac{1}{s} \left[ - \frac{2 \pi \bar{\Pi}_{\rm UV}}{25} \cdot \frac{\sigma_{\beta}}{s^\delta}   +  \left( \frac{10}{21} -4 \left( \bar{\Pi}_{\rm UV} - 2\bar{\pi}_{\rm UV}  \right) \right) \bar{\pi} - 4 \left( \frac{1}{5} + \bar{\pi}_{\rm UV} \right) \bar{\Pi} \right], \nl \\
\frac{d \bar{\Pi}}{ds} &\sim& \frac{\bar{\Pi}_{\rm UV}}{\sigma_\beta^{\bar{\Delta}} \theta_0 s^{2-{\bar{\Delta}}\delta}} + \frac{1}{s} \left[  \frac{\pi \bar{\Pi}_{\rm UV}}{10} \cdot \frac{\sigma_\beta}{s^{\delta}}   - \left( \frac{2}{3} +4 \left( 2\bar{\Pi}_{\rm UV} - \bar{\pi}_{\rm UV}  \right)  \right)\bar{\Pi} + 4 \bar{\Pi}_{\rm UV} \bar{\pi} \right],
\ee
and the leading terms of $(\bar{\pi},\bar{\Pi})$ is obtained by solving these ODEs.
The solution is given by
\begin{small}
\be
\bar{\pi}(s) &\sim&
- \frac{\left(\frac{1}{3} - 4 \bar{\Pi}_{\rm UV} +4\bar{\pi}_{\rm UV}-\bar{\Delta}\delta \right) \bar{\pi}_{\rm UV} + \frac{4}{5} \bar{\Pi}_{\rm UV}}{\left(\frac{31}{21}-4 \bar{\Pi}_{\rm UV} +8 \bar{\pi}_{\rm UV} -\bar{\Delta} \delta \right) \left(\frac{1}{3} - 8 \bar{\Pi}_{\rm UV} + 4\bar{\pi}_{\rm UV} - \bar{\Delta}\delta  \right) + 16 \left(\bar{\pi}_{\rm UV}+\frac{1}{5}\right) \bar{\Pi}_{\rm UV}} \cdot \frac{1}{\sigma_\beta^{\bar{\Delta}} \theta_0 s^{1-\bar{\Delta}\delta}} \nl
&& + \frac{2\pi}{25} \cdot \frac{ \left(\frac{8}{3} + 8 \bar{\Pi}_{\rm UV} + 6 \bar{\pi}_{\rm UV} - \delta \right) \bar{\Pi}_{\rm UV}}{\left(\frac{2}{3} + 8 \bar{\Pi}_{\rm UV} - 4 \bar{\pi}_{\rm UV} - \delta \right) \left(\frac{10}{21} -4 \bar{\Pi}_{\rm UV} +8 \bar{\pi}_{\rm UV}+\delta   \right) - 16 \left(\bar{\pi}_{\rm UV}+\frac{1}{5}\right) \bar{\Pi}_{\rm UV}} \cdot \frac{\sigma_\beta}{s^\delta}, \\ \nl
\bar{\Pi}(s) &\sim&
- \frac{\left( \frac{31}{21} -4 \bar{\Pi}_{\rm UV} + 4 \bar{\pi}_{\rm UV} - \bar{\Delta}\delta \right) \bar{\Pi}_{\rm UV}}{\left(\frac{31}{21}-4 \bar{\Pi}_{\rm UV} +8 \bar{\pi}_{\rm UV} -\bar{\Delta}\delta \right) \left(\frac{1}{3} - 8 \bar{\Pi}_{\rm UV}-4\bar{\pi}_{\rm UV} - \bar{\Delta}\delta \right) + 16 \left(\bar{\pi}_{\rm UV}+\frac{1}{5}\right) \bar{\Pi}_{\rm UV}}  \cdot \frac{1}{\sigma_\beta^{\bar{\Delta}} \theta_0 s^{1-\bar{\Delta}\delta}} \nl 
&& +\frac{\pi}{10} \cdot  \frac{\left(\frac{10}{21} - \frac{4}{5} \bar{\Pi}_{\rm UV} + 8 \bar{\pi}_{\rm UV} + \delta   \right) \bar{\Pi}_{\rm UV}}{\left(\frac{2}{3} + 8 \bar{\Pi}_{\rm UV}- 4 \bar{\pi}_{\rm UV} - \delta  \right) \left( \frac{10}{21} -4 \bar{\Pi}_{\rm UV} + 8 \bar{\pi}_{\rm UV} + \delta  \right) - 16 \left(\bar{\pi}_{\rm UV}+\frac{1}{5}\right) \bar{\Pi}_{\rm UV}} \cdot \frac{\sigma_\beta}{s^\delta}.
\ee
\end{small} 
Then, we consider the linearized equations given by
\be
\frac{d \delta \bar{\pi}}{ds} &\sim& \frac{1}{s} \left[ \left( \frac{10}{21} -4 \left( \bar{\Pi}_{\rm UV} - 2\bar{\pi}_{\rm UV}  \right) \right) \delta \bar{\pi} - 4 \left( \frac{1}{5} + \bar{\pi}_{\rm UV} \right) \delta \bar{\Pi} \right], \\
\frac{d \delta \bar{\Pi}}{ds} &\sim& \frac{1}{s} \left[ - \left( \frac{2}{3} +4 \left( 2\bar{\Pi}_{\rm UV} - \bar{\pi}_{\rm UV}  \right)  \right) \delta \bar{\Pi} + 4 \bar{\Pi}_{\rm UV} \delta \bar{\pi} \right].
\ee
These generate transmonomials coupling with the integration constants, which are obtained as 
\be
 \delta \bar{\pi}(s) && \sim \nl
&& \begin{dcases}
     \frac{2+7 \left( \bar{\pi}_{\rm UV} + \bar{\Pi}_{\rm UV}  \right)- \sqrt{(7 \bar{\pi}_{\rm UV}+2)^2+ 49 \bar{\Pi}_{\rm UV}^2-14 \left( 7 \bar{\pi}_{\rm UV}+ \frac{4}{5} \right) \bar{\Pi}_{\rm UV}}}{14 \bar{\Pi}_{\rm UV}} \cdot \frac{\sigma_1}{s^{\rho_1}}  + \frac{\sigma_2}{s^{\rho_2}} &  \mbox{if \ $\bar{\Pi}_{\rm UV} \ne 0$} \\
  \frac{7(1+5\bar{\pi}_{\rm UV})}{5(2+7\bar{\pi}_{\rm UV})} \cdot \frac{\sigma_1}{s^{\rho_1}} + \frac{\sigma_2}{s^{\rho_2}} &  \mbox{if \ $\bar{\Pi}_{\rm UV} = 0$} 
   \end{dcases}, \nl \\
 \delta \bar{\Pi}(s) && \sim \nl
  && \begin{dcases}
       \frac{\sigma_1}{s^{\rho_1}} +\frac{ \frac{7}{5}(2+10 \bar{\pi}_{\rm UV})}{2+ 7(\bar{\pi}_{\rm UV}+ \bar{\Pi}_{\rm UV})  -  \sqrt{(7 \bar{\pi}_{\rm UV}+2)^2+ 49 \bar{\Pi}_{\rm UV}^2- 14 (7 \bar{\pi}_{\rm UV} + \frac{4}{5}) \bar{\Pi}_{\rm UV}}} \cdot \frac{\sigma_2}{s^{\rho_2}} &  \mbox{if \ $\bar{\Pi}_{\rm UV} \ne 0$} \\
  \frac{\sigma_1}{s^{\rho_1}}  &  \mbox{if \ $\bar{\Pi}_{\rm UV} = 0$} 
   \end{dcases}, \nl
\ee
where $\sigma_i$ is the integration constant, and  $\rho_i$ is given by 
\be
&& (\rho_1,\rho_2) = \nl
&& \left( \frac{2}{21} - 6 \left( \bar{\pi}_{\rm UV} - \bar{\Pi}_{\rm UV} \right) + \frac{2}{7} \sqrt{(7 \bar{\pi}_{\rm UV}+2)^2+49 \bar{\Pi}_{\rm UV}^2-\frac{14}{5} (35 \bar{\pi}_{\rm UV}+4) \bar{\Pi}_{\rm UV}}, \right. \nl
  && \left. \quad \frac{2}{21} - 6 \left( \bar{\pi}_{\rm UV} - \bar{\Pi}_{\rm UV} \right) - \frac{2}{7} \sqrt{(7 \bar{\pi}_{\rm UV}+2)^2+49 \bar{\Pi}_{\rm UV}^2-\frac{14}{5} (35 \bar{\pi}_{\rm UV}+4) \bar{\Pi}_{\rm UV}} \right). \label{eq:rho_i}
\ee
Remind that some of asymptotic form of transport coefficients includes $\log \beta$, e.g.
$\beta^2 \log \beta\sim - \delta (\sigma_\beta s^{-\delta})^2 \log s$ in $\tilde{G}_{\pi,\Pi}(\beta)$.
Thus, the $\log$-type transmonomial should be taken into account.

\subsection{Class II $({\frak R}_{4,5})$} \label{sec:trans_early2}
In the case of Class II, the collision and non-collision parts are balanced with each other in the UV limit.
In this case, the collision part has to be proportional to $1/s$, and the leading order of inverse temperature needs to be
\be
\beta(s) \sim \frac{\sigma_\beta}{s^\delta}, \qquad \delta = \frac{1}{\bar{\Delta}},
\ee
where $\sigma_\beta$ is the integration constant.
Eqs.(\ref{eq:dpidtau_fm_UV2}) and (\ref{eq:dPIdtau_fm_UV2}) consequently gives a constraint for $\sigma_\beta$, and the solution is given by
\be
\sigma_{\beta} =
\begin{dcases}
  \left[ \frac{6(70-55\bar{\Delta}+9\bar{\Delta}^2) \theta_0}{35 \bar{\Delta} (3-\bar{\Delta})} \right]^{-\frac{1}{\bar{\Delta}}} & \mbox{for} \quad {\frak R}_4 \\
  \left[ \frac{2(6-\bar{\Delta}) \theta_0}{3 \bar{\Delta}} \right]^{-\frac{1}{\bar{\Delta}}} & \mbox{for} \quad {\frak R}_5
\end{dcases}.
\ee
Substituting this solution into Eqs.(\ref{eq:dpidtau_fm_UV2}) and (\ref{eq:dPIdtau_fm_UV2}) gives the leading term of $(\bar{\pi},\bar{\Pi})$, and it is given by
\be 
\frac{d \bar{\pi}}{ds} &\sim& \frac{\bar{\pi}}{\sigma_\beta^{\bar{\Delta}} \theta_0 s} + \frac{1}{s} \left[ - \frac{2 \pi \bar{\Pi}_{\rm UV}}{25} \cdot \frac{\sigma_{\beta}}{s^\delta}   +  \left( \frac{10}{21} -4 \left( \bar{\Pi}_{\rm UV} - 2\bar{\pi}_{\rm UV}  \right) \right) \bar{\pi} - 4 \left( \frac{1}{5} + \bar{\pi}_{\rm UV} \right) \bar{\Pi} \right], \nl \\
\frac{d \bar{\Pi}}{ds} &\sim& \frac{\bar{\Pi}}{\sigma_\beta^{\bar{\Delta}} \theta_0 s} + \frac{1}{s} \left[  \frac{\pi \bar{\Pi}_{\rm UV}}{10} \cdot \frac{\sigma_\beta}{s^{\delta}}   - \left( \frac{2}{3} +4 \left( 2\bar{\Pi}_{\rm UV} - \bar{\pi}_{\rm UV}  \right)  \right)\bar{\Pi} + 4 \bar{\Pi}_{\rm UV} \bar{\pi} \right].
\ee
The solution is given by
\be
&& \bar{\pi}(s) \sim   -\frac{42 \pi  \bar{\Pi }_{\rm UV} \left[ \left(24 \bar{\Pi}_{\rm UV}+3 \bar{\pi}_{\rm UV}-3 \delta +5\right) \sigma_\beta^{\bar{\Delta}} \theta_0 -3\right] \sigma_\beta^{\bar{\Delta}+1} \theta_0}{{\cal Z}}  \cdot \frac{1}{s^\delta}, \\ \nl
&& \bar{\Pi}(s) \sim  \frac{3 \pi  
  \bar{\Pi }_{\text{UV}} \left[\left(84 \bar{\Pi}_{\text{UV}}-840 \bar{\pi}_{\text{UV}}-105 \delta -50\right) \sigma_\beta^{\bar{\Delta}} \theta_0 - 105\right] \sigma_\beta^{\bar{\Delta}+1} \theta_0}{ 2 {\cal Z}}  \cdot \frac{1}{s^{\delta}},
\ee
where
\be
&& {\cal Z} = \nl
&& 5 \left[-36 \left(560 \bar{\pi }_{\rm UV}+105 \delta -18\right) \bar{\Pi}_{\rm UV}+5 \left(12 \bar{\pi }_{\rm UV}+3 \delta -2\right) \left(168 \bar{\pi }_{\rm UV}+21 \delta +10\right)+10080 \bar{\Pi }_{\rm UV}^2\right] \sigma_\beta ^{2 \bar{\Delta}} \theta_0^2 \nl
&& +150 \left(-126 \bar{\Pi }_{\rm UV}+126 \bar{\pi }_{\rm UV}+21 \delta -2\right) \sigma_\beta^{\bar{\Delta}} \theta_0 + 1575.
\ee
In order to obtain other transmonomials, we solve the linearized equation,
\be
\frac{d \delta \bar{\pi}}{ds} &\sim& \frac{\delta \bar{\pi}}{\sigma_\beta^{\bar{\Delta}} \theta_0 s} + \frac{1}{s} \left[  \left( \frac{10}{21} -4 \left( \bar{\Pi}_{\rm UV} - 2\bar{\pi}_{\rm UV}  \right) \right) \delta \bar{\pi} - 4 \left( \frac{1}{5} + \bar{\pi}_{\rm UV} \right) \delta \bar{\Pi} \right], \\
\frac{d \delta \bar{\Pi}}{ds} &\sim& \frac{\delta \bar{\Pi}}{\sigma_\beta^{\bar{\Delta}} \theta_0 s} - \frac{1}{s} \left[  \left( \frac{2}{3} +4 \left( 2\bar{\Pi}_{\rm UV} - \bar{\pi}_{\rm UV}  \right)  \right) \delta \bar{\Pi} - 4 \bar{\Pi}_{\rm UV} \delta \bar{\pi} \right],
\ee
and obtain the solution as
\be
\delta \bar{\pi}(t) &=&
\begin{dcases}
  \frac{7(15-8 \bar{\Delta})}{105-65 \bar{\Delta}} \cdot \frac{\sigma_1}{s^{\rho_1}} + \frac{\sigma_2}{s^{\rho_2}} & \mbox{for} \quad {\frak R}_4 \\ 
    \frac{1071-205 \bar{\Delta}- 27 \sqrt{49 - \frac{\bar{\Delta}(70-67 \bar{\Delta})}{15}}}{14(90-23 \bar{\Delta})} \cdot \frac{\sigma_1}{s^{\rho_1}} + \frac{\sigma_2}{s^{\rho_2}} & \mbox{for} \quad {\frak R}_5 
\end{dcases}, \nl \\
\delta \bar{\Pi}(s) &=&
\begin{dcases}
  \frac{\sigma_1}{s^{\rho_1}} & \mbox{for} \quad {\frak R}_4 \\ 
  \frac{\sigma_1}{s^{\rho_1}} + \frac{1071-205 \bar{\Delta}- 27 \sqrt{49 - \frac{\bar{\Delta}(70-67 \bar{\Delta})}{15}}}{882-\frac{602}{5} \bar{\Delta}} \cdot \frac{\sigma_2}{s^{\rho_2}}   & \mbox{for} \quad {\frak R}_5 
\end{dcases}, \nl
\ee
where $\sigma_i$ is the integration constant and  $\rho_i$ is given by 
\be
(\rho_1,\rho_2) &=& \left( \frac{2}{21} - 6 ( \bar{\pi}_{\rm UV} - \bar{\Pi}_{\rm UV} )  - \frac{1}{\sigma_\beta^{\bar{\Delta}} \theta_0} + \frac{2}{7} \sqrt{ ( 2 + 7 \bar{\pi}_{\rm UV} )^2 + 49 \bar{\Pi}_{\rm UV}^2 - \frac{14}{5} (35 \bar{\pi}_{\rm UV} + 4 ) \bar{\Pi}_{\rm UV}}, \right. \nl
&& \left.\quad \frac{2}{21} - 6 ( \bar{\pi}_{\rm UV} - \bar{\Pi}_{\rm UV} )  - \frac{1}{\sigma_\beta^{\bar{\Delta}} \theta_0} - \frac{2}{7} \sqrt{ ( 2 + 7 \bar{\pi}_{\rm UV} )^2 + 49 \bar{\Pi}_{\rm UV}^2 - \frac{14}{5} (35 \bar{\pi}_{\rm UV} + 4 ) \bar{\Pi}_{\rm UV}} \right). \label{eq:rho_i2}
\ee
By the similar argument to the case of Class I, the $\log$-type transmonomial exists.

\bibliographystyle{ieeetr}
\bibliography{non_conformal}

\end{document}